\documentclass{jfm}
\usepackage{mathrsfs}
\usepackage{amsmath}
\usepackage{stmaryrd}
\usepackage{bbding}
\usepackage{dcolumn}
\usepackage{amsfonts}
\usepackage{amssymb}
\usepackage{psfrag}
\usepackage{wrapfig}
\usepackage{makeidx}
\usepackage{bm}
\usepackage{epsf}
\usepackage{epsfig}
\usepackage{setspace}
\usepackage{epstopdf}
\usepackage{psfrag}
\usepackage{color}
\usepackage{scalerel}  
\usepackage{hyperref}

\usepackage{graphicx}
\usepackage{caption}
\usepackage{subcaption}
\usepackage{overpic}
                                       
\newcommand\reallywidehat[1]{\arraycolsep=0pt\relax
\begin{array}{c}
	\stretchto{
		\scaleto{
			\scalerel*[\widthof{\ensuremath{#1}}]{\kern-.5pt\bigwedge\kern-.5pt}
			{\rule[-\textheight/2]{1ex}{\textheight}} 
		}{\textheight} %
	}{0.5ex}\\           
	#1\\                 
	\rule{-1ex}{0ex}
\end{array}
}
\newtheorem{rmk}{Remark}

\shorttitle{Non-equilibrium time-relaxation kinetic turbulence model}
\shortauthor{G. Cao, L. Pan, K. Xu, M. Wan and S. Chen}

\title{Non-equilibrium time-relaxation kinetic model for compressible turbulence modeling}

\author
{
    Guiyu Cao\aff{1,2},
    Liang Pan\aff{3},
    Kun Xu\aff{2,4},
    Minping Wan\aff{5,6}
    \corresp{\email{wanmp@sustech.edu.cn}}
    \and Shiyi Chen\aff{1,5,6}
    \corresp{\email{chensy@sustech.edu.cn}}
}
\affiliation
{
    \aff{1}Academy for Advanced Interdisciplinary Studies, Southern University of Science and Technology, Shenzhen, Guangdong 518055, PR China
    \aff{2}Department of Mathematics, Hong Kong University of Science and Technology, Clear Water Bay, Kowloon, Hong Kong
    \aff{3}Laboratory of Mathematics and Complex Systems, School of Mathematical Sciences, Beijing Normal University, Beijing 100875, PR China
    \aff{4}Shenzhen Research Institute, Hong Kong University of Science and Technology, Shenzhen, Guangdong 518057, PR China
    \aff{5}Department of Mechanics and Aerospace Engineering, Southern University of Science and Technology, Shenzhen, Guangdong 518055, PR China
    \aff{6}Guangdong-Hong Kong-Macao Joint Laboratory for Data-Driven Fluid Mechanics and Engineering Applications, Southern University of Science and Technology, Shenzhen, Guangdong 518055, PR China
}

\begin{document}

\maketitle

\begin{abstract}
For the first time, the non-equilibrium time-relaxation kinetic model (NTRKM) is proposed for compressible turbulence modeling on unresolved grids. 
Within the non-equilibrium time-relaxation framework, NTRKM is extended in the form of modified Bhatnagar-Gross-Krook model. 
Based on the first-order Chapman-Enskog expansion, NTRKM connects with the six-variable macroscopic governing equations. 
The first five governing equations correspond to the conservative laws in mass, momentum and total energy, while the sixth equation governs the evolution of unresolved turbulence kinetic energy $K_{utke}$. 
The unknowns in NTRKM, including turbulent relaxation time and source term, 
are determined by essential gradient-type assumption and standard dynamic modeling approach.
Current generalized kinetic model on unresolved grids consequently offers a profound mesoscopic understanding for one-equation subgrid-scale turbulence kinetic energy $K_{sgs}$ model in compressible large eddy simulation.
To solve NTRKM accurately and robustly, a non-equilibrium gas-kinetic scheme is developed, which succeeds the well-established gas-kinetic scheme for simulating Navier-Stokes equations.
Three-dimensional decaying compressible isotropic turbulence and temporal compressible plane mixing layer on unresolved grids are simulated to evaluate the generalized kinetic model and non-equilibrium gas-kinetic scheme.
The performance of key turbulent quantities up to second-order statistics confirms that NTRKM is comparable with the widely-used eddy-viscosity Smagorinsky model (SM) and dynamic Smagorinsky model (DSM).
Specifically, compared with the DNS solution in temporal compressible plane mixing layer, the performance of NTRKM is much closer with DSM and better than SM.
This study provides a workable approach for compressible turbulence modeling on unresolved grids, enriching the understanding of turbulence modeling within the non-equilibrium time-relaxation framework.
\end{abstract}


\section{Introduction}
Turbulence modeling on unresolved grids is an extremely challenging issue in turbulence community for decades \citep{pope2001turbulent}.
With the rapid increasing of computational power, the large eddy simulation (LES) \citep{smagorinsky1963general, lilly1967representation} gradually 
becomes the tractable workhorse for high-fidelity unsteady turbulence simulation.
To simulate turbulent flows on unresolved grids, LES solves the filtered Navier-Stokes (NS) equations with resolvable large-scale 
turbulent structures explicitly, while the unresolved structures are modeled through subgrid-scale (SGS) models \citep{sagaut2006large,garnier2009large}. 
The widely-used eddy-viscosity LES models in physical space mainly include zero-equation Smagorinsky-class models and one-equation SGS turbulence kinetic energy (TKE) models \citep{schumann1975subgrid,yoshizawa1985statistically}.

Smagorinsky model (SM) proposed by \citet{smagorinsky1963general} belongs to zero-equation eddy-viscosity model. 
SM models unresolved turbulent structures through gradient-type assumption between the SGS stress and the resolved velocity gradient.
In practice, SM requires to adjust model coefficients according to the flow types, and suffers the dissipative performance near the wall, as well as SGS effect does not disappear in the laminar flow region \citep{deardorff1970numerical}. 
To deal with those drawbacks of SM, dynamic Smagorinsky model (DSM) \citep{germano1991dynamic,moin1991dynamic,lilly1992proposed,meneveau1996lagrangian} has been proposed for incompressible and compressible turbulence.
DSM allows the modeling coefficients to be computed locally on the basis of dynamic approaches. 
In recent two decades, modified zero-equation eddy-viscosity models have been constructed to be comparable with DSM while still keep the simple algebraic form, such as wall-adapting local eddy-viscosity model \citep{nicoud1999subgrid}, Vreman-type model \citep{vreman2004eddy}, $\sigma$-model \citep{nicoud2011using} and anisotropic minimum-dissipation model \citep{rozema2015minimum}.
By far, zero-equation eddy-viscosity models are the most commonly used class of LES models \citep{moser2021statistical}

Another branch to model the unresolved turbulent structures is deriving and modeling SGS turbulence kinetic energy equation.
\citet{schumann1975subgrid,yoshizawa1985statistically} have pioneered one-equation SGS turbulence kinetic energy models to incorporate history and non-local effects through transport equation of SGS turbulence kinetic energy $K_{sgs}$. 
These one-equation SGS turbulence kinetic energy models can be analogous to the workable one-equation Reynolds averaged Navier-Stokes (RANS) eddy-viscosity models \citep{wilcox1998turbulence}. 
As the grid filter width is taken as the characteristic modeling length scale, only the SGS turbulence kinetic energy equation is required to determine the eddy viscosity in LES models. 
One-equation SGS turbulence kinetic energy models have been extensively applied in incompressible LES \citep{krajnovic2002mixed,de2008localized}, which have shown  better performance in the prediction of turbulent flows.
Compared with the well-established research on compressibility correction for the unresolved TKE equation in the RANS simulation \citep{sarkar1991analysis, wilcox1998turbulence}, there only exists limited work on one-equation SGS turbulence kinetic energy models for compressible LES  \citep{yoshizawa1986statistical,pomraning2002dynamic,chai2012dynamic,cao2021three}.
Considering the compressibility effects can hardly be modeled in zero-equation eddy-viscosity model \citep{garnier2009large}, one-equation SGS turbulence kinetic energy models indeed offer the great promise in modeling compressible turbulent flows.

In the past decades, the gas-kinetic scheme (GKS) based on the Bhatnagar-Gross-Krook (BGK) model \citep{bhatnagar1954model,chapman1970mathematical} 
has been developed systematically for the computations from low speed flows to hypersonic ones \citep{xu2001gas,xu2015direct}.
Based on the time-dependent flux solver, including generalized Riemann problem  solver and gas-kinetic scheme \citep{li2016two,pan2016efficient}, 
a reliable framework was provided for developing the GKS into fourth-order accuracy.
More importantly, the high-order gas-kinetic scheme (HGKS) is as robust as the second-order one and works perfectly from the subsonic to hypersonic viscous heat conducting flows \citep{cao2018physical}.

With the advantage of finite volume GKS and HGKS, they have been naturally implemented in simulating turbulent flows, especially in the compressible regime.
For practical turbulent flows, \citet{hou1996lattice,chen2003extended} pioneered the turbulent relaxation time $\tau_t$ for BGK-type models in turbulence modeling within the equilibrium time-relaxation framework. 
Following the concept of turbulent relaxation time, the second-order GKS/HGKS coupled with $k - \omega$ SST model \citep{jiang2012implicit, righi2016gas, cao2019implicit}, S-A model \citep{pan2016gas}, Vreman-type LES model, and the hybrid RANS-LES method \citep{tan2018gas} have been performed in simulating high-Reynolds number turbulence.
These practical simulations have confirmed the accuracy and robustness of second-order GKS/HGKS coupled with traditional eddy-viscosity model.
In terms of low-Reynolds number turbulent flows, the second-order GKS/HGKS have been directly used as a DNS tool in simulating the canonical benchmarks \citep{fu2006numerical, liao2009gas, kumar2013weno, cao2019three, cao2021three, cao2022high}, such as compressible mixing layer, compressible homogeneous turbulence, turbulent channel flows, etc. 
HGKS shows special advantages in the supersonic turbulence due to its accuracy and super robustness, i.e., the supersonic isotropic turbulence with initial turbulent Mach number $Ma_{t0} = 2.0$ has been simulated successfully \citep{cao2021three}. 
Aiming to conduct the large-scale DNS, a parallel in-house code of HGKS has been developed \citep{cao2022high}.
Large-scale DNS up to $1024^3$ grids shows that the computational cost of HGKS is comparable with the high-order finite difference method \citep{bogey2004family}.

To construct one-equation SGS turbulence kinetic energy model for compressible LES within the time-relaxation framework, \citet{cao2019three,cao2021three} have systematically studied the high-fidelity 
DNS and delicate coarse-graining analysis on decaying compressible isotropic turbulence.
This paper aims to complete the compressible one-equation $K_{sgs}$ model for LES.
Firstly, non-equilibrium time-relaxation kinetic model (NTRKM) is extended in the form of modified BGK model.
NTRKM can offer an mesoscopic understanding for transport equation of the compressible SGS turbulence kinetic energy.
To reasonably maintain the accurate and robust numerical performance of HGKS,  finite volume non-equilibrium gas-kinetic scheme is designed when solving NTRKM.
Comparable with the widely-used eddy-viscosity SM and DSM, the decaying compressible isotropic turbulence  (DCIT) \citep{samtaney2001direct,cao2019three} and temporal compressible plane mixing layer (TCPML) \citep{sandham1991three, vreman1997large, pantano2002study} are simulated to assess current generalized kinetic model and corresponding non-equilibrium gas-kinetic scheme. These two cases are main engines to drive the development of compressible turbulence models.

The organization of this paper is as follows.
In \S \ref{sec:NTRKM_model}, NTRKM for compressible turbulence modeling is presented.
\S \ref{sec:NTRKM_gks} constructs finite volume non-equilibrium gas-kinetic scheme for NTRKM.
{\it{Posteriori}} tests on DCIT and TCPML are conducted in \S \ref{sec:NTRKM_cases}.
Conclusion and discussion are drawn in \S \ref{sec:NTRKM_conclusion}.

\section{Non-equilibrium time-relaxation kinetic model for compressible turbulence modeling}\label{sec:NTRKM_model}
In this section, NTRKM on unresolved grids will be proposed. 
The first-order Chapman-Enskog expansion provides the link between NTRKM and corresponding macroscopic governing equations with six macroscopic variables. 
Phenomenologically, the unknown turbulent relaxation time and source term can be modeled through the gradient-type assumption and dynamic modeling approach.

\subsection{Bhatnagar-Gross-Krook time-relaxation kinetic model}
For molecular transport and collision, the simplification of Boltzmann equation without external force is given by the BGK model \citep{bhatnagar1954model}
\begin{equation} \label{boltamann_bgk_eq}
	\begin{aligned}
		\frac{\partial f}{\partial t} + u_i \frac{\partial f}{\partial x_i} = \frac{g - f}{\tau},
	\end{aligned}
\end{equation}
where $f(\boldsymbol{x},t,\boldsymbol{u},\xi)$ is the number density of molecular at position $\boldsymbol{x} = (x_1,x_2,x_3)^T$ and molecular velocity $\boldsymbol{u} = (u_1, u_2, u_3)^T$ at time $t$ with internal degrees of freedom $\xi$.
The relation between distribution function $f(\boldsymbol{x},t,\boldsymbol{u},\xi)$ and macroscopic variables, such as mass, momentum and total energy can be obtained by taking moments in molecular velocity of $f(\boldsymbol{x},t,\boldsymbol{u},\xi)$ \citep{xu2001gas,xu2015direct}. 
The left hand side of BGK model denotes the free transport process, and the right hand side is the time-relaxation collision term.  
The collision term in BGK model shows simple relaxation process from $f(\boldsymbol{x},t,\boldsymbol{u},\xi)$ to a local equilibrium state $g$, with a molecular relaxation time $\tau$ which is related to the molecular viscosity $\mu$ and heat conduction coefficient $\kappa$ (see Appendix B \citep{xu2015direct}). 
The local equilibrium state $g$ is a Maxwellian distribution
\begin{equation}\label{maxwellian}
	\begin{aligned} 
		g = \rho (\frac{\lambda}{\pi})^{\frac{N + 3}{2}}  e^{- \lambda [(u_i - U_i)^2 + \xi^2]},
	\end{aligned}
\end{equation}
where $\rho$ is the density, $\lambda = m_o / (2k_BT)$ as $m_o$ is the molecular mass, $k_B$ the Boltzmann constant and $T$ the temperature, $U_i$ denotes the macroscopic velocity in $x_i$ direction. 
For three-dimensional equilibrium diatomic gas, the total number of degree of freedom in $\xi$ is $N = 2$, accounting for the two rotational modes $\xi^2 = \xi_1^2 + \xi_2^2$.
The specific heat ratio $\gamma$ is determined by $\gamma = (N + 5)/(N + 3)$. Zeroth-order Chapman-Enskog expansion \citep{chapman1970mathematical} with $f = g$ offers the Euler equations. 
NS equations can be derived with first-order truncation of Chapman-Enskog expansion 
\begin{equation}\label{ce_expansion_eq}
	\begin{aligned} 
		f = g - \tau (\frac{\partial{g}}{\partial t} + u_i \frac{\partial g}{\partial x_i}).
	\end{aligned}
\end{equation}
For Euler and NS equations, the second-order and high-order GKS 
based on BGK model (see equation \eqref{boltamann_bgk_eq}) has been systematically developed \citep{xu2001gas,pan2016efficient}. 
The well-established second-order GKS/HGKS presents its accurate and robust numerical performance from low speed flows to hypersonic ones \citep{xu2015direct}.

\begin{figure}
	\centering
	\begin{overpic}[width=0.4\textwidth]{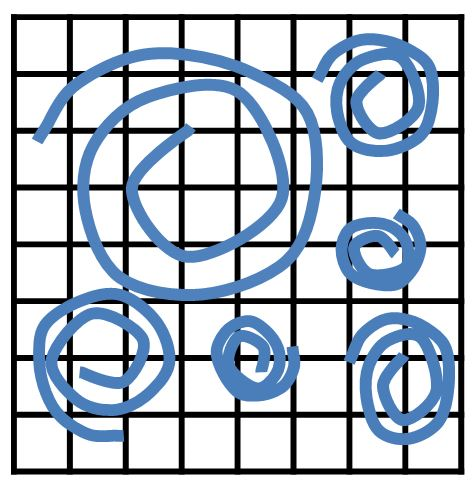}
		\put(-7,92){$(a)$}
	\end{overpic}	
	\quad \qquad
	\begin{overpic}[width=0.402\textwidth]{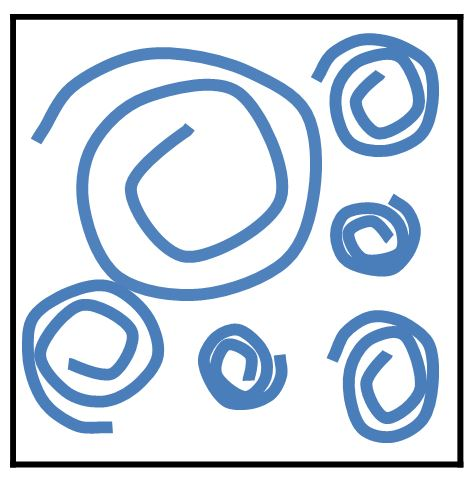}
		\put(-7,92){$(b)$}
	\end{overpic}
	\caption{\label{sketch_unresolved} Illustration of two-dimensional turbulence structure on $(a)$ resolved $8^2$ grids (DNS eliminates turbulence models entirely) and $(b)$ corresponding unresolved $1$ grid (turbulence modeling is required). Black lines form grids and bule lines represent "turbulent eddies".}
\end{figure}
\subsection{Non-equilibrium time-relaxation kinetic model}
Numerically, the unresolved state or resolved state on a 
numerical cell depends on the ratio of spatial-temporal resolution of numerical simulation to the local characteristic scale of flow field. 
Spatial-temporal resolution is mainly determined by the grid resolution and corresponding time step (determined by CFL condition \citep{courant1928partiellen}), as well as the accuracy of numerical scheme.
In terms of characteristic scale, setting the boundary layer as an example, thickness of boundary layer is reasonable characteristic scale in laminar boundary layer \citep{white2006viscous}, while smallest eddy scale as Kolmogorov scale acts as intrinsic characteristic scale in turbulent boundary layer \cite{kim1987turbulence}. 
When the numerical spatial-temporal resolution is not adequate for resolving the local characteristic-scale structures, the turbulence modeling is required.
By contrast, DNS resolves full scales above Kolmogorov scale, eliminating turbulence models entirely.

Figure \ref{sketch_unresolved} illustrates the comparison of two-dimensional turbulence structure in "turbulent eddies" on resolved $8^2$ grids (see figure \ref{sketch_unresolved}$(a)$) and corresponding unresolved $1$ grid (see figure \ref{sketch_unresolved}$(b)$). 
As demonstrated in figure \ref{sketch_unresolved}$(b)$, the unresolved grid means that the grid and corresponding time step is not fine enough to resolve the local smallest turbulence structure with fixed numerical scheme.
Unresolved grids definitely leads to the lost of turbulent information due to inevitable space and time averaging process when updating the macroscopic variables (similar as the averaging process in finite volume scheme \citep{xu2015direct}). 
The key point for turbulence modeling is to model the unresolved turbulence structure through additional non-trivial quantities on unresolved grids, i.e., non-trivial turbulent frequency governed by the stochastic differential equation \citep{pope2001turbulent}.
The trivial quantities are mass, momentum and total energy which are governed by the conservative laws, without contributing non-trivial information to the unresolved turbulent structures.
Subsequently, on unresolved grids (see figure \ref{sketch_unresolved}$(b)$), the non-trivial unresolved turbulence kinetic energy $K_{utke}$ and its quantitative dynamic evolution will be proposed for modeling the unresolved turbulent process.

\begin{figure}
	\centering
	\includegraphics[width=0.6\textwidth]{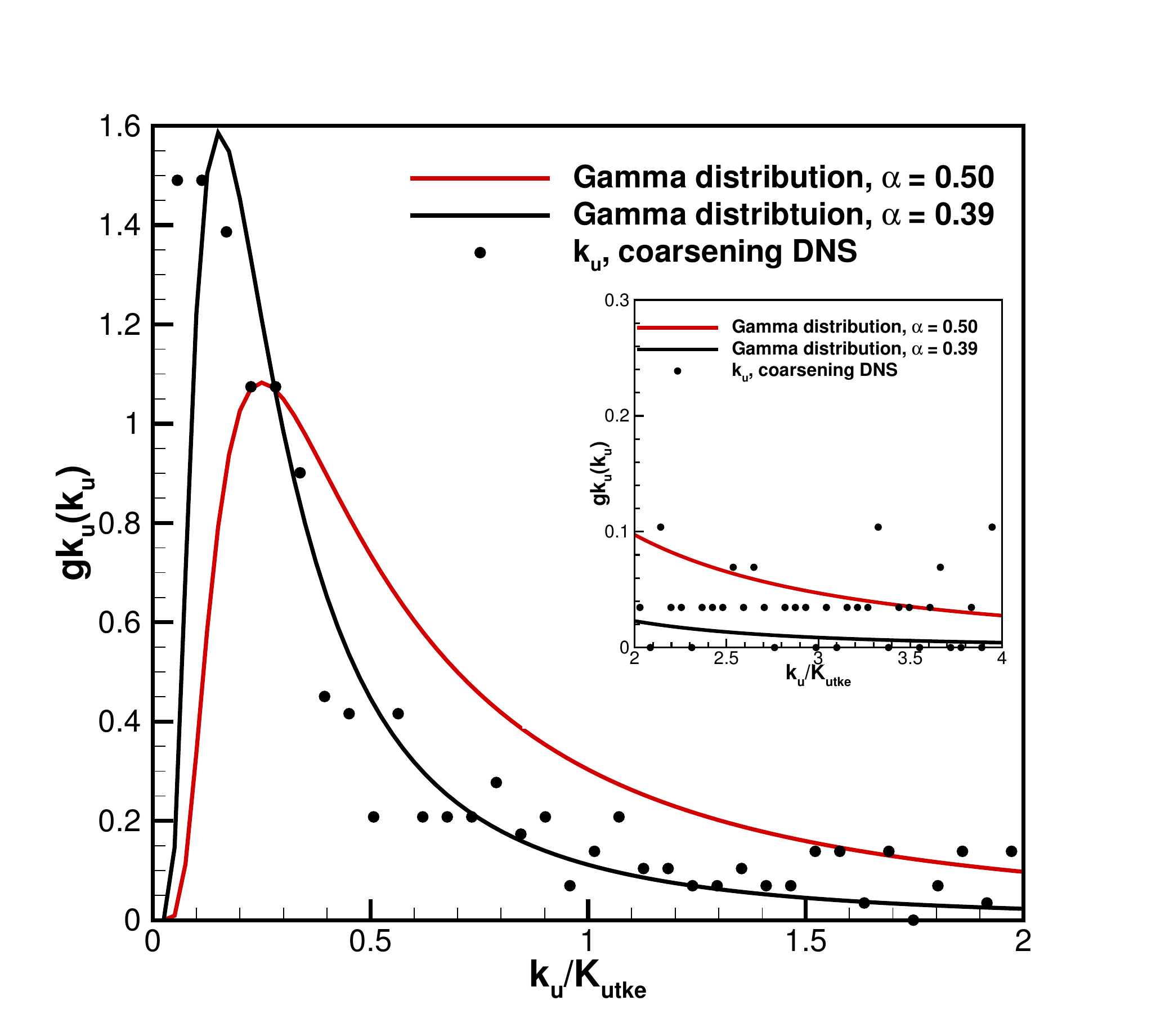}
	\caption{\label{fitting_gamma} $gk_u(k_u)$ from coarsening DNS \citep{cao2021three} at $t /\tau_{to} = 0.5$, $8^3$ resolved grids are coarsened to $1$ unresolved grids with Box filter \citep{vreman1994realizability}. $\tau_{to}$ is the large-eddy-turnover time.}	
\end{figure}
To model turbulence on unresolved grids, the non-equilibrium double time-relaxation kinetic model is proposed in the form of modified BGK model 
\begin{equation}\label{boltamann_bgk_neq}
	\begin{aligned}
		\frac{\partial f}{\partial t} + u_i \frac{\partial f}{\partial x_i}  = \frac{f^{eq} - f}{\tau + \tau_t} + \frac{g - f^{eq}}{\tau_{\ast}} \equiv \frac{f^{eq} - f}{\tau + \tau_t} + Q_s,
	\end{aligned}
\end{equation}
where $f(\boldsymbol{x},t,\boldsymbol{u},\xi, k_u)$ is the generalized number density of molecular on unresolved grids, $k_u$ the sample-space variable corresponding to unresolved TKE $K_{utke}$, $\tau_t$ the turbulent relaxation time, $\tau_{\ast}$  the newly-defined relaxation time from $f^{eq}$ to $g$, and $Q_s$  the secondary relaxation term. 
Turbulence equilibrium state $f^{eq}$ is introduced with Maxwellian distribution $g$ for resolved flow variables and Gamma distribution $gk_u$ for unresolved TKE as 
\begin{equation}\label{intermediate_feq}
	\begin{aligned}
		f^{eq} = g \boldsymbol{\cdot} gk_u = \rho (\frac{\lambda}{\pi})^{\frac{N + 3}{2}}  e^{- \lambda [(u_i - U_i)^2 + \xi^2]} \boldsymbol{\cdot} \frac{1}{\Gamma(\alpha)} (\frac{\alpha}{K_{utke}})^{\alpha} k_u^{\alpha - 1} e^{-\frac{\alpha k_u}{K_{utke}}},\\
	\end{aligned}
\end{equation}
where $\boldsymbol{\cdot}$ denotes the multiplication. 
We assume $gk_u (k_u)$ is the Gamma distribution, with non-negative shape parameter $\alpha$, 
mean $Mean(k_u) = K_{utke}$, variance $Var(k_u) = K_{utke}^2 / \alpha$, where $K_{utke}$ is the total unresolved TKE on unresolved grids. 
In Jayesh-Pope model \citep{pope2001turbulent}, Gamma distribution is the stationary distribution of turbulence frequency for statistically stationary isotropic turbulence. 
In NTRKM, the distribution of $K_{utke}$ on unresolved grids is chosen as Gamma distribution (see equation \eqref{intermediate_feq}), intuitively.
Thence, the double-relaxation process is named as non-equilibrium kinetic model, as the unresolved turbulence information $K_{utke}$ participates in the non-equilibrium relaxation process. 
Validity of Gamma distribution on $K_{utke}$ is conducted through the following coarse-graining process.  
Based on the previous DNS study on $384^3$ grids \citep{cao2021three}, the distribution of $K_{utke}$ through coarsening DNS solution is presented in figure \ref{fitting_gamma}.
$K_{utke}$ is normalized as $1$ in figure \ref{fitting_gamma}.
Gamma distribution with the parameter $\alpha = 0.50$ is the canonical distribution.
Gamma distribution equipped with the parameter $\alpha = 0.39$ fits well with sample-space $k_u$ from coarsening DNS data. 
In thick-tail region of figure \ref{fitting_gamma}, there exists apparent deviation between the fitted Gamma distribution and coarsening DNS data.
This deviation implies that the intense events in the compressible isotropic turbulence \citep{wang2013cascade} are hard to be modeled by Gamma distribution.
The optimal choice of distribution for $K_{utke}$ on unresolved grids still requires to be investigated.
However, the form of $gk_u$ does not affect the evolution of $K_{utke}$ with the subsequent finite volume non-equilibrium gas-kinetic scheme.
Since non-equilibrium gas-kinetic scheme acts as a hydrodynamic solver, only the total $K_{utke}$ gets involved in the updating process instead of $k_u$ in equation \eqref{intermediate_feq}. 
By contrast, when the kinetic solver is applied in updating the distribution function $f(\boldsymbol{x},t,\boldsymbol{u},\xi, k_u)$ directly on unresolved grids, i.e., unified gas-kinetic scheme (UGKS) \citep{xu2010unified}, the form of $gk_u$ will contribute to the evolution of $f(\boldsymbol{x},t,\boldsymbol{u},\xi, k_u)$.
If NTRKM is solved by kinetic solver, the distribution of $K_{utke}$ requires to be modeled much carefully. 
As a starter, turbulence equilibrium state $f^{eq}$ (see equation \eqref{intermediate_feq}) has been proposed for constructing the non-trivial quantity on unresolved grid, namely, depicting the $k_u$ for unresolved "turbulent eddies" as illustrated in figure \ref{sketch_unresolved}$(b)$.

The relation between macroscopic variables as mass $\rho$, momentum $(\rho U_1, \rho U_2, \rho U_3)$, total energy $\rho E$, 
and unresolved turbulence kinetic energy $\rho K_{utke}$ with the generalized distribution function $f(\boldsymbol{x},t,\boldsymbol{u},\xi, k_u)$ on unresolved grids is given by
\begin{equation}\label{macro_vars_vector}
	\begin{aligned} 
		\boldsymbol{Q} 
		= \int \boldsymbol{\psi} f \text{d} \Xi =
		\begin{pmatrix}
			\rho,  \rho U_1, \rho U_2, \rho U_3, \rho E, \rho K_{utke}
		\end{pmatrix}^T,
	\end{aligned}
\end{equation}
where $\boldsymbol{\psi} = (1, u_1, u_2, u_3, \displaystyle\frac{1}{2}(u_1^2 + u_2^2 + u_3^2 + \xi^2) + k_u, k_u)^T$ and $\text{d} \Xi = \text{d}u_1 \text{d}u_2 \text{d}u_3  \text{d}\xi \text{d}k_u$. 
As the $K_{utke}$ is introduced to model unresolved turbulent process quantitatively, one more constraint has to be imposed on current NTRKM to self-consistently determine all unknowns. 
This additional constraint is the $K_{utke}$ relaxation. 
Since only mass, momentum and total energy are conserved during collisions, the compatibility condition for the collision term becomes
\begin{equation}\label{source_vector}
	\begin{aligned}
		\boldsymbol{S} = \int \boldsymbol{\psi} (\frac{f^{eq} - f}{\tau + \tau_t} + Q_s)  \text{d} \Xi = (0,0,0,0,0,S_t)^T.
	\end{aligned}
\end{equation}
Unknown source term $S_t$ in equation \eqref{source_vector} can be modeled through relaxation model.
This relaxation process is analogy to well-established non-equilibrium kinetic model for multi-temperature flows \citep{xu2008multiple} as
\begin{equation}\label{source_time_relaxation}
	\begin{aligned}
		S_t = \frac{\rho (K_{utke}^{eq} - K_{utke})}{\tau_{\ast}}.
	\end{aligned}
\end{equation}
Conceptually, the equilibrium unresolved TKE $K_{utke}^{eq}$ and relaxation time $\tau_{\ast}$ in equation \eqref{source_time_relaxation} 
can be modeled on unresolved grids.
However, these two unknowns require profound {\it{priori}} knowledge and physical understanding of turbulence within the non-equilibrium time-relaxation framework.
As far as the authors know, modeling $K_{utke}^{eq}$ and $\tau_{\ast}$ in equation \eqref{source_time_relaxation} directly is pretty challenging for current stage of turbulence studies. 
To overcome this barrier, a comparison between the $\rho K_{utke}$ equation derived from the first-order Chapman-Enskog expansion on NTRKM and compressible $K_{sgs}$ equation from the compressible LES \citep{cao2021three} will be conducted.
Consequently, source term $S_t$ can be modeled in an alternative standard paradigm.
After modeling $S_t$, the dynamic evolution of non-trivial quantity $K_{utke}$ can be determined by equation \eqref{boltamann_bgk_neq} and equation \eqref{source_vector} quantitatively.

In contrast to the BGK model \eqref{boltamann_bgk_eq}, the right-hand-side collision operator in NTRKM equation \eqref{boltamann_bgk_neq} contains two terms corresponding to two-level collisions on unresolved grids. 
The relaxation process has been extended as $f \to f^{eq} \to g$, and the process from $f^{eq} \to g$ may take a much longer time $\tau_{\ast}$ than that of process from $f \to f^{eq}$ by $(\tau + \tau_t)$. 
On unresolved grids, for the first collision process, the information of unresolved $K_{utke}$ is memorized by the intermediate turbulence equilibrium distribution $f^{eq}$; for the second collision process, $K_{utke}$ is released into resolved kinetic energy and internal energy by $\tau_{\ast}$. 
The total energy assignment in two-level collision process can be classified as intermediate turbulence equilibrium state $\rho E = \rho \boldsymbol{U}^2/2 + \rho e + \rho K_{utke}$, and Maxwellian equilibrium state $\rho E = \rho \boldsymbol{U}^2/2 + \rho e$. 
$\boldsymbol{U} = (U_1, U_2, U_3)^T$ is the resolved macroscopic velocity vector with the definition of $\boldsymbol{U^2} = U_1^2 + U_2^2 + U_3^2$.
$e = (N + 3)RT/2$ is the internal energy in which $R$ is the gas constant. 
As figure \ref{sketch_unresolved}$(b)$ approaches to figure \ref{sketch_unresolved}$(a)$, the unresolved grids approach to resolved ones, meaning that the grid and time step is fine enough to resolve the smallest spatial-temporal characteristic structures.
Under this approaching process, the unresolved turbulent information approaches to disappear, so the non-trivial modeling information on unresolved turbulence structures would be eliminated automatically. 
This limit implies $K_{utke} \to 0$, turbulent relaxation time $\tau_t \to 0$, turbulence equilibrium state $f^{eq} \to g$, and $\tau_{\ast} \to \tau$.
Therefore, this limit leads the NTRKM as equation \eqref{boltamann_bgk_neq} to be consistent with the BGK model as equation \eqref{boltamann_bgk_eq} on resolved grid. 
This asymptotic process also indicates turbulent relaxation time $\tau_t$ depends on grid resolution and unresolved $K_{utke}$, which shed lights on the modeling unknown turbulent relaxation time $\tau_t$.

\subsection{Models for turbulent relaxation time $\tau_t$ and source term $S_t$} \label{subsec:closure}
To overcome the barrier of modeling the turbulent relaxation time $\tau_t$ and source term $S_t$ directly in kinetic model, the corresponding macroscopic governing equations from the NTRKM will be derived by Chapman-Enskog expansion.
NTRKM to macroscopic equations can be regarded as a projection, which essentially bridge the unknowns in NTRKM with the particular terms in macroscopic governing equations.

Using the turbulence equilibrium state denoted as equation \eqref{intermediate_feq}, with the frozen of $K_{utke}$ exchange, the first-order Chapman-Enskog expansion gives
\begin{equation}
	\begin{aligned}
		f = f^{eq} - (\tau + \tau_t) (\frac{\partial{f^{eq}}}{\partial t} + u_i \frac{\partial f^{eq}}{\partial x_i}).
	\end{aligned}
\end{equation}
From which the corresponding macroscopic governing equations in three dimension have been derived for the first time, as shown in Appendix \ref{appA}, namely
\begin{equation}\label{tgks_macro_formula1}
	\begin{aligned}
		\rho_{,t} + (\rho U_j)_{,j} = 0, 
	\end{aligned}
\end{equation}
\begin{equation}\label{tgks_macro_formula2}
	\begin{aligned}
		(\rho U_i)_{,t} + (\rho U_i U_j + p \delta_{ij})_{,j} = (\tau_{ij})_{,j}, 
	\end{aligned}
\end{equation}
\begin{equation}\label{tgks_macro_formula3}
	\begin{aligned}
		(\rho E)_{,t} + ((\rho E + p) U_j)_{,j} = (U_i \tau_{ij}  + q_j)_{,j}, 
	\end{aligned}
\end{equation}
\begin{equation}\label{tgks_macro_formula4}
	\begin{aligned}
		(\rho K_{utke})_{,t} + (\rho K_{utke} U_j)_{,j} = (U_j \tau_{st} + qk_j)_{,j} + S_t,
	\end{aligned}
\end{equation}
where $p$ is pressure related to the resolved temperature 
$p = \rho R T = \rho / (2 \lambda)$, the total energy $\rho E = \rho(\boldsymbol{U}^2 + 3RT + NRT)/2 + \rho K_{utke}$, and $\delta_{ij}$ is the Kronecker symbol. 
The viscous stress term  in equation \eqref{tgks_macro_formula2} is denoted by
\begin{equation}\label{general_stress1}
	\begin{aligned}
		\tau_{ij} &= (\mu + \mu_t)(U_{i,j} + U_{j,i} - \frac{2}{3} U_{k,k} \delta_{ij}) + \eta_t U_{k,k} \delta_{ij} -  \frac{2}{N + 3} \tau_{st} \delta_{ij}, 
	\end{aligned}
\end{equation}
with
\begin{equation}\label{general_stress2}
	\begin{aligned}
		\tau_{st} &= (\tau + \tau_t) S_t,
	\end{aligned}
\end{equation}
where molecular viscosity $\mu = \tau p$, turbulent eddy viscosity $\mu_t = \tau_t p$, turbulent bulk viscosity $\eta_t =  2N (\mu + \mu_t) / [3(N + 3)]$, 
and the last term $\tau_{st}$ results from the source term $S_t$. 
Typically, beyond the filtered compressible momentum equation in NS equations \citep{chai2012dynamic},
the generalized viscous stress $\tau_{ij}$ on unresolved grids in equation \eqref{general_stress1} contains additional terms.
These terms are turbulent bulk viscosity term and term related to the energy interaction from source term $S_t$.
Unresolved turbulence structure contributes to the generalized viscous stress so that $\tau_{ij}$ on unresolved grids deviates from the linear constitutive relation in equation \eqref{general_stress1}.
The heat conduction term in equation \eqref{tgks_macro_formula3} reads
\begin{equation}\label{general_q}
	\begin{aligned}
		q_j = (\kappa + \kappa_t) T_{,j} + qk_{j},
	\end{aligned}
\end{equation}
where molecular thermal conductivity is $\kappa = (N + 5) \tau p R /2$, turbulent thermal conductivity $\kappa_t = (N + 5) \tau_t p R /2$. 
Appendix \ref{appA} shows that Prandtl number $Pr = 1$, and turbulent Prandtl number $Pr_t = 1$. 
When recovering the realistic laminar and turbulent Prandtl number, a similar modification in the energy transport \citep{xu2001gas} should be implemented. 
The modification will be presented briefly in \S \ref{sec:NTRKM_gks}.
As presented in equation \eqref{tgks_macro_formula4} , the term $qk_j$ is related to the governing equation of $K_{utke}$ as
\begin{equation}\label{general_qj}
	\begin{aligned}
		qk_{j} = (\mu + \mu_t) (K_{utke})_{,j}.
	\end{aligned}
\end{equation}
In summary, the first five governing equations in equations \eqref{tgks_macro_formula1}-\eqref{tgks_macro_formula3} correspond to the conservative laws in mass, momentum and total energy with generalized constitutive relationship of stress as equations \eqref{general_stress1}-\eqref{general_stress2} and heat conduction term as equations \eqref{general_q}-\eqref{general_qj}.
While the sixth equation \eqref{tgks_macro_formula4} governs the evolution of $K_{utke}$. 
So far, the unclosed terms in NTRKM are the turbulent relaxation time $\tau_t$, and the source term $S_t$. 
In the following part, $\tau_t$ and $S_t$ will be modeled through the comparison amomg equations \eqref{tgks_macro_formula1}-\eqref{tgks_macro_formula4} and transport equations of the compressible SGS turbulence kinetic energy $K_{sgs}$.

With the Favre filtering process (i.e., $\widetilde{U_i}$ denotes Favre average of $U_i$), the compressible $\overline{\rho} K_{sgs}$ transport equation \citep{cao2021three} 
for compressible LES has been derived as 
\begin{equation}\label{rhok_les}
\left\{
	\begin{aligned}
		&(\overline{\rho} K_{sgs})_{,t} + (\overline{\rho} K_{sgs} \widetilde{U}_j)_{,j} = P_{sgs} - D_{sgs} + \Pi_{sgs} + T_{sgs},  \\
		&P_{sgs} = - \tau_{ij}^{sgs} \widetilde{S}_{ij}, \\
		&D_{sgs} = \overline{\sigma_{ij} U_{i,j}} - \overline{\sigma}_{ij} \widetilde{U}_{i,j}, \\
		&\Pi_{sgs} = \overline{p U_{k,k}} - \overline{p} \widetilde{U}_{k,k}, \\
		&T_{sgs} = [- \frac{1}{2}\overline{\rho} (\widetilde{U_i U_i U_j} - \widetilde{U_i U_i} \tilde{U}_j) + \tau_{ij}^{sgs} \widetilde{U}_i \\
		& \qquad + (\overline{\sigma_{ij} U_i} - \overline{\sigma}_{ij} \widetilde{U}_i) - \overline{\rho} R (\widetilde{T U_j} - \tilde{T} \widetilde{U}_j)]_{,j},
	\end{aligned}
\right.
\end{equation}
where $P_{sgs}$ is the production term, $D_{sgs}$ the total dissipation term, $\Pi_{sgs}$ the pressure-dilation transfer, and last term $T_{sgs}$ the sum of SGS diffusion terms.
In equation \eqref{rhok_les}, SGS turbulence kinetic energy $\overline{\rho} K_{sgs} = \tau_{kk}^{sgs}/2 = \overline{\rho} (\widetilde{U_k U_k} - \widetilde{U}_k \widetilde{U}_k) /2$, SGS stress $\tau_{ij}^{sgs} = \overline{\rho} (\widetilde{U_i U_j} - \widetilde{U}_i \widetilde{U}_j)$, $\widetilde{S}_{ij} = (\widetilde{U}_{i,j} + \widetilde{U}_{j,i})/2$, and $\sigma_{ij} = \mu \big(U_{i,j} + U_{j,i} - 2 U_{k,k} \delta_{ij}/3)$.
The total SGS dissipation rate $D_{sgs}$ can be decomposed into two parts, namely, SGS solenoidal dissipation rate $\varepsilon_s^{sgs}$ and SGS dilational dissipation rate $\varepsilon_d^{sgs}$ as
\begin{equation}\label{dissipation}
\left\{
	\begin{aligned}
		D_{sgs} &= \epsilon_s^{sgs} + \epsilon_d^{sgs}, \\
		\varepsilon_s^{sgs} &= \overline{\mu} (\widetilde{\omega_i \omega_i} - \tilde{\omega}_i \tilde{\omega}_i), \\
		\varepsilon_d^{sgs} &= \frac{4}{3} \overline{\mu}(\widetilde{U_{k,k}^2} - \widetilde{U}_{k,k}^2),
	\end{aligned}
\right.
\end{equation}
where $\omega_i =  \epsilon_{ijk} U_{k,j}$ is the vorticity and $\tilde{\omega}_i =  \epsilon_{ijk} \widetilde{U}_{k,j}$ the filtered one with the permutation symbol $\epsilon_{ijk}$.
Comparing the governing equation of $\rho K_{utke}$ in equation \eqref{tgks_macro_formula4} with the exact $\overline{\rho} K_{sgs}$ equation in equation \eqref{rhok_les}, it is seen that source term $S_t$ in equation \eqref{tgks_macro_formula4} is the net effect of SGS production term, SGS dissipation term and the SGS pressure-dilation transfer as
\begin{equation}\label{source_term}
	\begin{aligned}
		S_t \equiv P_{sgs} - D_{sgs} + \Pi_{sgs}.
	\end{aligned}
\end{equation}
Consequently, current NTRKM as equation \eqref{boltamann_bgk_neq} provides an mesoscopic understanding in transport equation of the compressible SGS turbulence kinetic energy $K_{sgs}$ as equation \eqref{rhok_les}.
The non-trivial quantity $K_{utke}$ is proposed for modeling the unresolved turbulence structures (see figure \ref{sketch_unresolved}$(b)$), and the governing equation of $K_{utke}$ is responsible for the evolution of unresolved turbulent process.
This macroscopic description is consistent with the projection of NTRKM, namely, the double-relaxation kinetic model can be regarded as the mesoscopic understanding of one-equation SGS $K_{sgs}$ model.
We stress that the $K_{utke}$ on unresolved grids will be defaultly regarded as SGS turbulence kinetic energy $K_{sgs}$ on filtered grid.
Especially, in the following modeling and simulations, grid filter width adopts as the effective grid length of control volume (see \S \ref{subsec:dcit} and \S \ref{subsec:tcpml}) in finite-volume numerical scheme, so $K_{sgs}$ can be treated as $K_{utke}$ by default.  
Similarly, the SGS variables are treated equivalently as unresolved variables without special statement.

As presented in equation \eqref{general_stress1}, the connection between eddy viscosity $\mu_t$ and turbulent relaxation time $\tau_t$ is given by $\tau_t = \mu_t/p$. 
As shown in figure \ref{sketch_unresolved}$(b)$, $\tau_t$ can be explained as the relaxation time for the turbulent eddies \citep{chen2003extended}.
The larger turbulent relaxation time originates from the strong non-equilibrium turbulence process, i.e., eddies transport and collision on unresolved grids.
Following the seminal modeling strategy \citep{yoshizawa1986statistical, chai2012dynamic}, turbulent relaxation time $\tau_t$ and SGS stress $\tau_{ij}^{sgs}$ can be modeled as
\begin{equation}\label{mut_tauij_closure1}
	\begin{aligned} 
		\tau_t = \frac{C_s \overline{\Delta} \overline{\rho} K_{utke}^{\frac{1}{2}}}{p}, 
	\end{aligned}
\end{equation}
\begin{equation}\label{mut_tauij_closure2}
	\begin{aligned} 
		\tau_{ij}^{sgs} = - 2 C_s \overline{\Delta} \overline{\rho} K_{utke}^{\frac{1}{2}} \widetilde{S}_{ij}^{\ast} + \frac{2}{3} \overline{\rho} K_{utke} \delta_{ij},
	\end{aligned}
\end{equation}
where $C_s$ is the model coefficient, $\overline{\Delta}$ the grid filter width, $\widetilde{S}_{ij}^{\ast} = \widetilde{S}_{ij} - \widetilde{S}_{kk} \delta_{ij}/3$ the traceless tensor of $\widetilde{S}_{ij}$.
When correcting the total energy transport to recover the realistic turbulent Prandtl number $Pr_t$, the dynamic Prandtl number $Pr_t$ can be modeled as 
\begin{equation}
	q_j = - C_s \overline{\Delta} \overline{\rho} K_{utke}^{\frac{1}{2}} \tilde{T}_{,j} / Pr_t.
\end{equation}
As shown in equation \eqref{mut_tauij_closure1}, with the aid of essential gradient-type assumption, turbulent relaxation time $\tau_t$ has been closed in NTRKM.
After modeling the SGS stress $\tau_{ij}^{sgs}$, the SGS production term in source term in equation \eqref{source_term} is modelled correspondingly.
For the left unknowns in source term $S_t$, the models of SGS dissipation rate and SGS pressure-dilation transfer read \citep{chai2012dynamic}
\begin{equation}\label{chai_sgs_closure1} 
	\begin{aligned}
		\epsilon_s^{sgs} = \frac{C_{\epsilon s} \overline{\rho} K_{utke}^{\frac{3}{2}}}{\overline{\Delta}}, 
	\end{aligned}
\end{equation}
\begin{equation}\label{chai_sgs_closure2} 
	\begin{aligned}
		\epsilon_d^{sgs} = \frac{C_{\epsilon d} \overline{\rho} Ma_k^2 K_{utke}^{\frac{3}{2}}}{\overline{\Delta}}, 
	\end{aligned}
\end{equation}
\begin{equation}\label{chai_sgs_closure3} 
	\begin{aligned}
		\Pi_{sgs} = C_{\Pi} \overline{\Delta}^2 \overline{p_{,j}} (\widetilde{U}_k)_{j,k},
	\end{aligned}
\end{equation}
where $Ma_k^2 = 2K_{utke}/(\gamma R T)$ is the unresolved TKE Mach number. 
In terms of determining the unknown model coefficients, current paper follows the standard dynamic approach \citep{germano1991dynamic,moin1991dynamic,chai2012dynamic}. 
In equations \eqref{mut_tauij_closure1} - \eqref{chai_sgs_closure3}, model coefficients $C_s$, $C_{\Pi}$ and $Pr_t$ can be dynamically computed through Germano identity \citep{germano1991dynamic,moin1991dynamic}.
Additionally, $C_{\epsilon s}$ and $C_{\epsilon d}$ can be obtained by the analogy 
between the grid-filter-level SGS stress and the resolved stress across the test filter level \citep{menon1996high, chai2012dynamic}. 
The detailed derivation of all dynamic model coefficients and necessary remarks are presented in Appendix \ref{appB}.

Turbulent relaxation time $\tau_t$ and source term $S_t$ have been modeled on the basis of equations \eqref{source_term}-\eqref{chai_sgs_closure3} with essential gradient-type assumption and standard dynamic approaches.
In the subsequent section, instead of solving equations \eqref{tgks_macro_formula1}-\eqref{tgks_macro_formula4} with the traditional finite-volume hydrodynamic solver, the NTRKM as equations \eqref{boltamann_bgk_neq} is solved directly with the flux function provided by the time-dependent integral solution in the spirit of well-established gas-kinetic scheme \citep{xu2001gas, xu2015direct}.

\section{Non-equilibrium gas-kinetic scheme for generalized kinetic model}\label{sec:NTRKM_gks}
In this section, to maintain the accurate and robust numerical performance of HGKS \citep{pan2016efficient, cao2018physical}, the finite volume non-equilibrium gas-kinetic scheme is proposed to solve NTRKM.

For finite volume method, the key procedure is updating the macroscopic flow variables inside each control volume through the numerical fluxes. 
Taking moments of the NTRKM as equation \eqref{boltamann_bgk_neq} and integrating with respect to control volume on unresolved grids, the finite volume scheme can be expressed as
\begin{equation}\label{semi}
	\begin{aligned}
		\frac{\text{d}(\boldsymbol{Q}_{ijk})}{\text{d}t}=-\frac{1}{|\Omega_{ijk}|}\sum_{s=1}^6\mathbb{F}_{s}(t) + \boldsymbol{S}_{ijk},
	\end{aligned}
\end{equation}
where $\boldsymbol{Q}_{ijk}$ is the cell averaged macroscopic variables as equation \eqref{macro_vars_vector}, $\boldsymbol{S}_{ijk}$ is cell averaged source term as equation \eqref{source_vector} with $S_t$ modeled through equations \eqref{source_term}-\eqref{chai_sgs_closure3}.
The control volume $\Omega_{ijk}=[(x_1)_i-\Delta x_1/2, (x_1)_i+\Delta x_1/2] \boldsymbol{\cdot} [(x_2)_j - \Delta x_2/2, (x_2)_j+\Delta x_2/2] \boldsymbol{\cdot} [(x_3)_k-\Delta x_3/2, (x_3)_k+\Delta x_3/2]$,
$|\Omega_{ijk}|$ is the volume of $\Omega_{ijk}$ and $\mathbb{F}_{s}(t)$ is the time-dependent numerical flux across the cell interface $\Sigma_{s}$. 
The numerical flux $\mathbb{F}_{s}(t)|_{x_1}$ in $x_1$ direction (at cell interface $(x_1)_{i + 1/2}$) is given as example
\begin{equation}\label{flux_x}
	\begin{aligned}
		\mathbb{F}_{s}(t)|_{x_1}
		&=\iint_{\Sigma_{s}|_{x_1}}
		\boldsymbol{F} (\boldsymbol{Q}) \boldsymbol{\cdot} \boldsymbol{n} \text{d}\sigma \\
		&= \sum_{m,n=1}^2\omega_{mn}
		\int \boldsymbol{\psi} u_1
		f(\boldsymbol{x}_{i+1/2,j_m,k_n},t,\boldsymbol{u},\xi,k_u)\text{d}\Xi\Delta x_2 \Delta x_3,
	\end{aligned}
\end{equation}
where $\boldsymbol{n}$ is the outer normal direction. 
The Gaussian quadrature is used over the cell interface for equation \eqref{flux_x}, where $\omega_{mn}$ is the quadrature weight,
$\boldsymbol{x}_{i+1/2,j_m,k_n}=[(x_1)_{i+1/2},(x_2)_{j_m},(x_3)_{k_n}]^T$, and
$[(x_2)_{j_m},(x_3)_{k_n}]$ is the quadrature point of cell
interface $[(x_2)_j - \Delta x_2/2, (x_2)_j+\Delta x_2/2] \boldsymbol{\cdot} [(x_3)_k-\Delta x_3/2, (x_3)_k+\Delta x_3/2]$. 
When constructing the numerical fluxes, the secondary relaxation term $Q_s$ in 
equation \eqref{boltamann_bgk_neq} is not considered, and the effect of $Q_s$ is taken into account as the source term in equation \eqref{semi}.
The gas distribution function $f(\boldsymbol{x}_{i+1/2,j_m,k_n},t,\boldsymbol{u},\xi,k_u)$ in the local coordinate can be obtained by the integral solution of equation \eqref{boltamann_bgk_neq} as 
\begin{equation} \label{formal_solution}
	\begin{aligned}
		f(\boldsymbol{x}_{i+1/2,j_m,k_n},t,\boldsymbol{u},\xi,k_u) &= \frac{1}{(\tau + \tau_t)}\int_0^t
		f^{eq}(\boldsymbol{x}',t',\boldsymbol{u},\xi, k_u)e^{-(t-t')/(\tau + \tau_t)}\text{d}t'\\
		&+e^{-t/(\tau + \tau_t)}f_0(-\boldsymbol{u}t,\xi, k_u),
	\end{aligned}	
\end{equation}
where $\boldsymbol{x}'=\boldsymbol{x}_{i+1/2,j_m,k_n}-\boldsymbol{u}(t-t')$
is the trajectory of molecular on unresolved grids, $f_0$ the initial gas
distribution function, and $f^{eq}$ the corresponding turbulence equilibrium state in the form of equation \eqref{intermediate_feq}. 
Along the line of GKS \citep{xu2001gas,cao2018physical}, for the multi-dimensional kinetic solver, $f^{eq}$ and $f_0$ can be constructed as
\begin{equation}\label{const_feq}
	\begin{aligned}
		f^{eq} = f^{eq}_0(1 + \overline{a}_1 x_1 + \overline{a}_2 x_2 + \overline{a}_3 x_3 + \overline{A} t),
	\end{aligned}
\end{equation}
and
\begin{equation}\label{const_f0}
	\begin{aligned}
		f_0 =
		\begin{cases}
			f^{eq}_l [1 +  (a_1^l x_1 + a_2^l x_2 + a_3^l x_3) - (\tau + \tau_t) (a_1^l u_1 + a_2^l u_2 + a_3^l u_3 + A_l)], &x \leq 0, \\
			f^{eq}_r [1 +  (a_1^r x_1 + a_2^r x_2 + a_3^r x_3) - (\tau + \tau_t) (a_1^r u_1 + a_2^r u_2 + a_3^r  u_3 + A_r)], &x > 0,
		\end{cases}
	\end{aligned}
\end{equation}
where $f^{eq}_l$ and $f^{eq}_r$ are the initial gas distribution functions on both sides of a cell interface $\Sigma_{s}$.
$f^{eq}_0$ is the initial turbulence equilibrium state located at the cell interface, which can be determined through the compatibility condition
\begin{equation}
\begin{aligned}
	\int \boldsymbol{\psi} f^{eq}_0 \text{d}\Xi= \int_{u_1>0} \boldsymbol{\psi} f^{eq}_l \text{d}\Xi + \int_{u_1<0} \boldsymbol{\psi} f^{eq}_r \text{d}\Xi.
\end{aligned}
\end{equation}
Substituting $f^{eq}$ (see equation \eqref{const_feq}) and $f_0$ (see equation \eqref{const_f0}) into equation \eqref{formal_solution}, the time-dependent gas distribution function at the Gaussian point is evaluated as
\begin{equation}\label{formalsolution_neq}
	\begin{aligned}
	&f(\boldsymbol{x}_{i+1/2,j_m,k_n},t,\boldsymbol{u},\xi,k_u) = (1 - e^{-t/(\tau + \tau_t)}) f^{eq}_0  \\
	&+ [(t + \tau + \tau_t) e^{-t\tau} - \tau] (\overline{a}_1 u_1 + \overline{a}_2 u_2 + \overline{a}_3 u_3) f^{eq}_0 	\\
	&+ [t - (\tau + \tau_t) + (\tau + \tau_t) e^{-t (\tau + \tau_t)}]  \overline{A} f^{eq}_0   \\
	&+ e^{-t/(\tau + \tau_t)} f^{eq}_l [1 - (\tau + \tau_t + t) (a_1^l u_1 + a_2^l u_2 + a_3^l u_3) - (\tau + \tau_t) A_l] H(u) \\
	&+ e^{-t/(\tau + \tau_t)} f^{eq}_r [1 - (\tau + \tau_t + t) (a_1^r u_1 + a_2^r u_2 + a_3^r u_3) - (\tau + \tau_t) A_r] (1 - H(u)).
\end{aligned}
\end{equation}
With the relation of macroscopic variables and turbulence equilibrium distribution function $f^{eq}$, the spatial mesoscopic coefficients $\overline{a}_1$, $a_1^l$, $\cdots$, $a_3^l$, $a_3^r$ and temporal mesoscopic coefficients $\overline{A}$, $A_l$, $A_r$ in equation \eqref{formalsolution_neq} can be determined and details are presented in Appendix \ref{appC}. 
Equation \eqref {formalsolution_neq} provides a gas evolution process from kinetic scale to hydrodynamic scale on unresolved grids, where both inviscid and viscous fluxes are recovered from a time-dependent and multi-dimensional gas distribution function at a cell interface.
This flux function couples the inviscid and all dissipative terms \citep{xu2001gas,cao2018physical}, 
and has advantages in comparison with traditional hydrodynamic solver in which the Riemann solver and central difference are used for the inviscid and viscous terms.
For Prandtl number fix, both the laminar Prandtl number $Pr$ and turbulent Prandtl number $Pr_t$ should be taken into consideration. 
Total energy flux $\boldsymbol{F}{(\rho E)}$ in equation \eqref{flux_x} should be modified as 
$\boldsymbol{F}^{new} {(\rho E)}= \boldsymbol{F}{(\rho E)} + \{(\mu Pr_t + \mu_t Pr)/[Pr Pr_t (\mu + \mu_t)] - 1\} q$, 
where the time-dependent heat flux can be evaluated precisely by 
$q = \int (u - U) \{[(u_i - U_i)^2 + \xi^2]/2 + k_u\} f d\Xi$.

The second-order accuracy in time can be achieved by one step integration, with the time-dependent  kinetic flux as equation \eqref{formalsolution_neq}. 
To achieve high-order accuracy in space and time, the fifth-order WENO-Z spatial reconstruction \citep{castro2011high} and two-stage fourth-order time discretization \citep{li2016two,pan2016efficient} are implemented. 
The characteristic reconstruction is applied to improve the robustness for compressible flows with strong discontinuities (i.e., DCIT) \citep{pan2020high}.
The characteristic variables are defined as $\boldsymbol{Q}_c = \boldsymbol{R}^{-1} \boldsymbol{Q}$, where $\boldsymbol{R}$ is 
the right eigenmatrix of Jacobian matrix at Gaussian quadrature point and details are given in Appendix \ref{appD}. 
When dealing with compressible flows without strong discontinuities, such as TCPML, the linear WENO spatial reconstruction based on conservative variables is adopted.
For source term in equation \eqref{semi}, the one-step forward Euler method is applied in two-stage updating process to guarantee the robustness.
Therefore, the finite volume non-equilibrium gas-kinetic scheme has been constructed with the second-order kinetic flux, fifth-order 
WENO-Z reconstruction, two-stage fourth-order time discretization and one-step forward Euler method for source term.
The current non-equilibrium gas-kinetic scheme has been well implemented in the in-house platform for turbulence 
simulation \citep{cao2022high}, and the {\it{posteriori}} tests on compressible turbulent flows will be presented in the following section.

\section{{\it{\textbf{Posteriori}}} tests} \label{sec:NTRKM_cases}
In this section, the decaying compressible isotropic turbulence \citep{samtaney2001direct,cao2019three} and 
temporal compressible plane mixing layer \citep{sandham1991three, vreman1997large, pantano2002study} are regarded as cornerstones to assess NTRKM and non-equilibrium gas-kinetic scheme.

\subsection{Decaying compressible isotropic turbulence}\label{subsec:dcit}
DCIT \citep{samtaney2001direct} is the building-block case to demonstrate the performance of modeling on compressible turbulence.
For the flow with discontinuities, we have
\begin{equation}
\begin{aligned}
	\tau + \tau_t = \frac{\mu + \mu_t}{p} + C_{num} \displaystyle|\frac{p_l-p_r}{p_l+p_r}| \text{d} t,
\end{aligned}
\end{equation}
where $p$ is the pressure at the cell interface, $p_l$ and $p_r$ the pressure on the left and right sides of the cell interface. 
$\text{d} t$ is the time step, and a fixed $C_{num} = 2.5$.
The reason for including artificial dissipation through the additional term in the molecular relaxation 
time $\tau$ and the turbulent relaxation time $\tau_t$ is to improves the numerical stability. 
As the earlier remark states, the grid filter width adopts as the grid length of control volume for 
DCIT, i.e., $\overline{\Delta} = \Delta x_1 = \Delta x_2 =\Delta x_3$ on equivalent spaced grids. 
Additionally, test filter width $\widehat{\overline\Delta}$ is set to twice the grid filter 
width $\overline{\Delta}$ when determining the dynamic model coefficients, 
namely $\widehat{\overline\Delta} = 2 \overline{\Delta}$.  DNS for  DCIT has been 
well studied using HGKS systematically \citep{cao2019three, cao2021three}.
In this section, following previous DNS set-up, LES on unresolved grids will be conducted directly.

\begin{table}
	\begin{center}
		\centering
		\begin{tabular}{ccccc}
			Case      &\text{grid size}    &$K_0$        &$I_0$   &$\kappa_{max} \eta_0$  \\
			DNS     &$512^3$             &0.5055       &0            &3.6       \\
			R$_1$     &$128^3$             &0.4931       &0.0151       &0.90     
		\end{tabular}
	\caption{\label{re72gridtable} Numerical parameters for DCIT.}
	\end{center}
\end{table}
In the computation, the initial Taylor microscale Reynolds number is $Re_{\lambda 0} = 72$ and the initial turbulent Mach number is fixed at $Ma_{t0} = 0.6$. 
The detailed initial conditions are set as previous work \citep{cao2019three}, and the periodic boundary condition for six macroscopic variables is used. 
A three-dimensional solenoidal random initial velocity field can be generated by a specified spectrum as
\begin{equation}\label{spectral}
	\begin{aligned}
		E(\kappa) = A_0 \kappa^4 \exp (- 2 \kappa^2/\kappa_0^2), 
	\end{aligned}
\end{equation}
with the fixed $A_0 = 0.00013$ and $ \kappa_0 = 8$.
After generating the initial velocity field on $512^3$ resolved grids, the filtered velocity fields can be obtained on unresolved grids, i.e., filtered flow fields on $128^3$ grids.
When filtering velocity field, the positive definite kernel of Box filter is adopted to guarantee the positive unresolved 
TKE \citep{vreman1994realizability}, thus the initial pointwise $K_{utke 0}$ on unresolved grids can be obtained as the initial condition for equation \eqref{tgks_macro_formula4}.
Table \ref{re72gridtable} shows the numerical parameters for DCIT of DNS and R$_1$, where $\kappa_{max}$ is maximum resolved wave number and $\eta_0$ is the Kolmogorov length scale.
$\left\langle K_0 \right\rangle$ is the initial ensemble resolved TKE in which $\left\langle \cdot \right\rangle$ denotes the spatial average on the whole computational domain. 
Turbulence intensity $I_0$ denotes the ratio of initial ensemble unresolved $\left\langle K_{utke 0} \right\rangle$ to the 
initial ensemble resolved TKE as $I_0 = \left\langle K_{utke 0}\right\rangle / \left\langle K_0 \right\rangle$.
Table \ref{re72gridtable} shows that the grid resolution meets the DNS criterion $\kappa_{max} \eta_0 \geq  2.71$ for DCIT \citep{cao2019three}.
Obviously, the grid resolution of R$_1$ is not adequate for DNS, which is regarded as compressible LES on unresolved grids.
The representative key statistical quantities, including the resolved root-mean-square density fluctuation $\rho_{rms}$
and resolved turbulence kinetic energy $K$, are given by
\begin{equation}\label{rho_k_prms1}
	\begin{aligned}
		\rho_{rms}= \left\langle (\rho -  \left\langle  \rho \right\rangle)^2 \right\rangle^{\frac{1}{2}}, 
	\end{aligned}
\end{equation}
\begin{equation}\label{rho_k_prms2}
	\begin{aligned}
		K= \frac{1}{2} \rho \boldsymbol{U}^2.
	\end{aligned}
\end{equation}
The ensemble budget of resolved $K$ is computed, which can be described approximately by \citep{sarkar1991analysis}
\begin{equation}\label{dkdt1}
	\begin{aligned}
		\frac{\text{d} \left\langle K \right\rangle}{\text{d}t} =  - \left\langle \varepsilon \right \rangle + \left\langle p U_{k,k} \right\rangle,
	\end{aligned}
\end{equation}
\begin{equation}\label{dkdt2}
	\begin{aligned}
		\varepsilon =\varepsilon_s + \varepsilon_d, 
	\end{aligned}
\end{equation}
where $\varepsilon_s = \mu \omega_i \omega_i$ is the resolved solenoidal dissipation rate, $\displaystyle\varepsilon_d= 4\mu U_{k,k}^2/3$ the resolved dilational dissipation rate without considering bulk viscosity, $p U_{k,k} $ the resolved pressure-dilation transfer.

\begin{figure}
	\centering
	\begin{overpic}[width=0.495\textwidth]{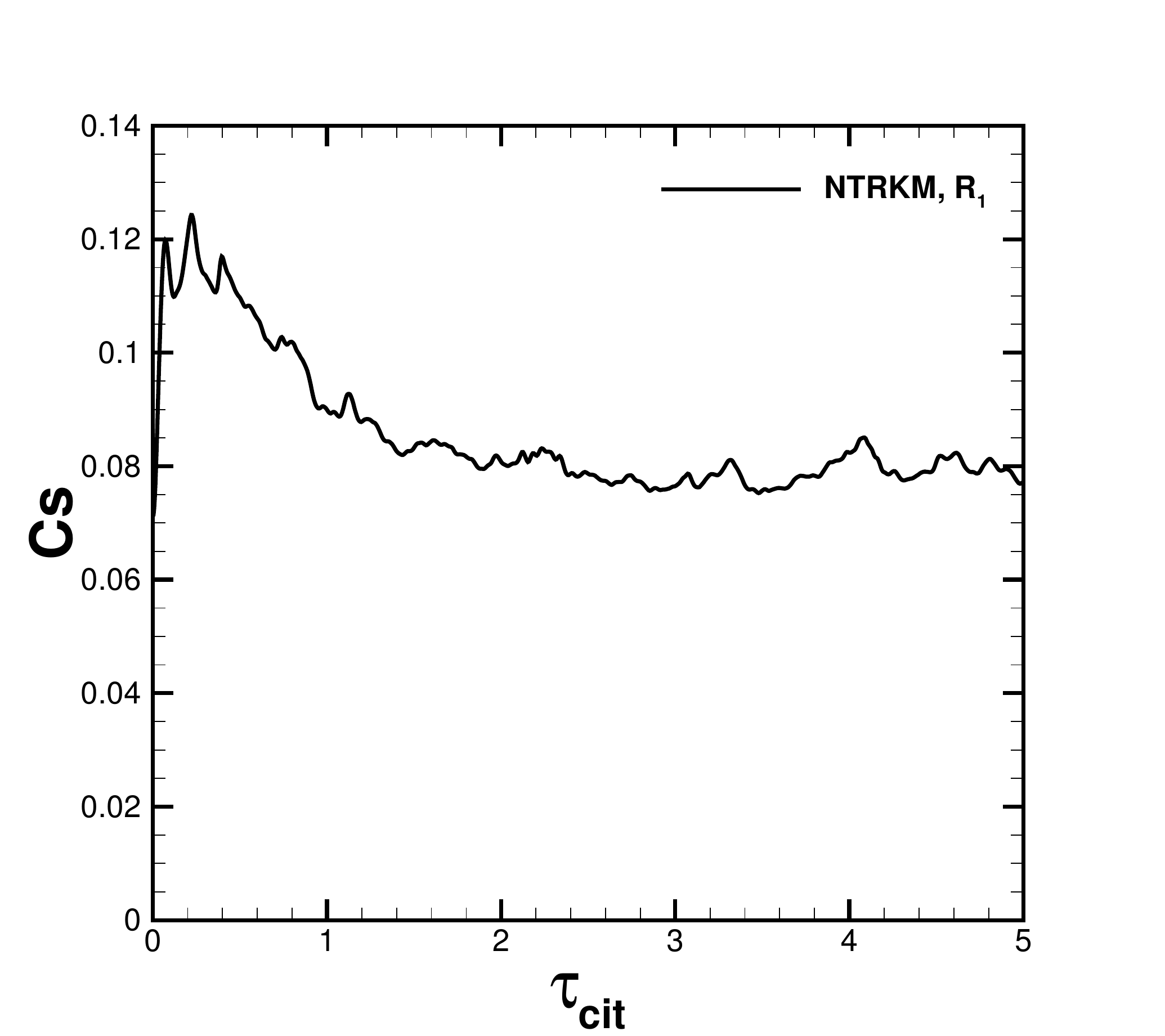}
		\put(-2,75){$(a)$}
	\end{overpic}	
    \begin{overpic}[width=0.495\textwidth]{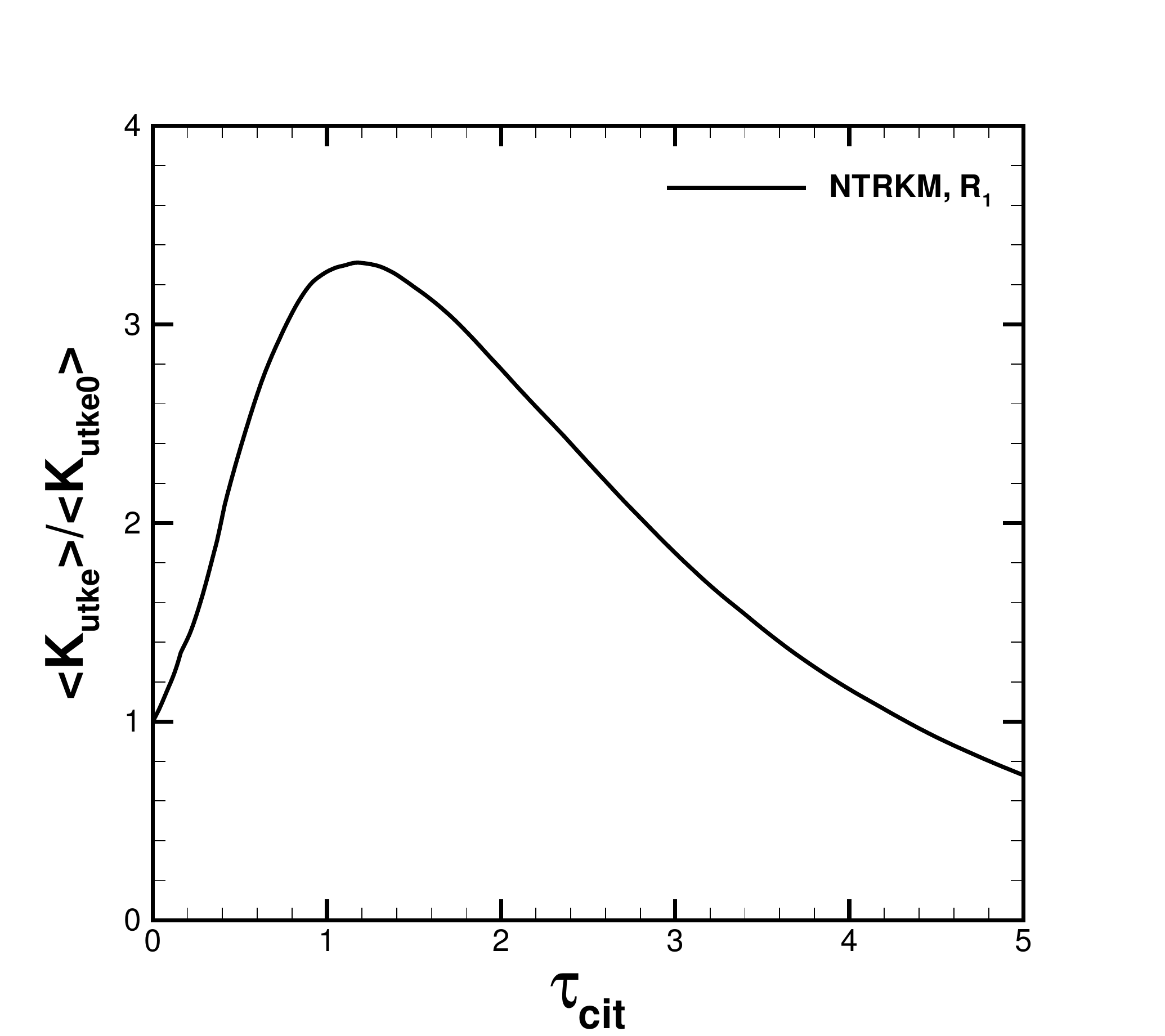}
    	\put(-2,75){$(b)$}
    \end{overpic}
\caption{\label{ntrkm_kutke_cs} Time history of $(a)$ dynamic coefficient $C_s$ and $(b)$ normalized ensemble unresolved $\left\langle K_{utke}\right\rangle / \left\langle K_{utke0}\right\rangle$ for case R$_1$ with NTRKM.}
\end{figure}
Figure \ref{ntrkm_kutke_cs} shows the time history of dynamic coefficient $C_s$ as equation \eqref{mut_tauij_closure1} for turbulent relaxation time $\tau_t$
and normalized ensemble $\left\langle K_{utke}\right\rangle / \left\langle K_{utke 0}\right\rangle$.
$\tau_{cit}=t/\tau_{to}$ is the normalized time and $\tau_{to}$ is the large-eddy-turnover time \citep{cao2019three}.
Firstly, model coefficient $C_s$ is presented to validate the implementation of dynamic modeling approach as Appendix \ref{appB_2}.
The empirical model coefficient $C_s$ is recommended as a fixed value $0.05$  \citep{yoshizawa1985statistically}.
Figure \ref{ntrkm_kutke_cs}$(a)$ shows that the dynamic coefficient $C_s$ in NTRKM fluctuates between $[0.06, 0.12]$ for case R$_1$. 
Current dynamic approach shows that $C_s$ reasonably depends on the grid resolution and the evolution of flow fields.
In figure \ref{ntrkm_kutke_cs}$(b)$, the normalized ensemble unresolved $\left\langle K_{utke}\right\rangle / \left\langle K_{utke 0}\right\rangle$ increases approximately within $\tau_{cit} \leq 1.5$ and decrease consecutively, which behaves similarly as previous literature \citep{chai2012dynamic}.
In equation \eqref{rhok_les}, the SGS production term represents the inter-scale transfer associated with the interaction of resolved and unresolved scales.
The SGS dissipation terms act as the sink of $K_{utke}$ in source term $S_t$.
The evolution of $\left\langle K_{utke}\right\rangle / \left\langle K_{utke 0}\right\rangle$ implies that the ensemble forward resolved energy cascade dominates at the early stage, and then the SGS dissipation terms dominate.
Figure \ref{ntrkm_kutke_cs}$(b)$ indicates that the intrinsic equilibrium assumption on $K_{utke}$, such as $\left\langle S_t\right\rangle \approx 0$ 
for zero-equation eddy-viscosity LES models 
\citep{lilly1967representation, germano1991dynamic, moin1991dynamic} is not valid, which confirms that the evolution of $K_{utke}$ on 
unresolved grids is crucial for compressible LES modeling \citep{yoshizawa1985statistically, chai2012dynamic}.

\begin{figure}
	\centering
	\begin{overpic}[width=0.495\textwidth]{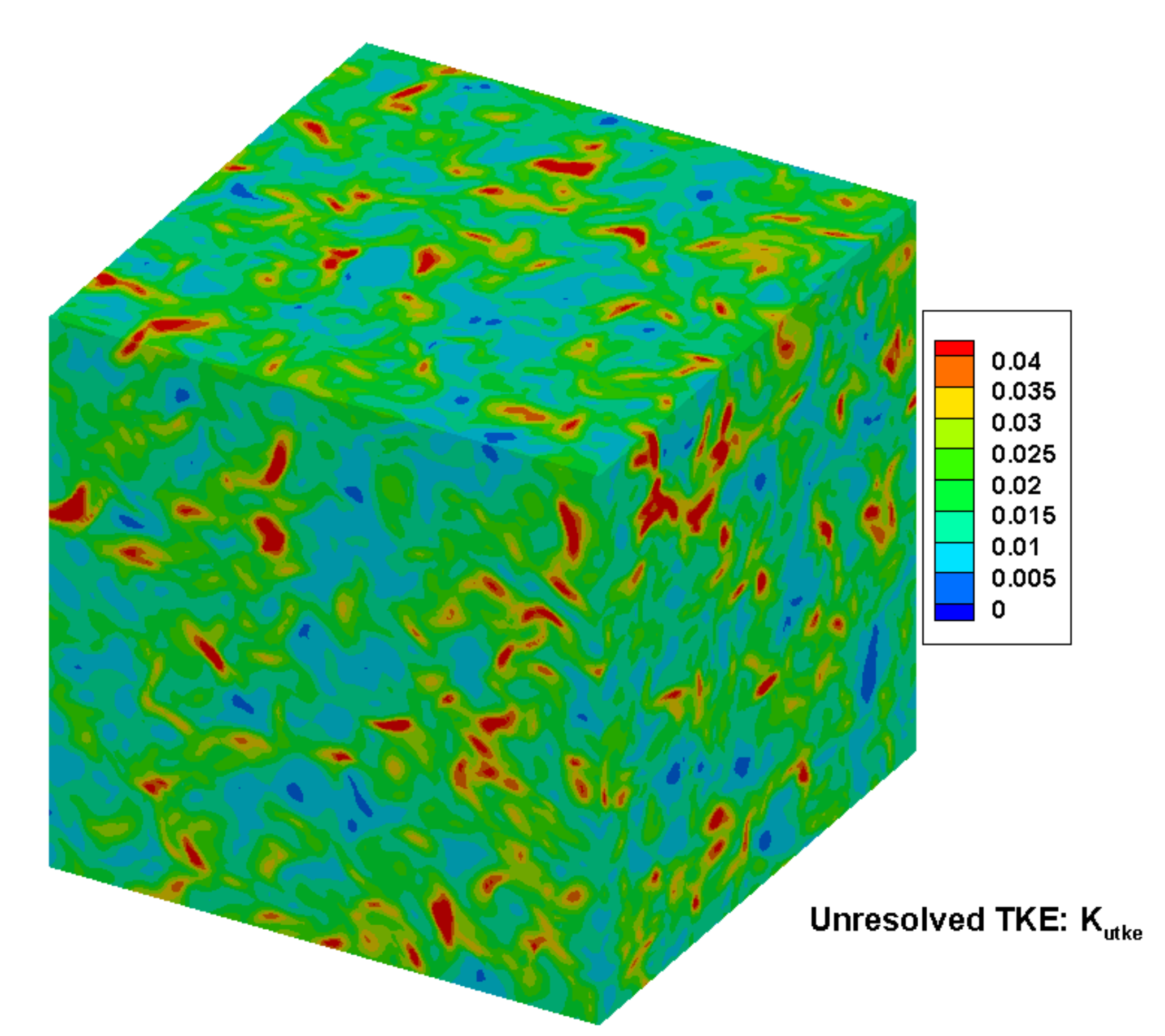}
		\put(-2,75){$(a)$}
	\end{overpic}	
	\begin{overpic}[width=0.495\textwidth]{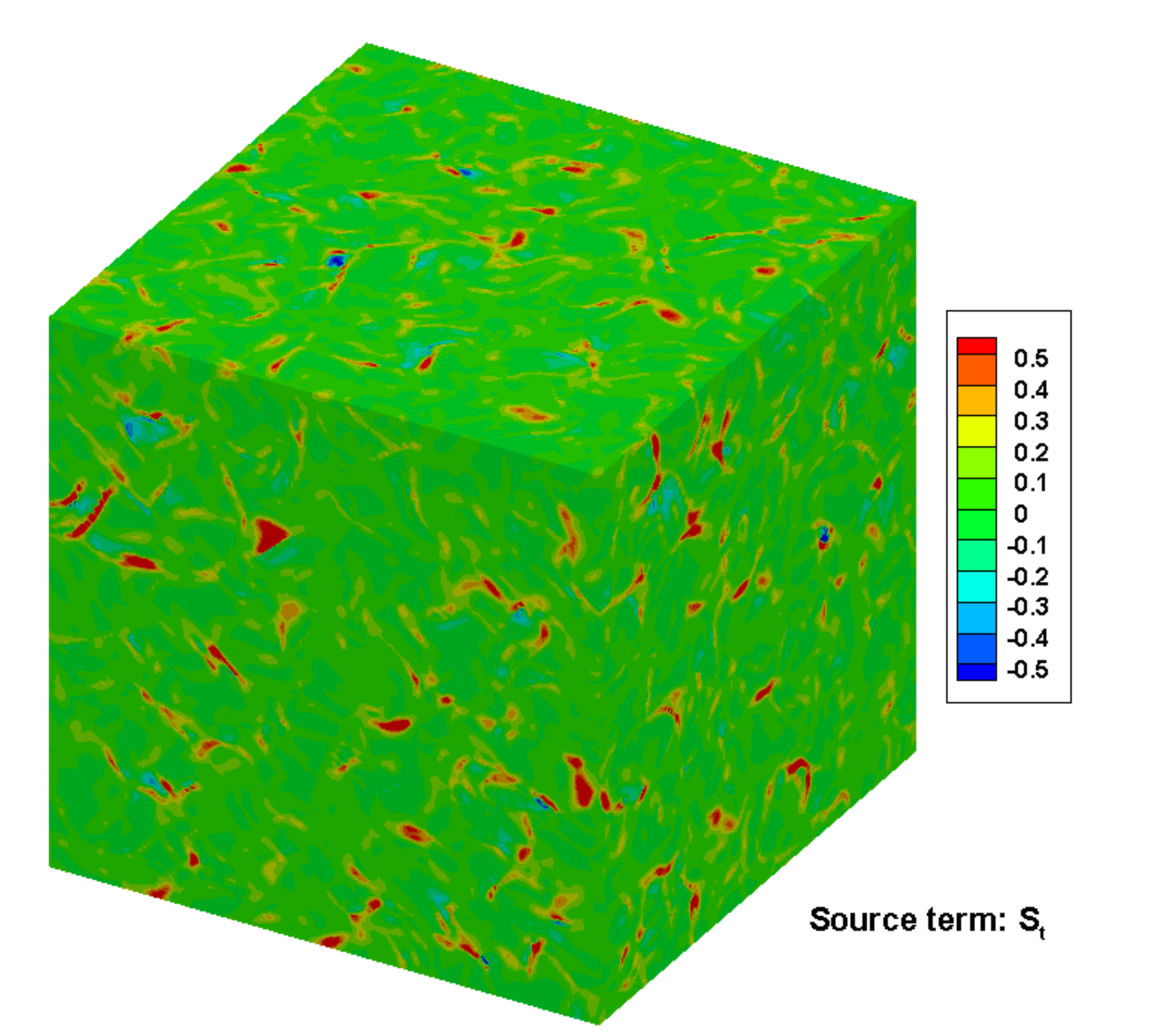}
		\put(-2,75){$(b)$}
	\end{overpic}
	\begin{overpic}[width=0.495\textwidth]{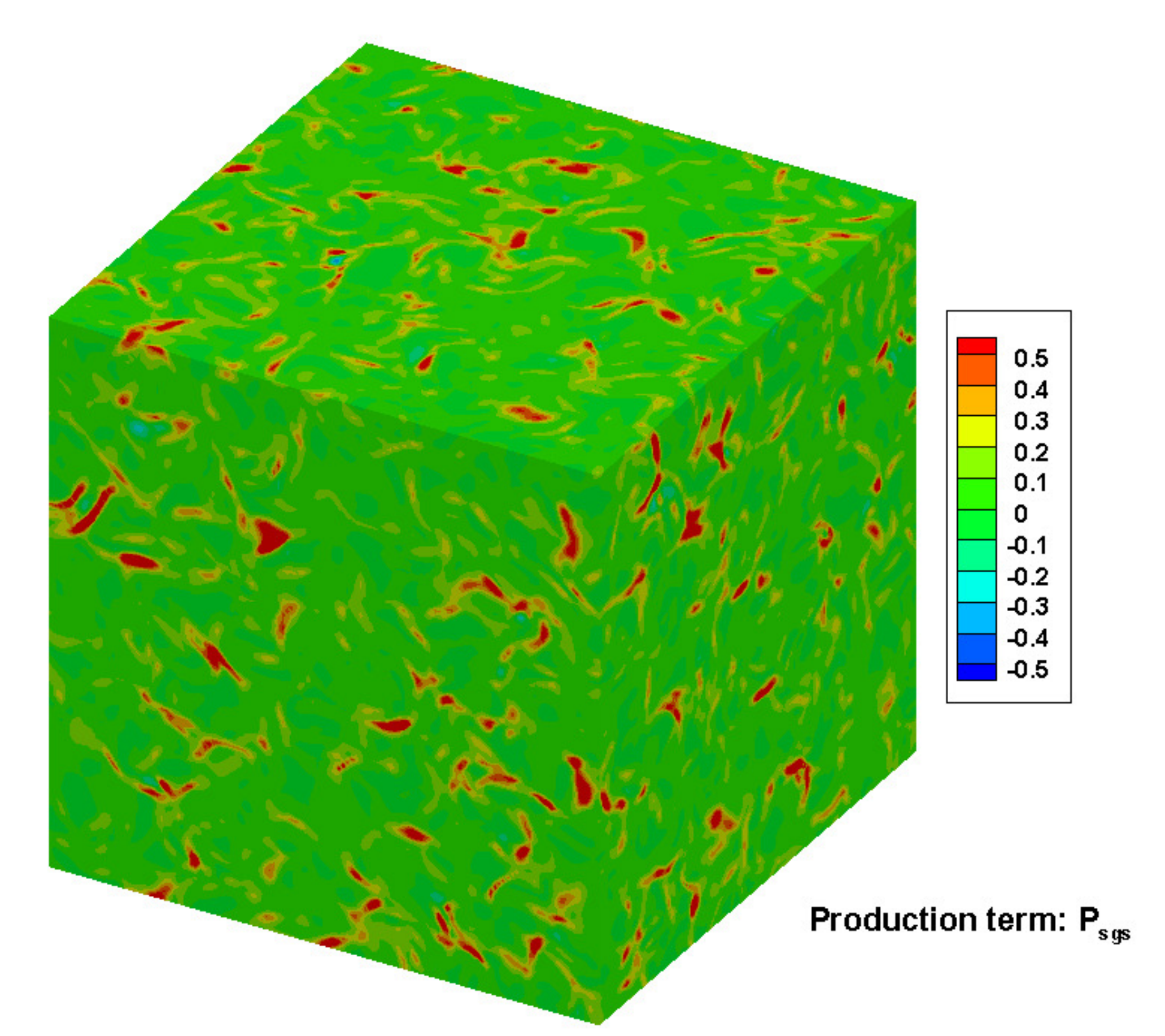}
		\put(-2,75){$(c)$}
	\end{overpic}	
	\begin{overpic}[width=0.495\textwidth]{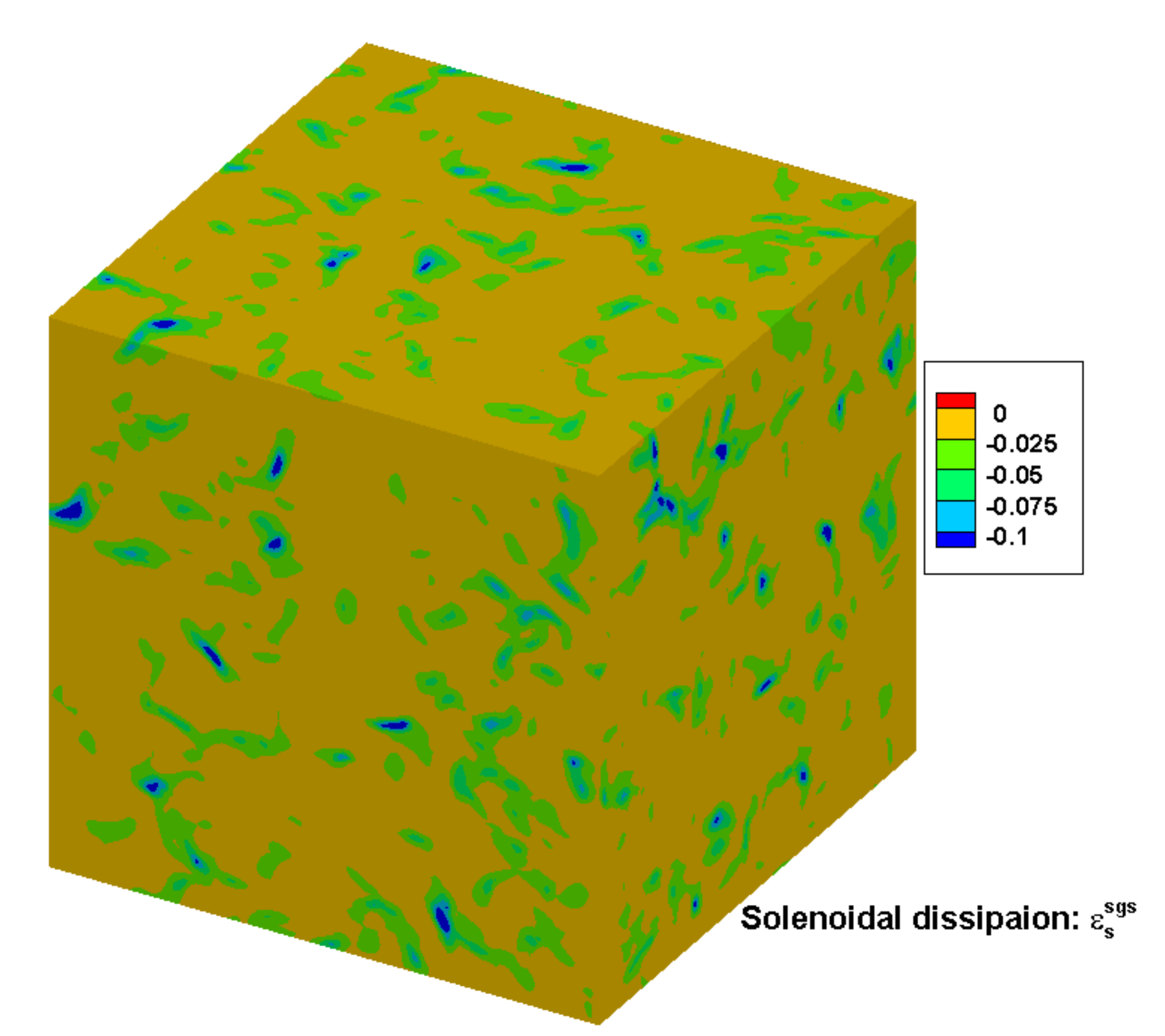}
		\put(-2,75){$(d)$}
	\end{overpic}
	\begin{overpic}[width=0.495\textwidth]{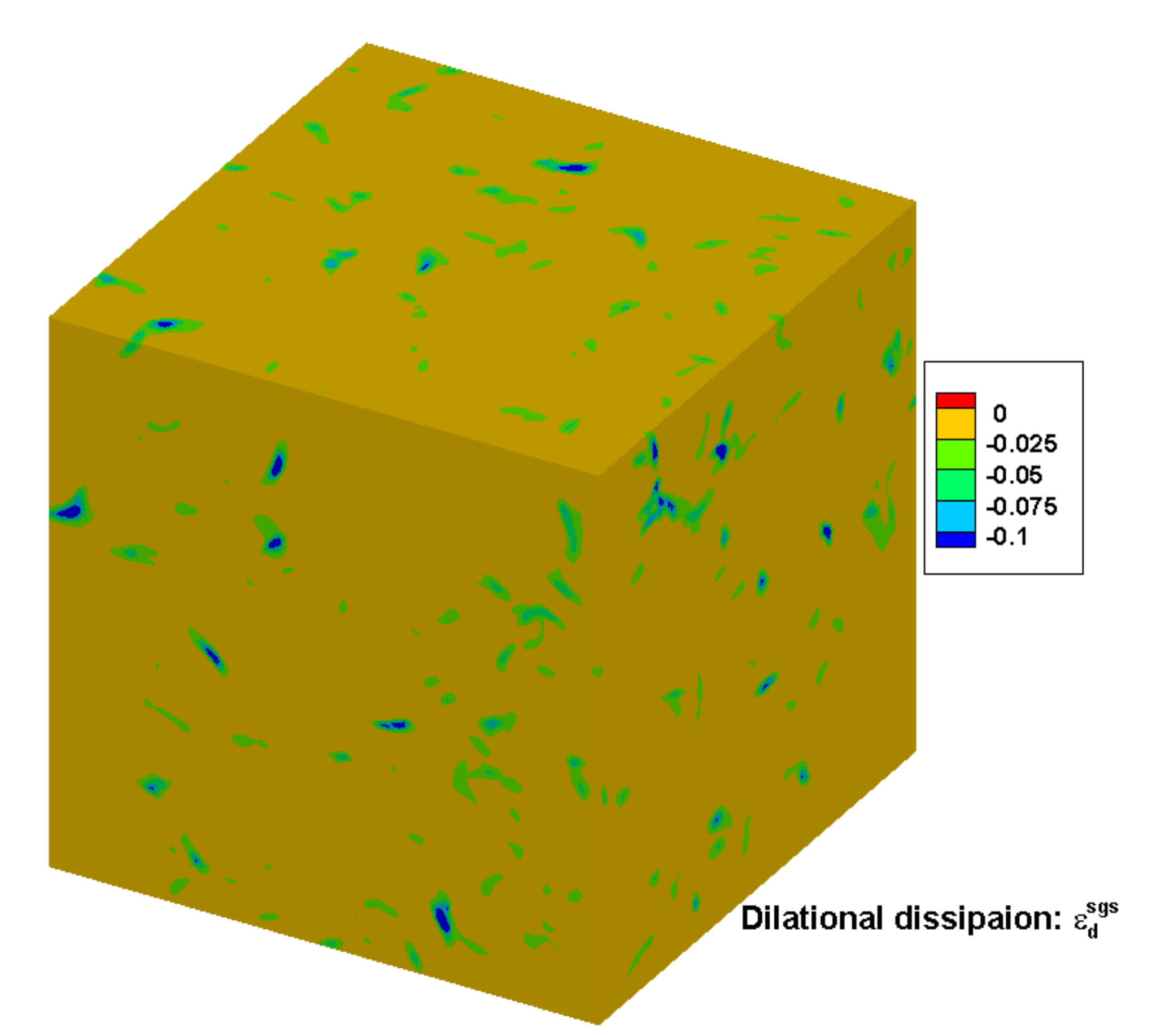}
		\put(-2,75){$(e)$}
	\end{overpic}	
	\begin{overpic}[width=0.495\textwidth]{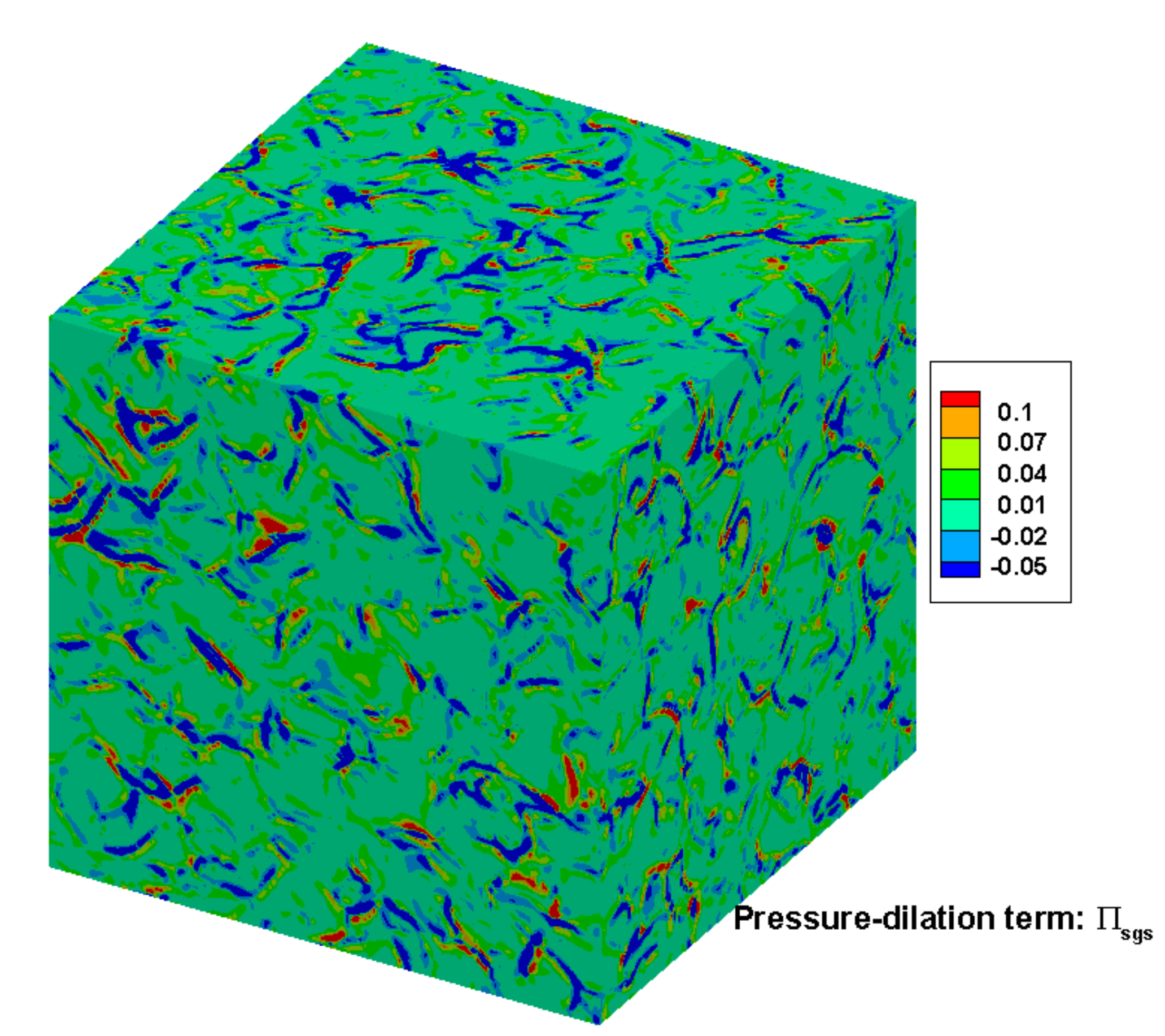}
		\put(-2,75){$(f)$}
	\end{overpic}
	\caption{\label{source_terms} Three-dimension contours of $(a)$ unresolved $K_{utke}$, $(b)$ source term $S_t$, and the components of source term $(c)$ $P_{sgs}$, $(d)$ $\varepsilon_s^{sgs}$, $(e)$ $\varepsilon_d^{sgs}$, $(f)$ $\Pi_{sgs}$ at $\tau_{cit} = 0.5$ for case R$_1$ with NTRKM.}
\end{figure}
Figure \ref{source_terms} shows the contours of unresolved $K_{utke}$, source term $S_t$, and the components of source term $P_{sgs}$, $\varepsilon_s^{sgs}$, $\varepsilon_d^{sgs}$, $\Pi_{sgs}$ at $\tau_{cit} = 0.5$ for case R$_1$. 
Figure \ref{source_terms}$(a)$ illustrates the contour of unresolved $K_{utke}$, and figure \ref{source_terms}$(b)$ confirms the ensemble positive source term $S_t$ which accounts for the increase of unresolved $K_{utke}$ in figure \ref{ntrkm_kutke_cs}$(b)$.
More specifically, figure \ref{source_terms}$(c)$ presents that the magnitude of ensemble unresolved production rate at $\tau_{cit} = 0.5$ is larger than that of ensemble unresolved dissipation rate and ensemble unresolved pressure-dilation transfer. 
Qualitatively, the negative unresolved dissipation rate and the high similarity between unresolved solenoidal dissipation rate and unresolved dilational dissipation rate are observed in figure \ref{source_terms}$(d)(e)$. 
Figure \ref{source_terms}$(f)$ shows that the magnitude and portion of negative unresolved pressure-dilation transfer are larger than the positive ones, which behave similarly as delicate {\it{priori}} coarse-graining analysis of compressible unresolved TKE budget \citep{cao2021three}.

\begin{figure}
	\centering
	\begin{overpic}[width=0.495\textwidth]{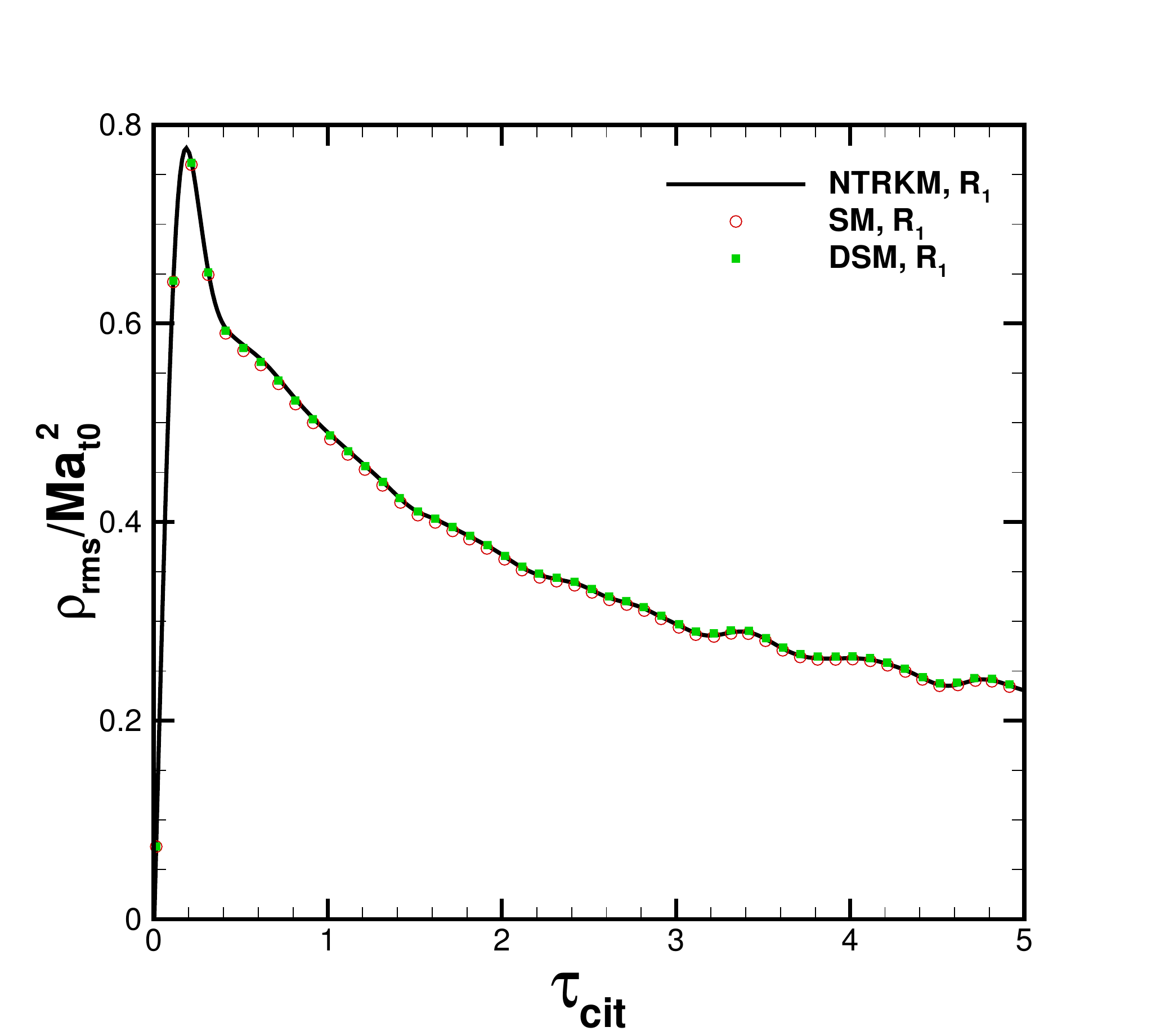}
		\put(-2,75){$(a)$}
	\end{overpic}	
	\begin{overpic}[width=0.495\textwidth]{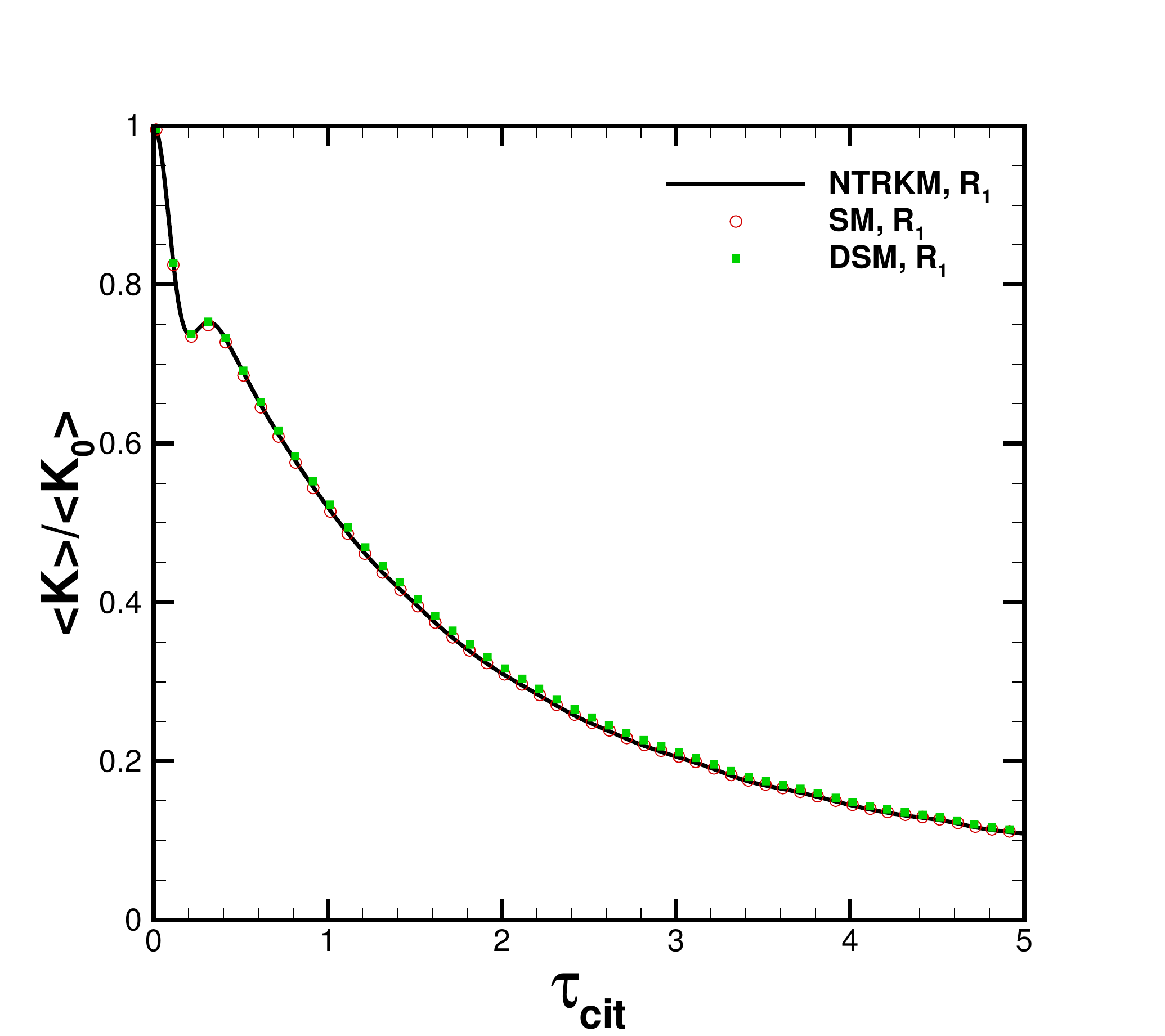}
		\put(-2,75){$(b)$}
	\end{overpic}
	\begin{overpic}[width=0.495\textwidth]{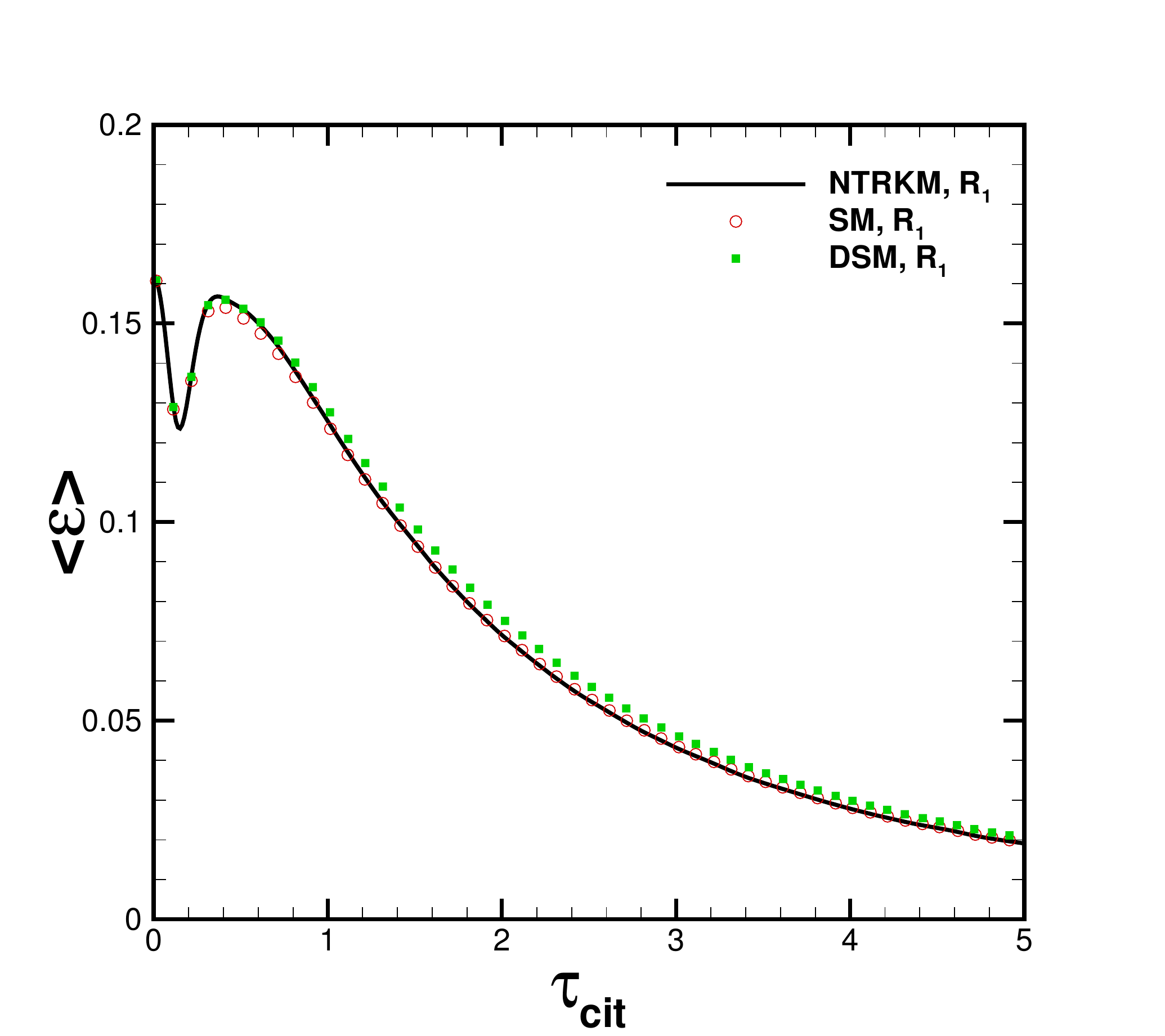}
		\put(-2,75){$(c)$}
	\end{overpic}	
	\begin{overpic}[width=0.495\textwidth]{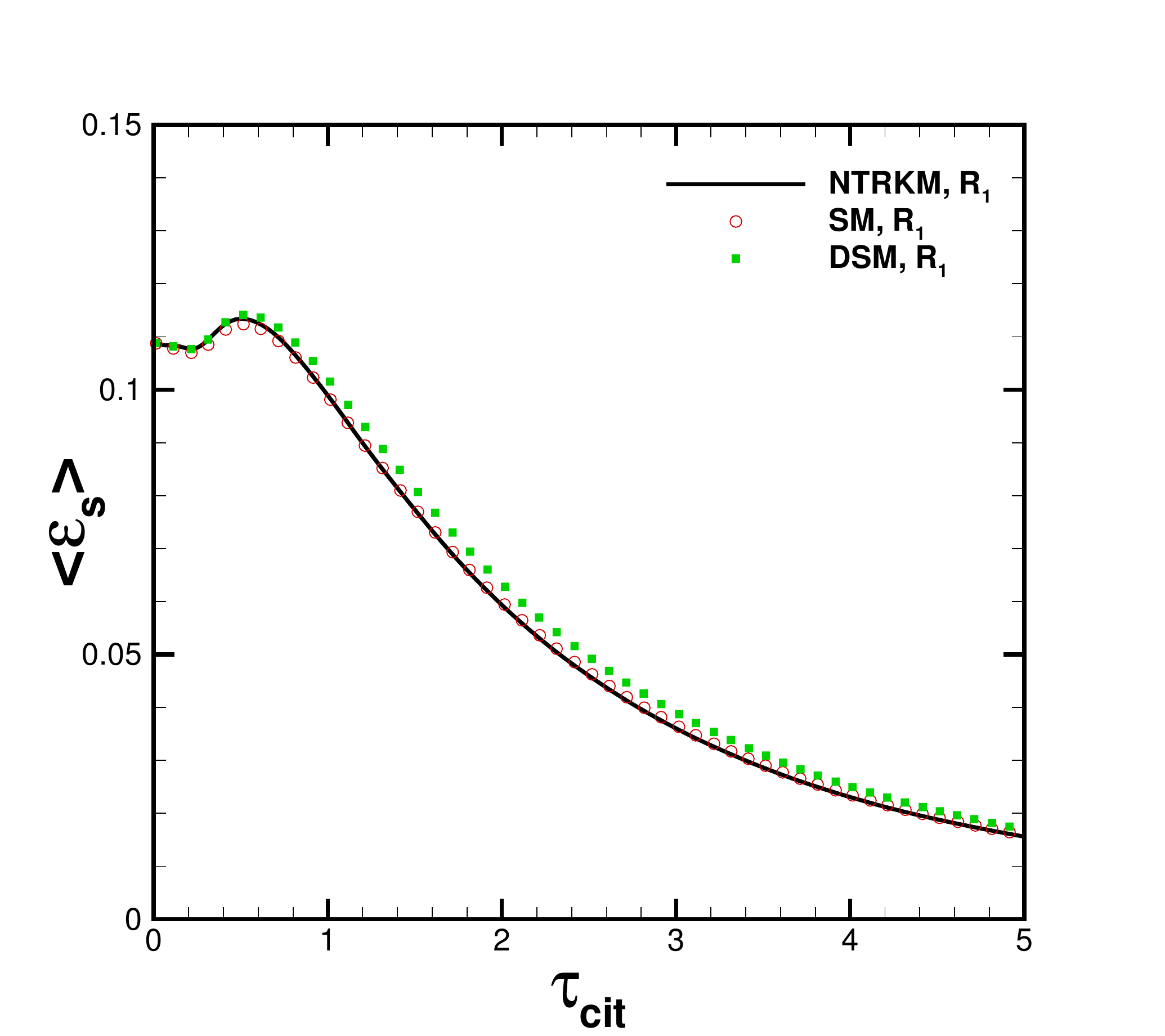}
		\put(-2,75){$(d)$}
	\end{overpic}
	\begin{overpic}[width=0.495\textwidth]{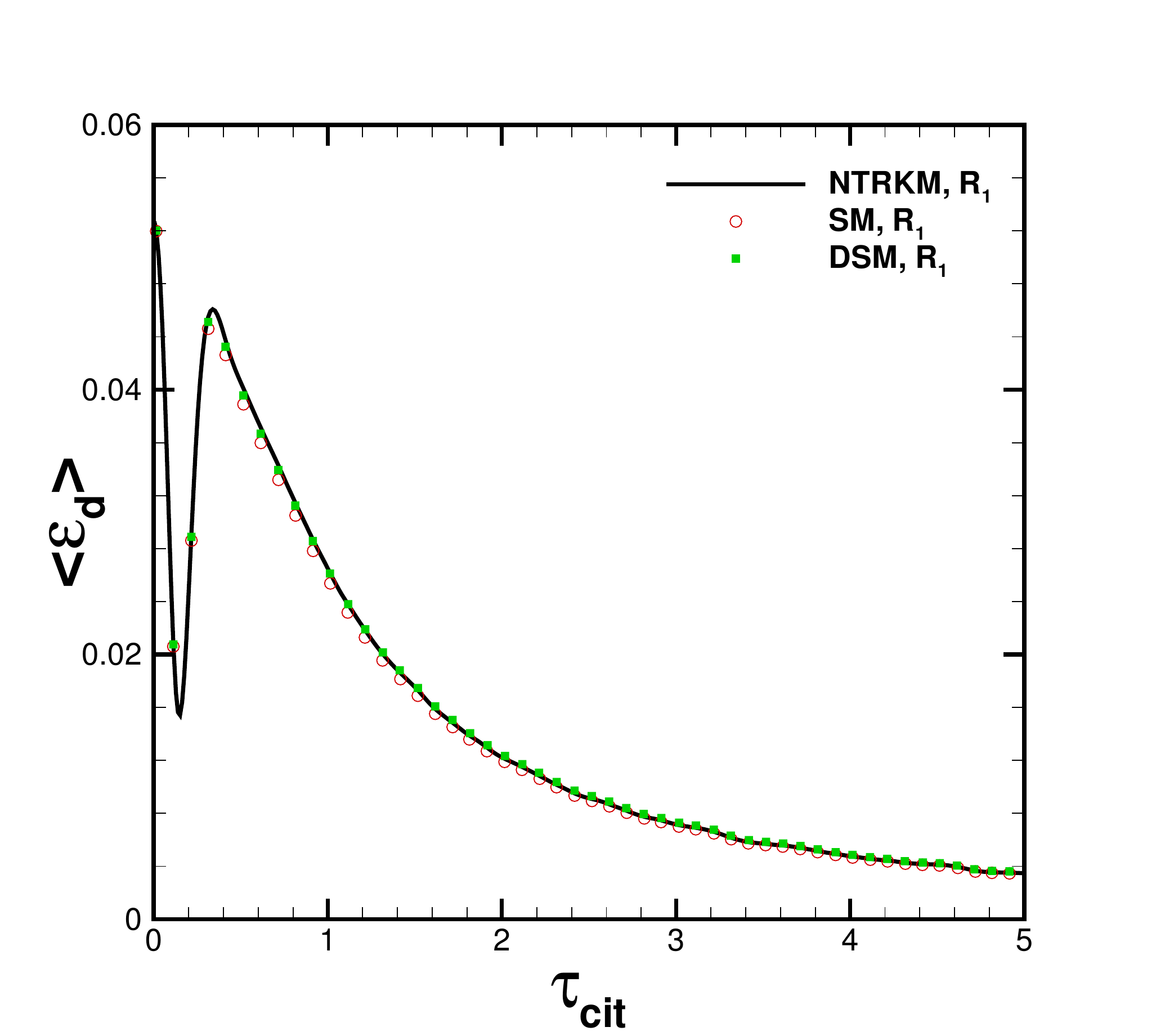}
		\put(-2,75){$(e)$}
	\end{overpic}	
	\begin{overpic}[width=0.495\textwidth]{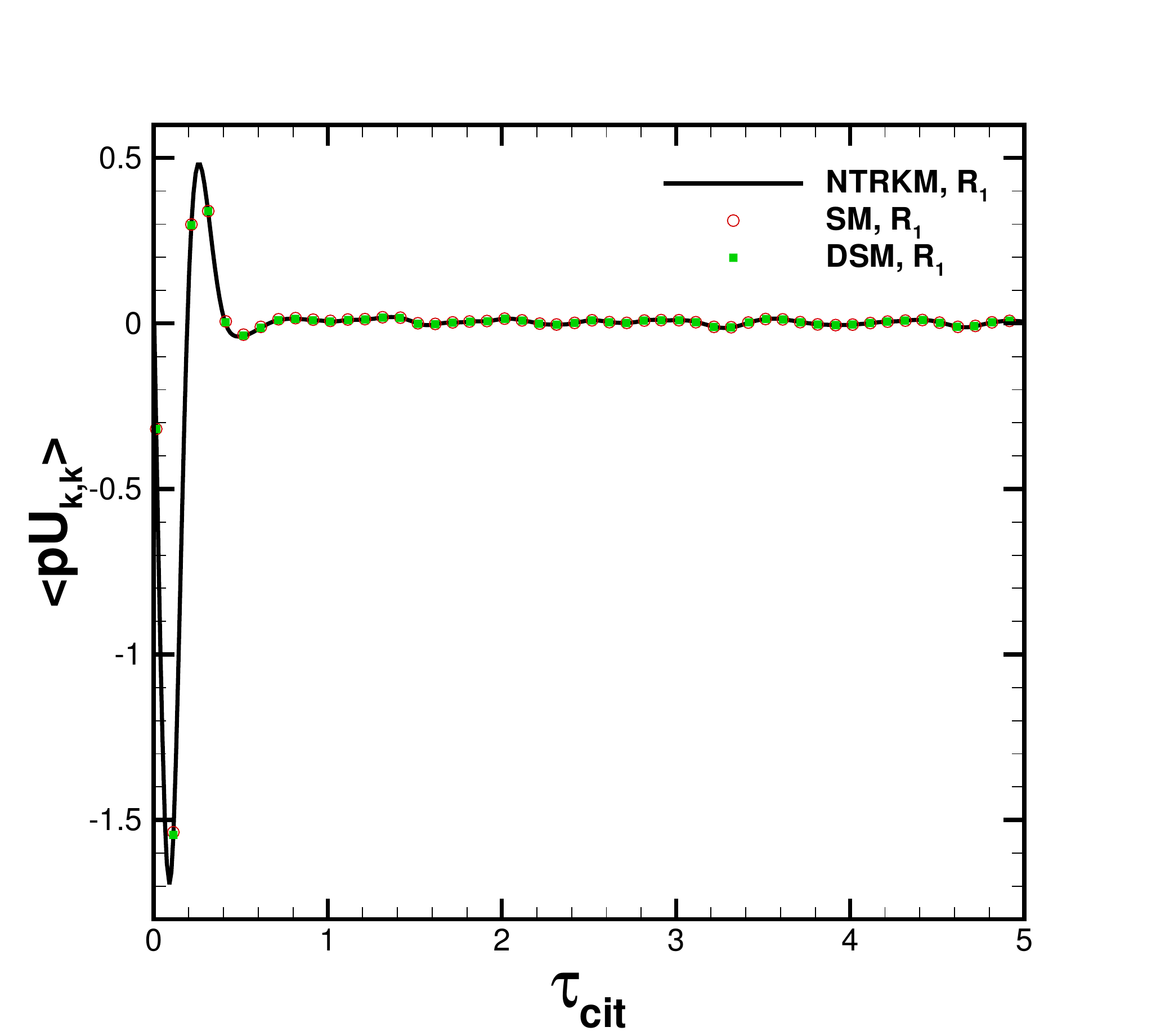}
		\put(-2,75){$(f)$}
	\end{overpic}
	\caption{\label{key_stastics} Time history of $(a)$ normalized resolved root-mean-square density fluctuation $\rho_{rms}/Ma_{t0}^2$, $(b)$ normalized ensemble resolved turbulence kinetic energy $\left \langle K \right \rangle /\left \langle K_0 \right \rangle$, $(c)$ ensemble resolved dissipation rate $\left\langle \varepsilon \right\rangle$, $(d)$ ensemble resolved solenoidal dissipate rate $\left\langle \varepsilon_s \right\rangle$, $(e)$ ensemble resolved dilational dissipation rate $\left\langle \varepsilon_d \right\rangle$, and $(f)$ ensemble resolved pressure-dilation transfer $\left\langle p \theta \right\rangle$ for case R$_1$.}
\end{figure}
To evaluate the performance of NTRKM and non-equilibrium gas-kinetic scheme, figure \ref{key_stastics} shows the key resolved statistical quantities as equations \eqref{rho_k_prms1}-\eqref{dkdt2} for case R$_1$.
Figure \ref{key_stastics} shows that the key statistical quantities of  NTRKM are comparable with those from the widely-used SM \citep{smagorinsky1963general} and DSM \citep{moin1991dynamic}.
SM and DSM are dealt with the equilibrium time-relaxation framework \citep{cao2019implicit}, which are also implemented in the in-house DNS code \citep{cao2022high} with modifying the $\tau$ to $\tau + \tau_t$ on unresolved grids.
Among these three compressible LES models, the heat flux is not modified, namely, $Pr = 1$ and $Pr_t = 1$ are treated fairly on all simulations.
In terms of density-weighted SM, the eddy viscosity takes $\mu_t = (C_{sm} \overline{\Delta})^2 \overline{\rho} |\widetilde{S}|$ 
with magnitude $|\widetilde{S}| = (2 \widetilde{S}_{ij} \widetilde{S}_{ij})^{\frac{1}{2}}$.
Coefficient $C_{sm}$ is fixed as $0.17$ deduced from the scaling law of TKE spectrum in high-Reynolds number isotropic turbulence \citep{lilly1967representation}. 
For density-weighted DSM, the eddy viscosity is determined as $\mu_t = C_{dsm} \overline{\Delta}^2 \overline{\rho} |\widetilde{S}|$, 
and the dynamic coefficient $C_{dsm}$ is computed through the dynamic technique presented in Appendix \ref{appB_1}.
The dynamic coefficient $C_{dsm}$ in density-weighted DSM evolves between $[0.016, 0.025]$ for case R$_1$. 
Thus, the dynamic coefficient $C_{dsm}$ is slightly smaller than the empirical static coefficient $C_{sm}^2 \approx 0.029$ in current DCIT, 
which accounts for the comparable performance between density-weighted SM and density-weighted DSM on unresolved grids.
Overall, the performance of key statistical turbulent quantities shows that current NTRKM is comparable with the widely-used eddy-viscosity SM and DSM.
Numerical performance of NTRKM and corresponding non-equilibrium gas-kinetic scheme (see \S \ref{sec:NTRKM_gks}) offers the confidence for simulating practical turbulence on unresolved grids.

\subsection{Temporal compressible plane mixing layer}\label{subsec:tcpml}
For temporal compressible plane mixing layer \citep{sandham1991three, vreman1997large, pantano2002study}, practical simulations on unresolved grids are conducted to further assess the performance of NTRKM and the non-equilibrium gas-kinetic scheme (see \S \ref{sec:NTRKM_gks}).
For the flow without strong discontinuities, the collision time  is given by
\begin{equation}
\begin{aligned}
	\tau + \tau_t = \frac{\mu + \mu_t}{p}.
\end{aligned}
\end{equation}
For TCPML, the grid filter width adopts as the effective grid length of control volume, i.e., $\overline{\Delta} = (\Delta x_1 \times \Delta x_2 \times \Delta x_3)^{1/3}$. 
Test filter width $\widehat{\overline\Delta}$ still keeps to twice the grid filter width $\overline{\Delta}$ in determining the dynamic model coefficients, namely $\widehat{\overline\Delta} = 2 \overline{\Delta}$.
In this section, DNS in TCPML will be validated firstly.
Then, LES studies restart from the filtered DNS solution on unresolved grids.

TCPML is initialized by a hyperbolic tangent profile for the streamwise velocity \citep{arun2019topology} as 
\begin{equation}\label{mixing_init_u}
\left\{
	\begin{aligned}
		U_1 &= \frac{1}{2}\Delta U \tanh(\frac{- x_2}{2 \delta_{\theta 0}}), \\
		U_2 &= 0, \\
		U_3 &= 0,
	\end{aligned}
\right.
\end{equation}
where $\Delta U = U_{lo} - U_{up}$, and initial momentum thickness $\delta_{\theta 0} = 1$ is adopted.
As equation \eqref{mixing_init_u} presents, two equal and opposite streamwise velocities are simulated as $- U_{up} = U_{lo} = 1$.
With the unity Prandtl number, Crocco-Busemann relation \citep{sandham1990numerical} gives the initial temperature profile as
\begin{equation}\label{mixing_init_t}
	\begin{aligned}
		\frac{T}{T_{\infty}} = 1 + Ma_{c}^2 \frac{\gamma - 1}{2} (1 - U_1^2). 
	\end{aligned}
\end{equation}
The initial density is set to a uniform value $\rho_{\infty}$.
The convective Mach number $Ma_{c} = 0.75$ and initial vorticity thickness-based Reynolds number $Re_{\omega 0} = 640$ are simulated as
\begin{equation}\label{mixing_mach1}
	\begin{aligned}
		Ma_{c} = \frac{\Delta U}{2 \sqrt{\gamma R T_{\infty}}}, 
	\end{aligned}
\end{equation}
\begin{equation}\label{mixing_mach2}
	\begin{aligned}
		Re_{\omega 0} = \frac{\rho_{\infty} \Delta U \delta_{\omega 0}}{\mu_{\infty}},
	\end{aligned}
\end{equation}
where $\mu_{\infty}$ is the reference viscosity corresponding to reference temperature $T_{\infty}$, and viscosity is determined by power law as $\mu(T) = \mu_{\infty} (T/T_{\infty})^{0.67}$ \citep{sandham1991three}.
With the uniform initial density, initial vorticity thickness can be estimated as $\delta_{\omega 0} = \Delta U/|\partial U_1/ \partial x_2|_{max}$ in which the maximum of denominator is reached in the centre plane.
$|\cdot|$ represents the absolute value.
The momentum thickness $\delta_{\theta}$ is defined as
\begin{equation}\label{mixing_momentum_thickness}
	\begin{aligned}
		\delta_{\theta} = \frac{1}{\rho_{\infty} (\Delta U)^2} \int_{-\infty}^{\infty} [\left\langle \rho U_1(x_2) \right\rangle - \left\langle \rho U_{lo} \right\rangle ][\left\langle \rho U_{up} \right\rangle - \left\langle \rho U_1(x_2) \right\rangle] \text{d} x_2,
	\end{aligned}
\end{equation}
where $\left \langle \cdot \right \rangle$ represents the plane average along the streamwise and spanwise directions, and $\delta_{\omega 0} \approx 4 \delta_{\theta 0}$ since the finite transverse domain $[-L_1/2, L_1/2]$.
For TCPML, the turbulent stress tensor $R_{ij}$ and anisotropy stress tensor $b_{ij}$ read
\begin{equation} \label{cml_reynolds_anisotropic1}
	\begin{aligned}
		b_{ij} = \frac{R_{ij} - \frac{2}{3} K_R \delta_{ij} }{2 K_R},
	\end{aligned}
\end{equation}
\begin{equation} \label{cml_reynolds_anisotropic2}
	\begin{aligned}
		R_{ij} = \frac{\left\langle\rho U_i^{'} U_j^{'}\right\rangle}{\left\langle \rho \right\rangle}, 
	\end{aligned}
\end{equation}
where $U_i^{'} = U_i - \left\langle \rho U_i \right\rangle /\left\langle \rho \right\rangle$,
$K_R$ the so-called resolved turbulence kinetic energy as $K_R = R_{ii}/2$. 
Anisotropy stress tensor $b_{ij}$ is an important characteristic of turbulence, especially for advanced turbulence closures \citep{pantano2002study}.
In following statistical process, $R_{ij}$ is integrated across mixing layer within $[-\delta_{\omega}(\tau_{ml}), \delta_{\omega}(\tau_{ml})]$, while $b_{ij}$ is integrated within $[-4 \delta_{\theta}(\tau_{ml}), 4 \delta_{\theta}(\tau_{ml})]$ with normalized time $\tau_{ml} = \Delta U t/\delta_{\theta 0}$.
To accelerate the transition process, the initial condition is specified by adding a random number to density, temperature, transverse and spanwise velocities at each mesh point \citep{sandham1991three}, i.e., $\rho_p = \rho_{\infty} + A_{s1} r_d e^{-x_2/(2\delta_{\theta 0})^2}$. $r_d$ is a random number uniformly distributed between $[-0.5, 0.5]$ and the amplitude $A_{s1} = 0.2$.
In terms of streamwise velocity, besides the random number, the artificial sinusoidal-type perturbation has been added as
\begin{equation}\label{mixing_init_u2}
	\begin{aligned}
		U_{1p} &= U_1 + U_1[A_{s1} + A_{s2} sin (\gamma_{s1} x_2)(B_{s1} + B_{s2})]r_d e^{-x_2/(2\delta_{\theta 0})^2}, 
	\end{aligned}
\end{equation} 
where $A_{s2} = 0.6$, $\gamma_{s1} = 0.25$, and $B_{s1} = A_{s3} [cos(\gamma_{s2} x_1) + cos(2 \gamma_{s2} x_1) + cos(4 \gamma_{s2} x_1)]$, $B_{s2} =  A_{s4} [cos(\gamma_{s2} x_1)cos(\gamma_{s2} x_3) + cos(2 \gamma_{s2} x_1)cos(2 \gamma_{s2} x_3) + cos(4 \gamma_{s2} x_1)cos(4 \gamma_{s2} x_3)$ with $A_{s3} = 0.2$, $A_{s4} = 0.4$ and $\gamma_{s2} = 0.235$.
The initial condition for primitive variables $(\rho_p, U_{1p}, U_{2p}, U_{3p}, T_p)^T$ can be obtained for DNS.
The computational domain is discretized uniformly in three directions.
Boundary conditions in the homogeneous streamwise and spanwise directions are periodic.
In the transverse direction, the non-reflective boundary condition of conservative variables is given according to one-dimensional 
Riemann invariants \citep{toro2013riemann}, whereas the outlet boundaries are used for the unresolved $K_{utke}$ (see equation \eqref{tgks_macro_formula4}).
\begin{table}
	\begin{center}		
		\centering
		\begin{tabular}{ccccc}
			Case      &$L_1 \times L_2 \times L_3$ &$N_1 \times N_2 \times N_3$  &$Ma_c$  &$Re_{\omega 0}$    \\
			Ref$_1$   &$172\delta_{\theta 0} \times 129\delta_{\theta 0} \times 86\delta_{\theta 0}$  &$256 \times 192 \times 128$             &0.70     &640      \\			
			Ref$_2$   &$314\delta_{\theta 0} \times 157\delta_{\theta 0} \times 78.5\delta_{\theta 0}$  &$512 \times 256 \times 128$             &0.75      &640      \\
			DNS       &$314\delta_{\theta 0} \times 157\delta_{\theta 0} \times 78.5\delta_{\theta 0}$ &$576 \times 384 \times 192$              
			&0.75     &640      \\
			M$_1$     &$314\delta_{\theta 0} \times 157\delta_{\theta 0} \times 78.5\delta_{\theta 0}$ &$144 \times 96 \times 48$              
			&0.75     &640     
		\end{tabular}
		\caption{\label{cml_parameters_table_dns} Numerical parameters for TCPML.}
	\end{center}
\end{table}
\begin{figure}
	\centering
	\begin{overpic}[width=0.495\textwidth]{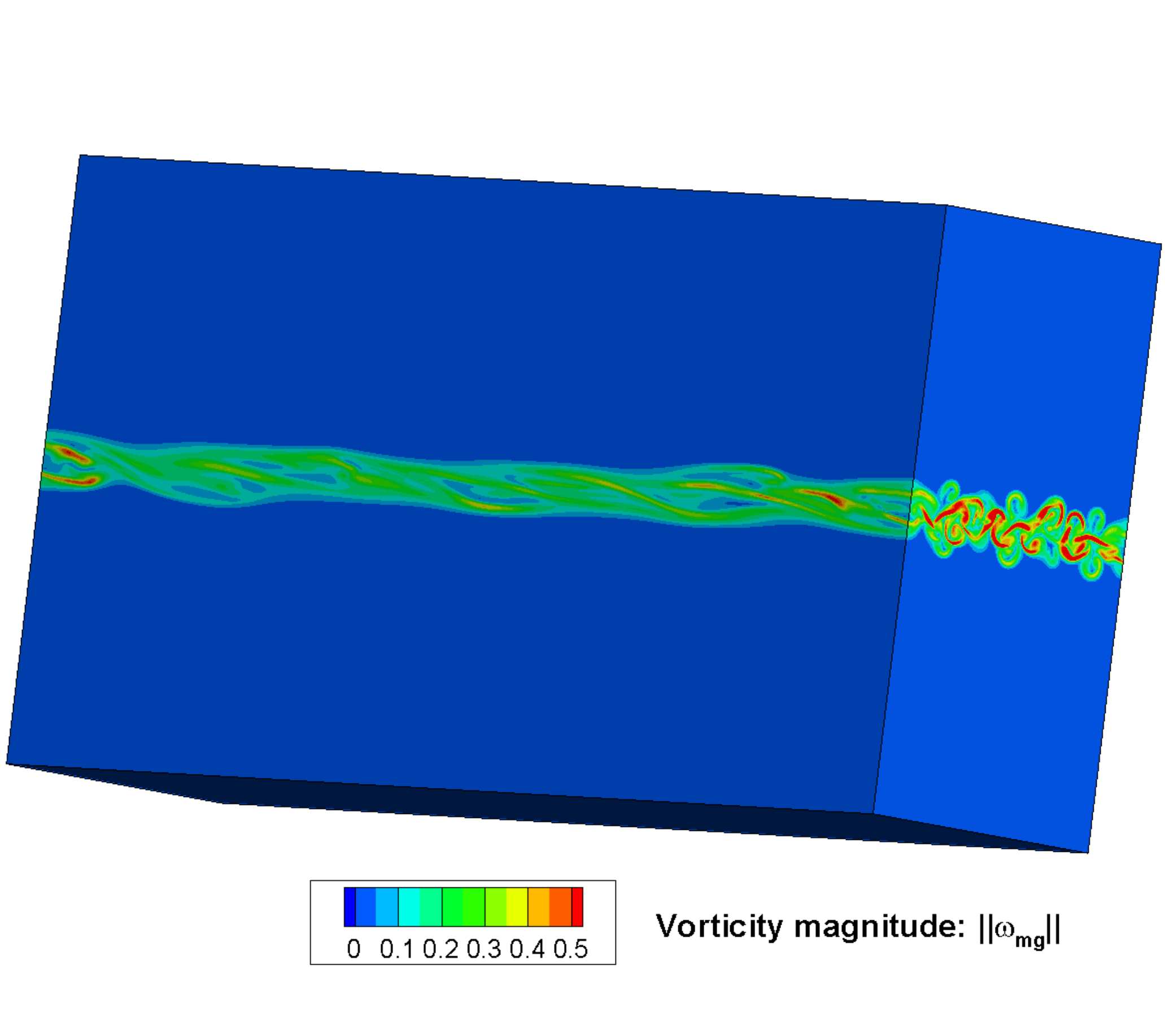}
		\put(-2,75){$(a)$}
	\end{overpic}	
	\begin{overpic}[width=0.495\textwidth]{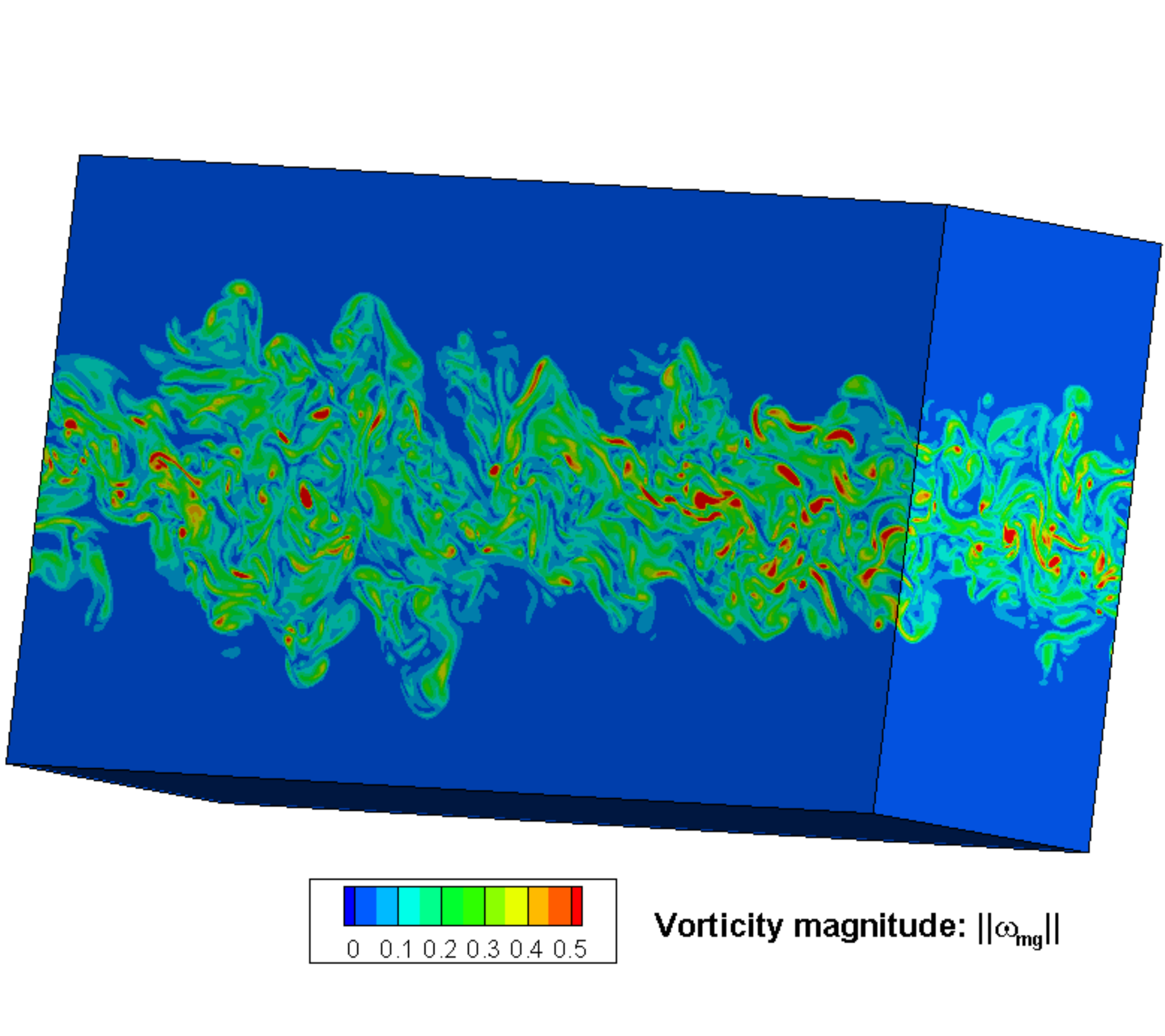}
		\put(-2,75){$(b)$}
	\end{overpic}
	\caption{\label{cml_dns_vorticity}  Contours in magnitude of vorticity at $(a)$ $\tau_{ml} = 400$ and $(b)$ $\tau_{ml} = 1400$ for DNS.}
\end{figure}

\begin{table} 
	\begin{center}
		\centering
		\begin{tabular}{ccccc}
			Case     &$\dot{\delta}_{\theta}/\dot{\delta}_{inc}$ &$Re_{\omega}$   &$Ma_t$     &$(b_{11}, b_{12}, b_{22})$ \\
			Ref$_1$  &0.675  &7790   &-           &(0.15, 0.15, -0.10)   \\			
			Ref$_2$  &0.589   &8160   &0.30       &(0.13, 0.13, -0.12)   \\
			DNS     &0.588   &8052   &0.28        &(0.14, 0.16, -0.07)  
		\end{tabular}
		\caption{\label{cml_quantities_table_dns} Key quantities for DNS in TCPML at $\tau_{ml} = 1400$.}
	\end{center}
\end{table}
\begin{figure}
	\centering
	\begin{overpic}[width=0.495\textwidth]{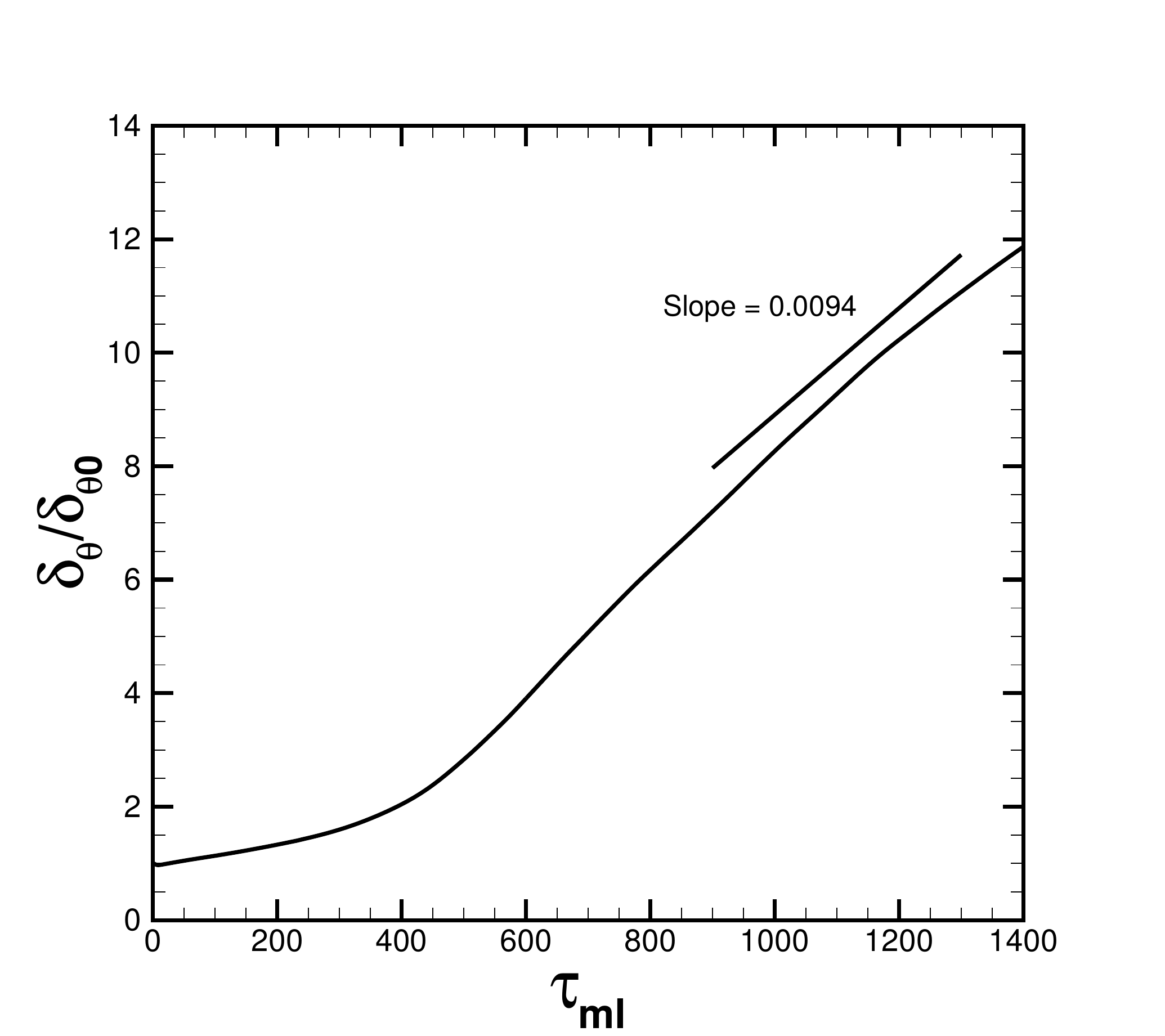}
		\put(-2,75){$(a)$}
	\end{overpic}	
	\begin{overpic}[width=0.495\textwidth]{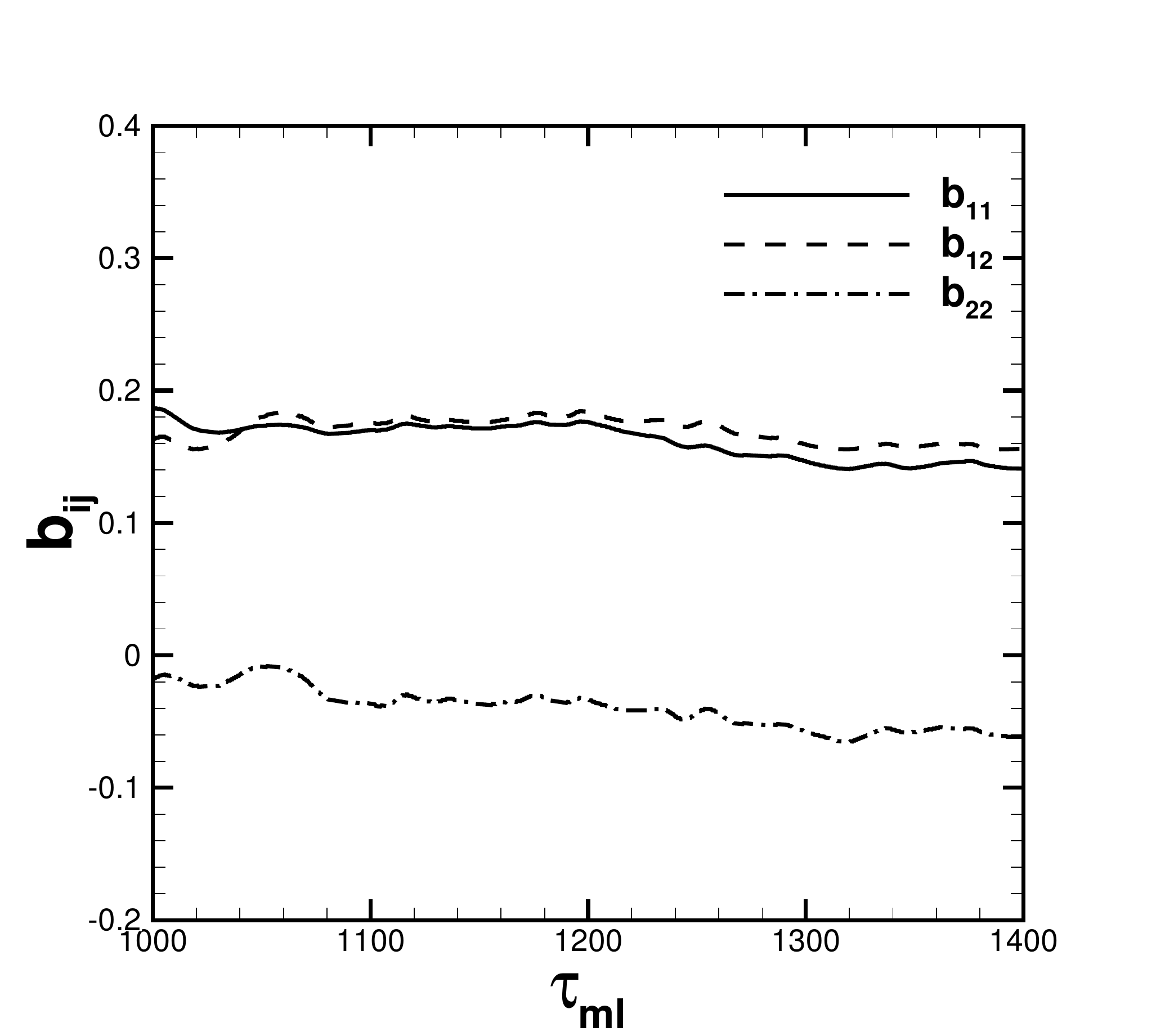}
		\put(-2,75){$(b)$}
	\end{overpic}
	\caption{\label{cml_dns_thetabij} Time history of $(a)$ normalized momentum thickness $\delta_{\theta}/\delta_{\theta 0}$ and $(b)$ evolution of anisotropy stress tensor $b_{ij}$ for DNS.}
\end{figure}
\begin{figure}
	\centering
	\begin{overpic}[width=0.495\textwidth]{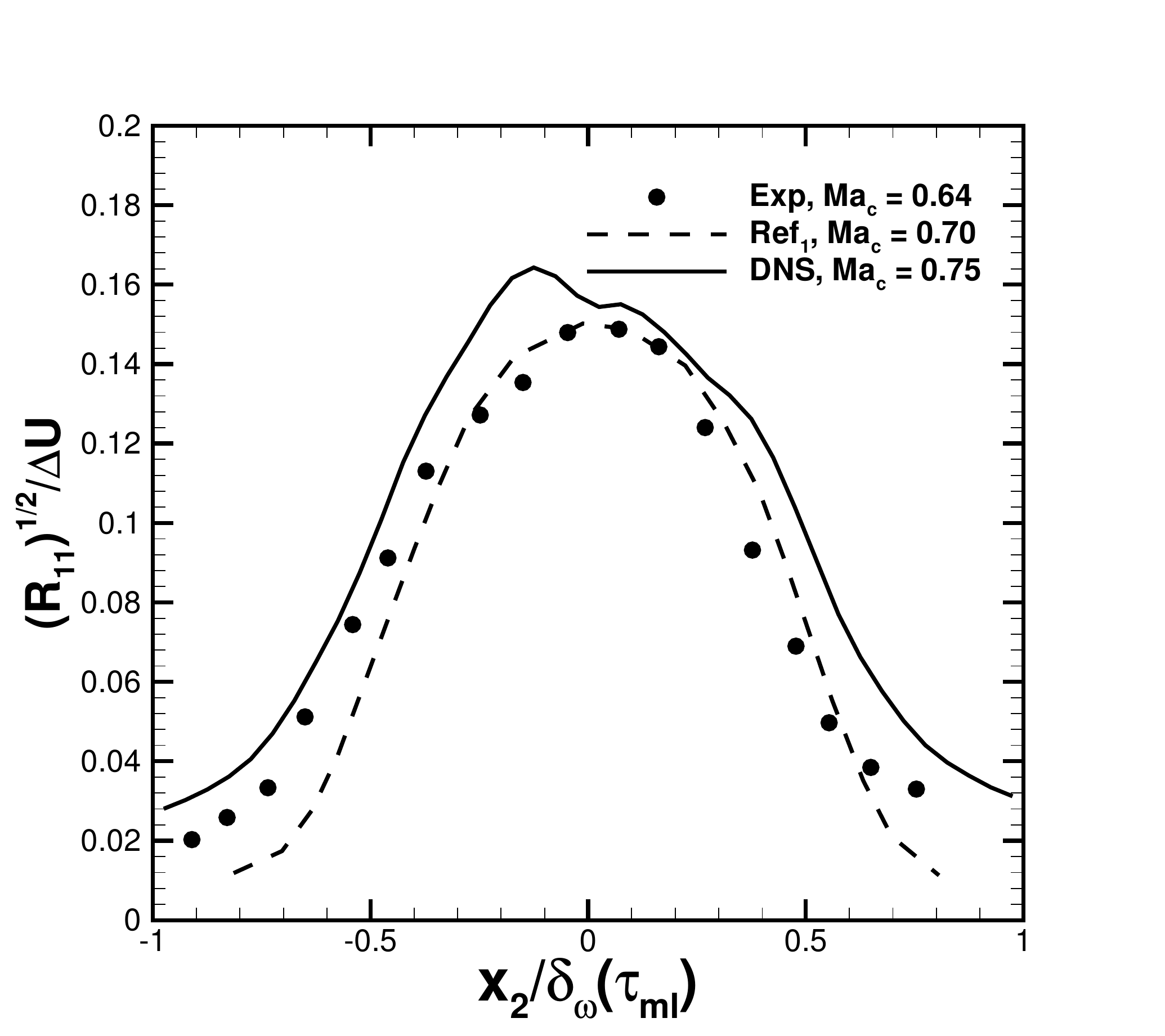}
		\put(-2,75){$(a)$}
	\end{overpic}	
	\begin{overpic}[width=0.495\textwidth]{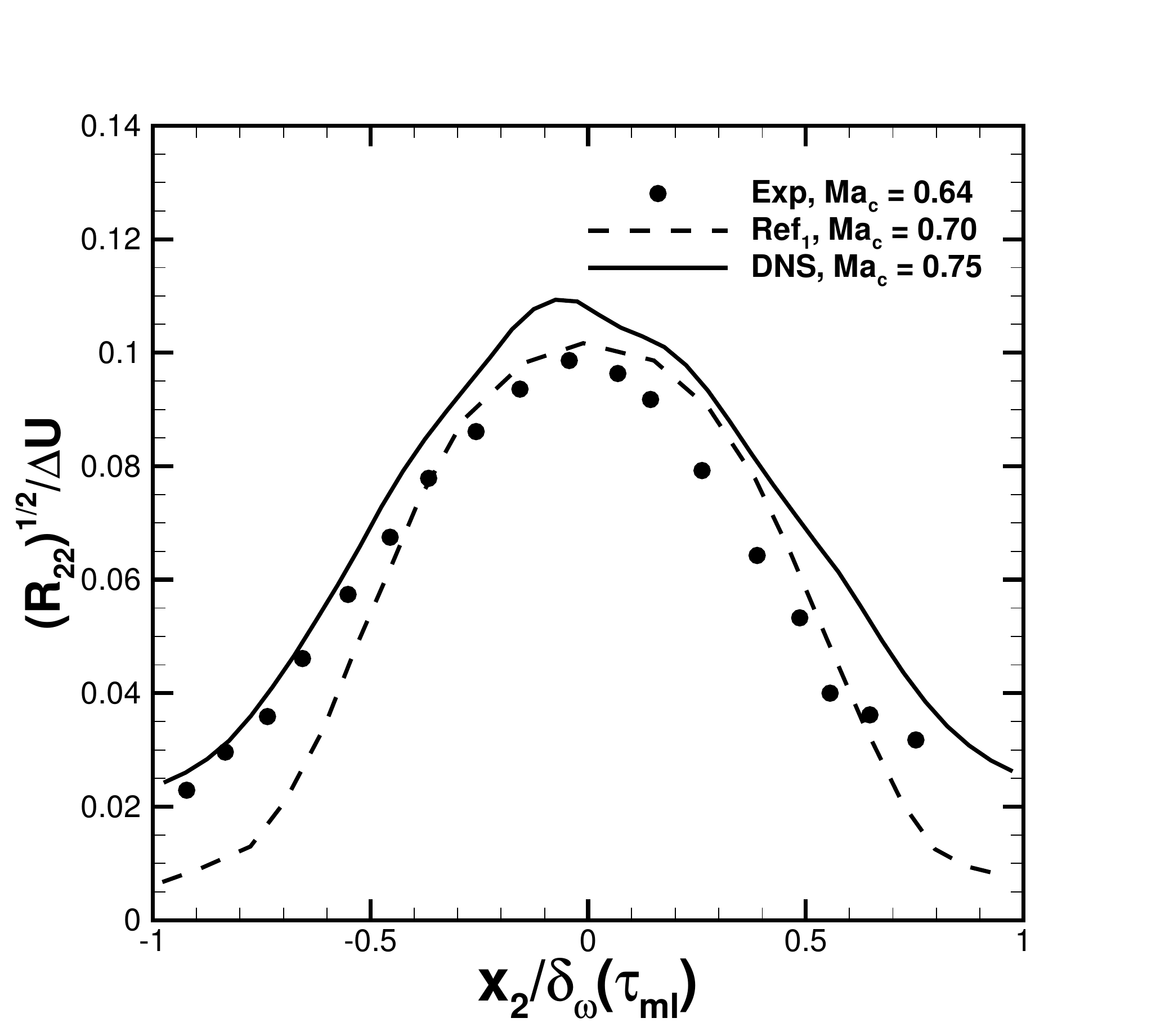}
		\put(-2,75){$(b)$}
	\end{overpic}
	\caption{\label{cml_dns_reynoldstress} Profiles of normalized turbulent stress $(a)$ $(R_{11})^{1/2}/\Delta U$ and $(a)$ $(R_{22})^{1/2}/\Delta U$ for experiment \citep{elliott1990compressibility}, Ref$_1$ \citep{pantano2002study} and DNS.}
\end{figure}
Table \ref{cml_parameters_table_dns} shows the numerical parameters for TCPML. Ref$_1$ was simulated by high-order finite difference method \citep{pantano2002study} with a smaller $Ma_c = 0.70$ and Ref$_2$ was simulated by WENO-enhanced GKS \citep{arun2019topology}.
The size effect of computational domain and grid convergence studies have been well studied in Ref$_2$.  
Compared with the WENO-enhanced GKS \citep{kumar2013weno}, the current in-house code is equipped with the genuine spatial-temporal HGKS \citep{pan2016efficient, cao2022high}.
Table \ref{cml_parameters_table_dns} shows that the same computational domain adopted and much finer grid are used by the current HGKS, which definitely guarantee the resolution of DNS.
Figure \ref{cml_dns_vorticity} shows the contours in magnitude of vorticity $\|\omega_{mg}\| = \sqrt{2 \omega_i \omega_i}$ at $\tau_{ml} = 400$ and $\tau_{ml} = 1400$.
Against the performance of $\|\omega_{mg}\|$ at transitional stage ($\tau_{ml} = 400$, see figure \ref{cml_dns_vorticity}$(a)$), the magnitude of vorticity not only enlarges thicker but also behaves more intermittently at the self-similarity stage ($\tau_{ml} = 1400$, see figure \ref{cml_dns_vorticity}$(b)$).

It is well known that compressibility suppresses the mixing layer growth rate $\dot{\delta}_{\theta} = \text{d} \delta_{\theta}/\text{d}\tau_{ml}$.
In table \ref{cml_quantities_table_dns}, the normalized growth rate $\dot{\delta}_{\theta}/\dot{\delta}_{inc} = 0.589$ agrees well with that in Ref$_2$ \citep{arun2019topology}, and reasonably smaller than that of Ref$_1$ (smaller $Ma_c = 0.70$ corresponding to larger normalized momentum thickness) \citep{pantano2002study}.
Incompressible growth rate $\dot{\delta}_{inc} = 0.016$ is chosen for the hyperbolic tangent profile as equation \eqref{mixing_init_u}. 
Growth rate in this paper is computed by the least-square method within $\tau_{ml} \in [1000, 1300]$ as shown in figure \ref{cml_dns_thetabij}$(a)$.
Table \ref{cml_quantities_table_dns} also presents the vorticity thickness-based Reynolds number $Re_{\omega}$ and the turbulent Mach number $Ma_t$ at $\tau_{ml} = 1400$.
Here, turbulent Mach number is defined as $Ma_t^2 = 2 K_R/(\gamma R T_{\infty})$.
Table \ref{cml_quantities_table_dns} shows that $Re_{\omega}$ and $Ma_t$ at the center plane are in good agreement for all cases.
(In Ref$_1$ and Ref$_2$, the ending of the simulation may be $\tau_{ml} = 600$).
Figure \ref{cml_dns_thetabij}$(b)$ shows the time history of anisotropy stress tensor $b_{ij}$ as equation \eqref{cml_reynolds_anisotropic1}.
We observe the the well-matched quasi-stationary profiles during the self-similarity stage.
Table \ref{cml_quantities_table_dns} shows the components of anisotropy stress tensor $(b_{11}, b_{12}, b_{22})$ at $\tau_{ml} = 1400$.
The large deviation in $b_{22}$ can be attributed to the differences in setting up the initial perturbation field \citep{arun2019topology}.
More specifically, figure \ref{cml_dns_reynoldstress} shows the profiles of normalized turbulent stress $(R_{11})^{1/2}/\Delta U$ and $(R_{22})^{1/2}/\Delta U$ (see equation \eqref{cml_reynolds_anisotropic2}).
Refereed solutions suggest the envelop for normalized turbulent stress in TCPML  correspond to $Ma_c = 0.75$.  
The reasonable deviation originates from the different convective Mach number, where the data of Ref$_1$ and experiment \citep{elliott1990compressibility}
correspond to $Ma_c = 0.70$ and  $Ma_c = 0.64$, respectively.
Overall, the current DNS results agree well with refereed numerical simulations. 
After obtaining the high-fidelity flow fields from DNS, M$_1$ with NTRKM, SM and DSM will be conducted on unresolved grids subsequently.

\begin{figure}
	\centering
	\begin{overpic}[width=0.495\textwidth]{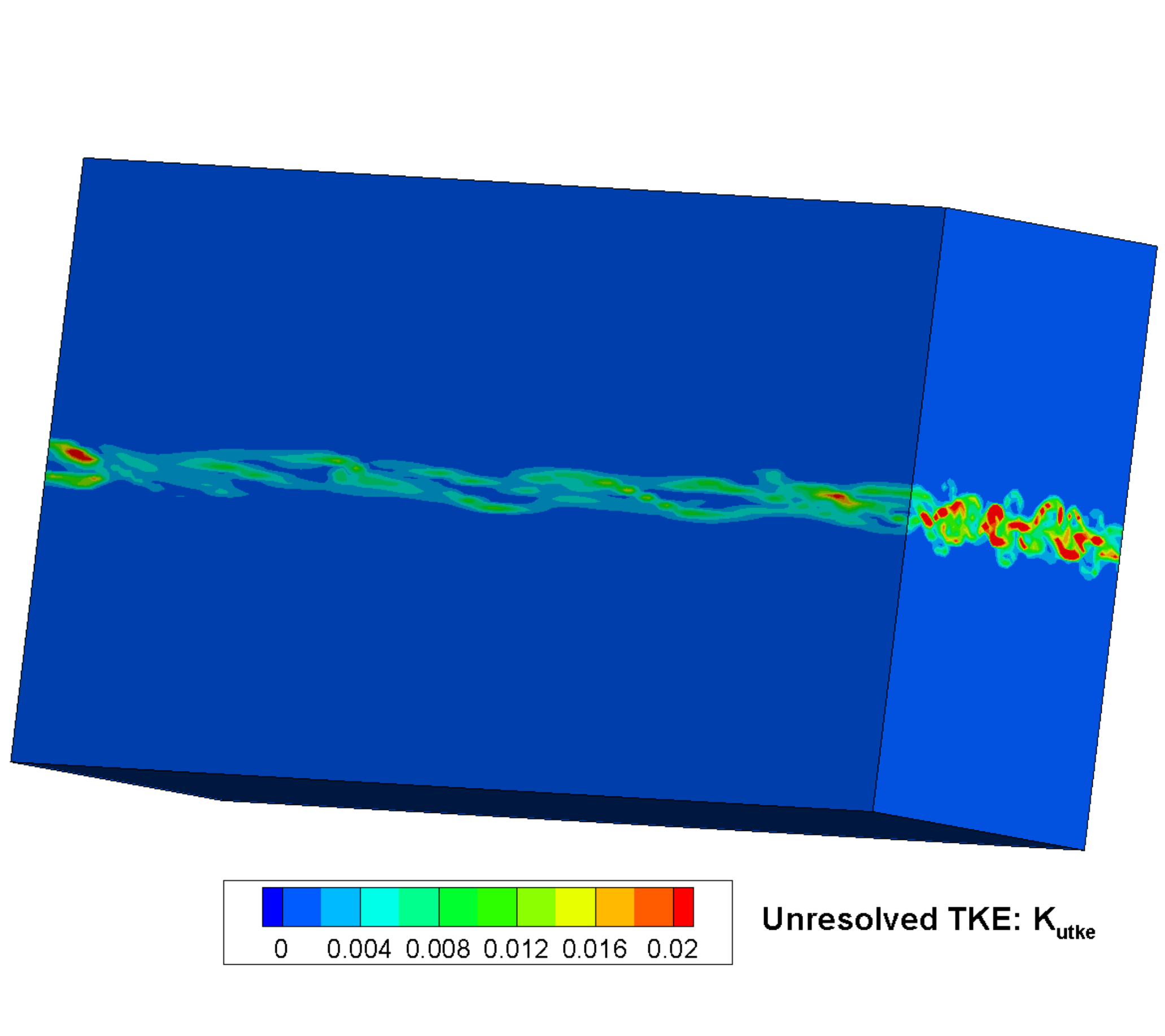}
		\put(-2,75){$(a)$}
	\end{overpic}	
	\begin{overpic}[width=0.495\textwidth]{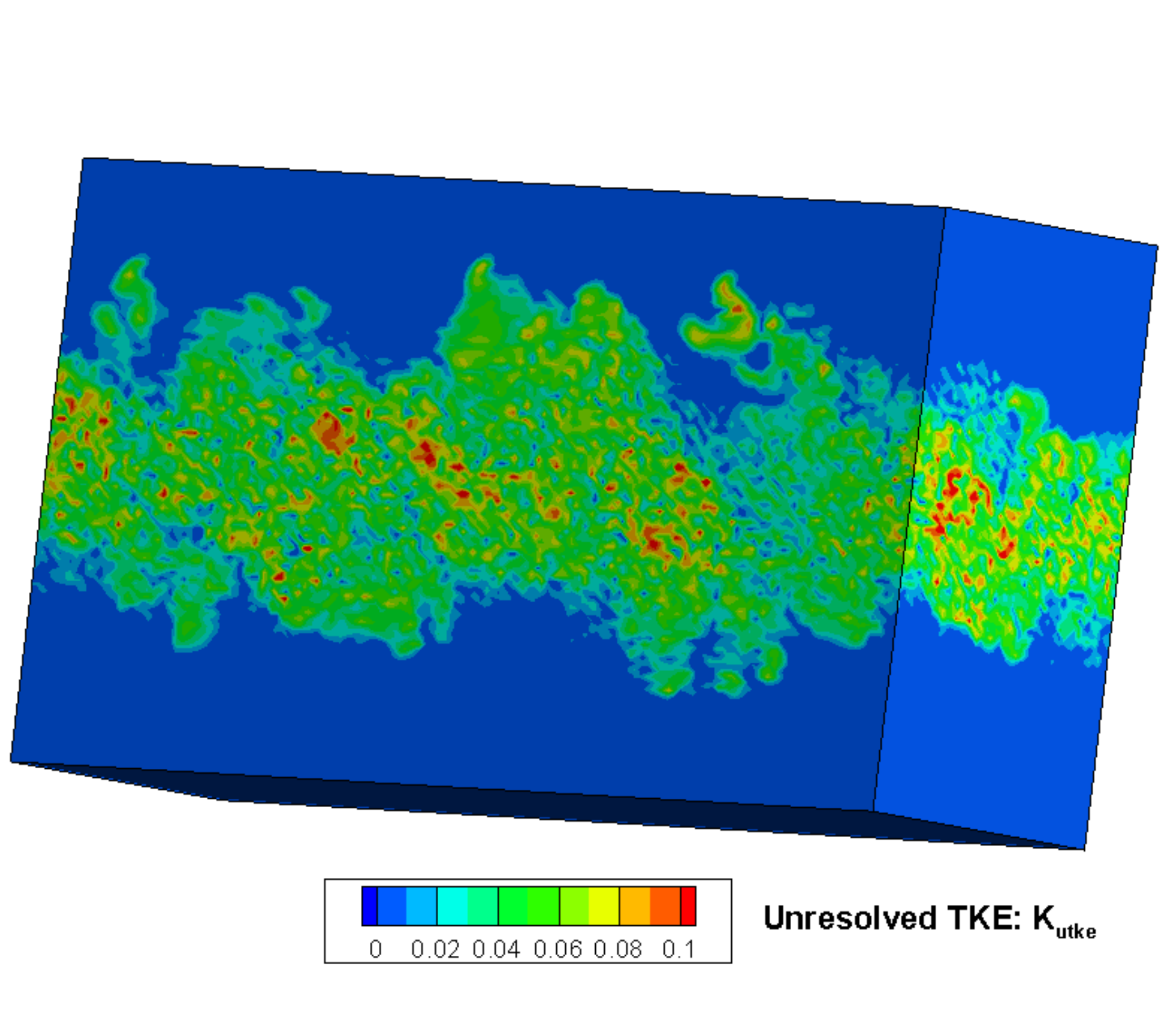}
		\put(-2,75){$(b)$}
	\end{overpic}
	\caption{\label{cml_ntrkm_ksgs}  Contours of unresolved $K_{utke}$ at $(a)$ $\tau_{ml} = 400$ and $(b)$ $\tau_{ml} = 1400$ for case M$_1$ with NTRKM.}
\end{figure}

Table \ref{cml_parameters_table_dns} shows that M$_1$ with NTRKM, SM and DSM are conducted on unresolved uniform grids $144 \times 96 \times 48$.
Box filter is used to generate the initial (restarted) six-variable flow field (see equations \ref{tgks_macro_formula1}-\ref{tgks_macro_formula4}) from the DNS solution at $\tau_{ml} = 400$, i.e., $4^3$ resolved grids are coarsen to $1$ unresolved grid. 
The same computational domain and boundary conditions are applied as the DNS.
In terms of SM, to guarantee the numerical stability and improve the dissipative behavior, \citet{vreman1997large} recommended $C_{sm} = 0.1$ for compressible mixing layer.
When implementing NTRKM, the minimum unresolved TKE is set as $\left\langle K_{utke0} \right\rangle/10000$, where the initial ensemble unresolved $\left\langle K_{utke0} \right\rangle = 0.00058$.
Figure \ref{cml_ntrkm_ksgs}$(a)$ shows the initial pointwise unresolved TKE.
The initial unresolved $K_{utke}$ is in a small magnitude and restricted in a narrow region.
From transitional stage to self-similarity stage (at $\tau_{ml} = 1400$), figure \ref{cml_ntrkm_ksgs}$(b)$ shows that the magnitude of unresolved $K_{utke}$ increase obviously, as well as entrain to a much wider region similar as the figure \ref{cml_dns_vorticity}$(b)$.
Again, the intrinsic equilibrium assumption on $K_{utke}$, such as $\left\langle S_t\right\rangle \approx 0$ for zero-equation eddy-viscosity LES models \citep{lilly1967representation, germano1991dynamic, moin1991dynamic} may not hold.

\begin{figure}
	\centering
	\begin{overpic}[width=0.495\textwidth]{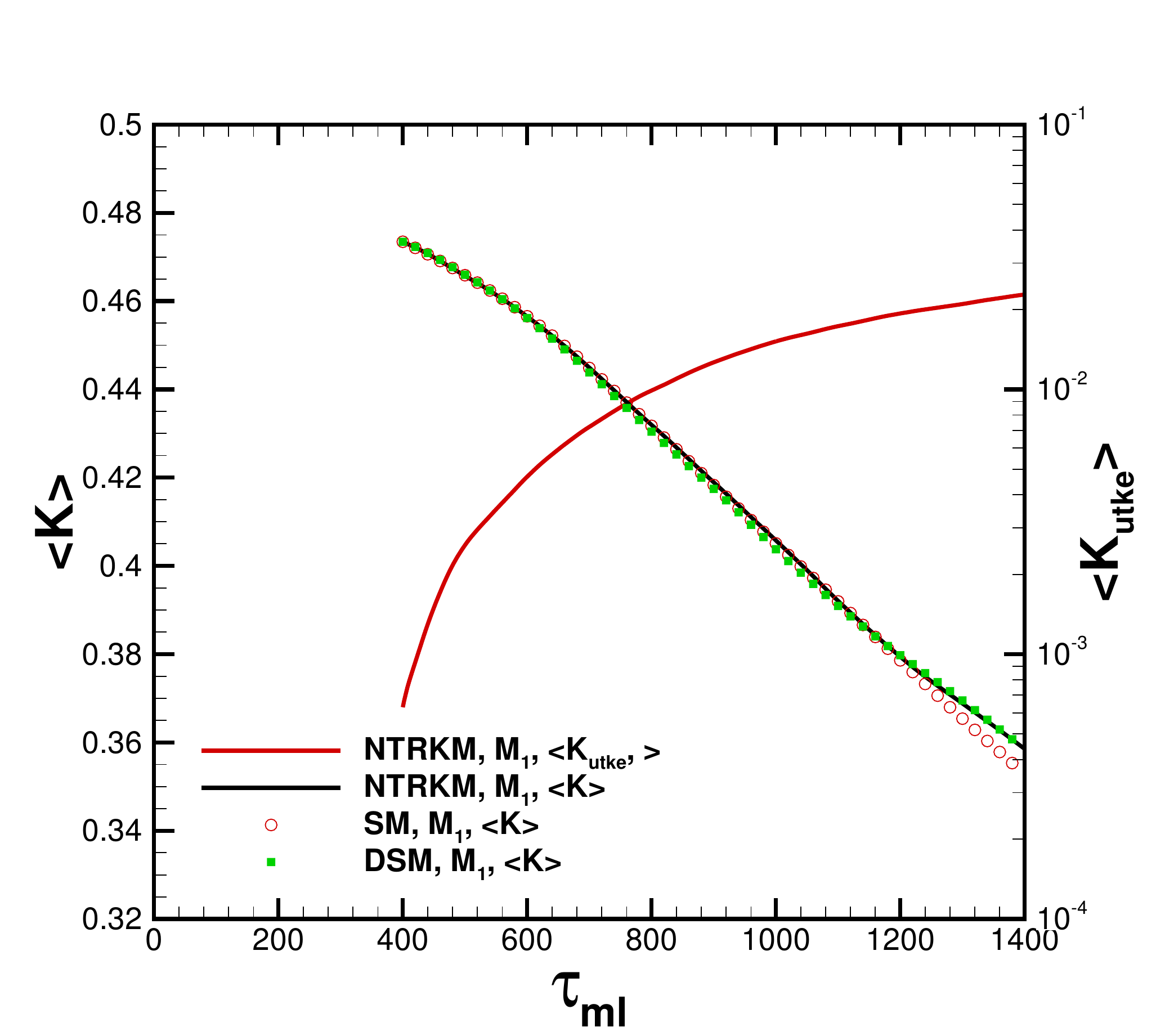}
		\put(-2,75){$(a)$}
	\end{overpic}	
	\begin{overpic}[width=0.495\textwidth]{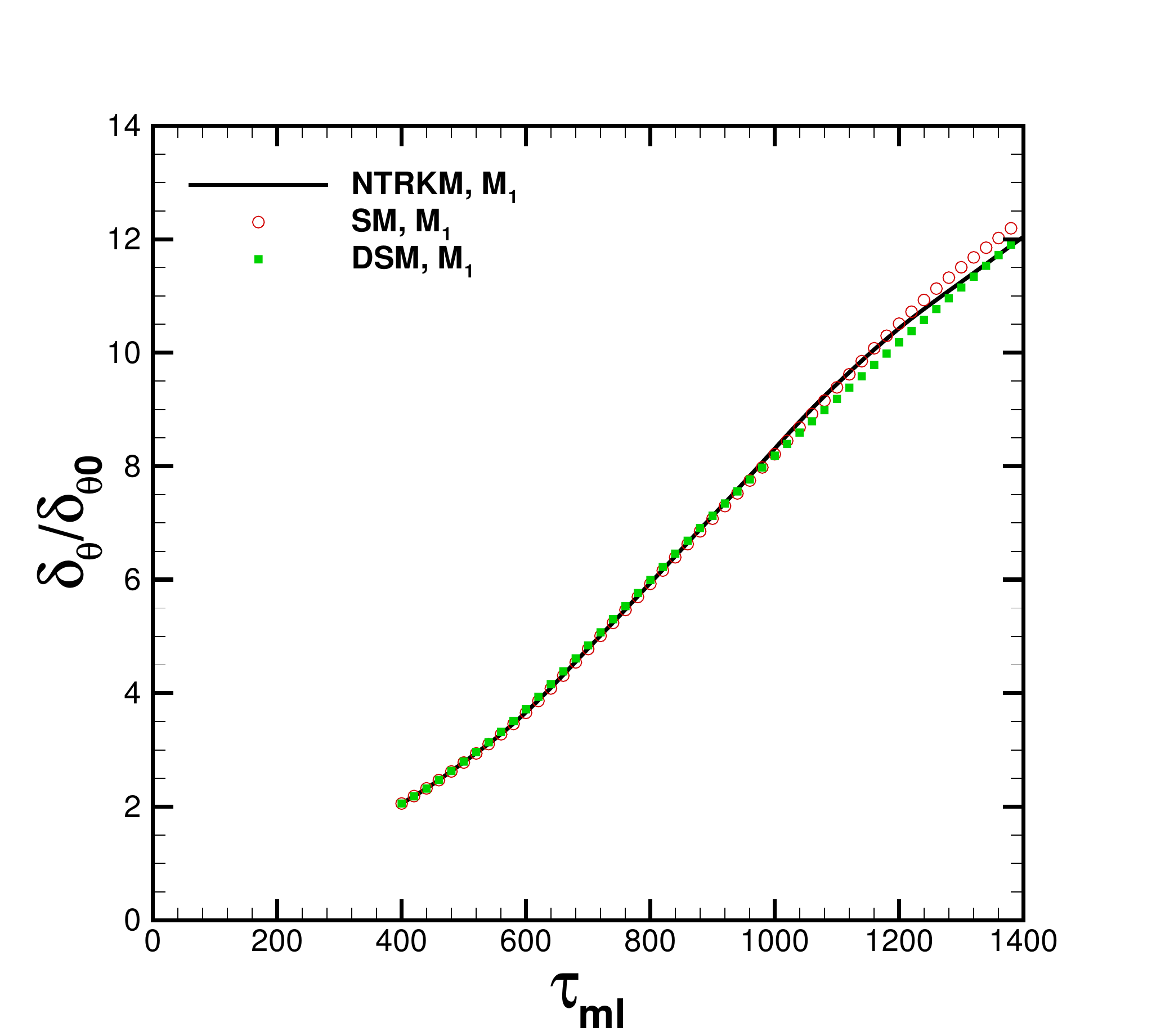}
		\put(-2,75){$(b)$}
	\end{overpic}
	\caption{\label{cml_les_Ktheta} Time history of $(a)$ ensemble unresolved $\left\langle K_{utke} \right \rangle$ (only from NTRKM) and ensemble resolved kinetic energy $\left\langle K \right \rangle$, and $(b)$ normalized momentum thickness $\delta_{\theta}/\delta_{\theta 0}$ for case M$_1$.}
	\vspace{-4mm}
\end{figure}

Figure \ref{cml_les_Ktheta}$(a)$ shows the time history of ensemble unresolved $\left\langle K_{utke} \right \rangle$ from NTRKM and ensemble resolved kinetic energy $\left\langle K \right \rangle$.
The ensemble unresolved TKE from NTRKM increases, while the ensemble resolved kinetic energy $\left\langle K \right\rangle$ decrease in the dissipative system.
Figure \ref{cml_les_Ktheta}$(b)$ also shows the evolution of normalized momentum thickness $\delta_{\theta}/\delta_{\theta 0}$.
Figure \ref{cml_les_Ktheta} shows that the performance of statistical quantities from NTRKM is much closer with that from DSM.
In table \ref{cml_quantities_table_les}, the normalized growth rate $\dot{\delta}_{\theta}/\dot{\delta}_{inc} = 0.611$ from NTRKM agrees well with that from DSM, slightly larger than that of DNS.
However, SM overestimates the normalized growth rate up to $20\%$.
At the center plane, table \ref{cml_quantities_table_les} also shows the vorticity thickness-based Reynolds number $Re_{\omega}$, and the turbulent Mach 
\clearpage
\begin{table}
	\begin{center}
		\centering
		\begin{tabular}{ccccc}
			Case M$_1$    &$\dot{\delta}_{\theta}/\dot{\delta}_{inc}$  &$Re_{\omega}$  &$Ma_t$   &$(b_{11}, b_{12}, b_{22})$ \\
			NTRKM  &0.611  &11912     &0.27    &(0.13, 0.16, -0.08)  \\
			SM     &0.694  &10106     &0.29    &(0.12, 0.16, -0.04)    \\
			DSM    &0.619  &11814     &0.30    &(0.15, 0.15, -0.08)   
		\end{tabular}
		\caption{\label{cml_quantities_table_les} Key quantities for LES in TCPML at $\tau_{ml} = 1400$.}
	\end{center} 
\end{table}
\hspace{-5mm}
number $Ma_t$ from LES at $\tau_{ml} = 1400$.
The vorticity thickness-based Reynolds numbers $Re_{\omega}$ based on three LES models are larger than that of DNS, where the turbulent Mach number $Ma_t$ from LES agrees well with that of DNS.
Components of anisotropy stress tensor at $\tau_{ml} = 1400$ agree well with each other in $b_{11}$ and $b_{12}$ (see equation \eqref{cml_reynolds_anisotropic1}).
The large deviation appears in $b_{22}$, as NTRKM solution coincides with the DSM solution.
Compared with DNS results in table \ref{cml_quantities_table_dns}, NTRKM and DSM outperform the SM on $b_{22}$.

\begin{figure}
	\centering
	\begin{overpic}[width=0.495\textwidth]{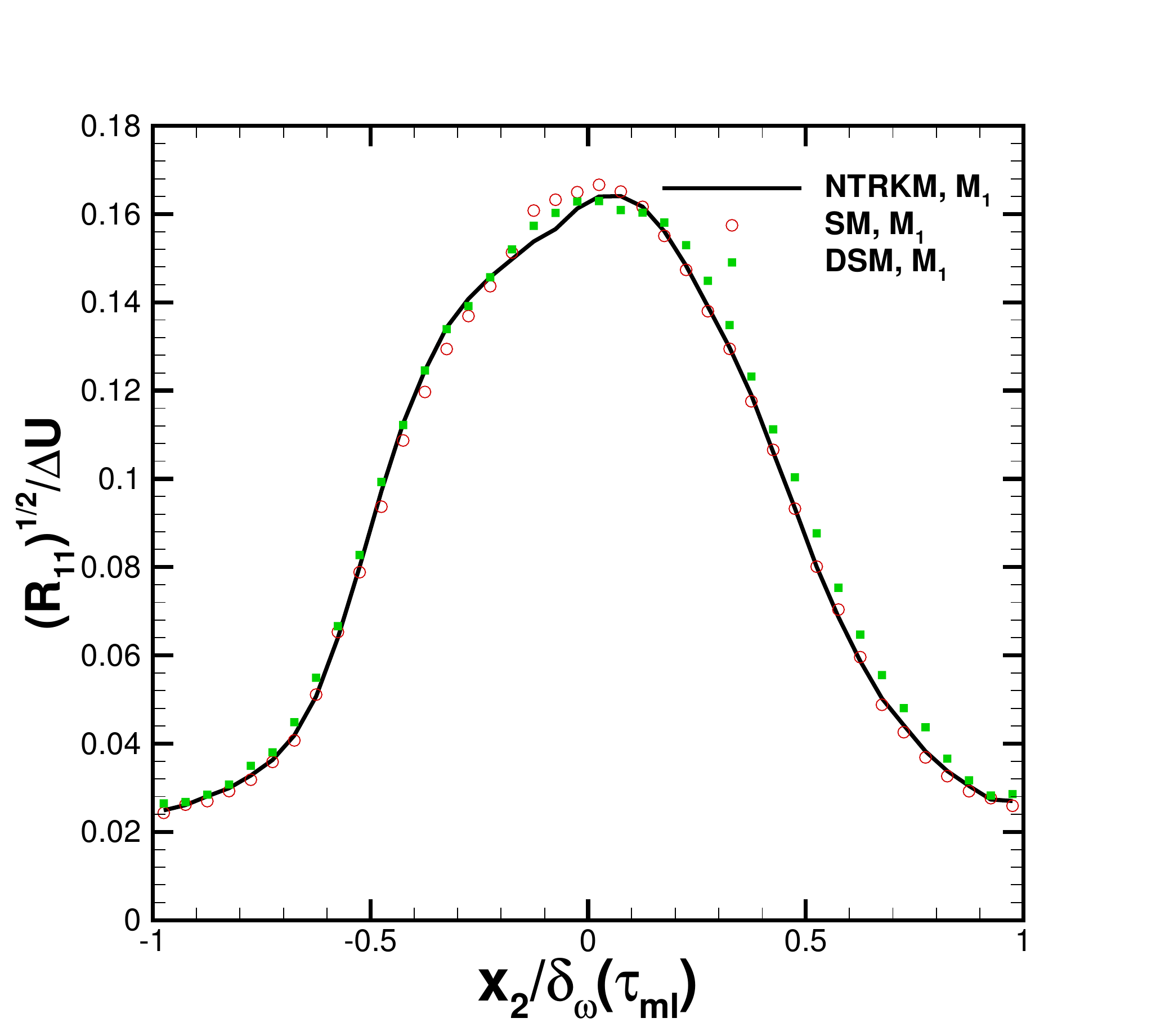}
		\put(-2,75){$(a)$}
	\end{overpic}	
	\begin{overpic}[width=0.495\textwidth]{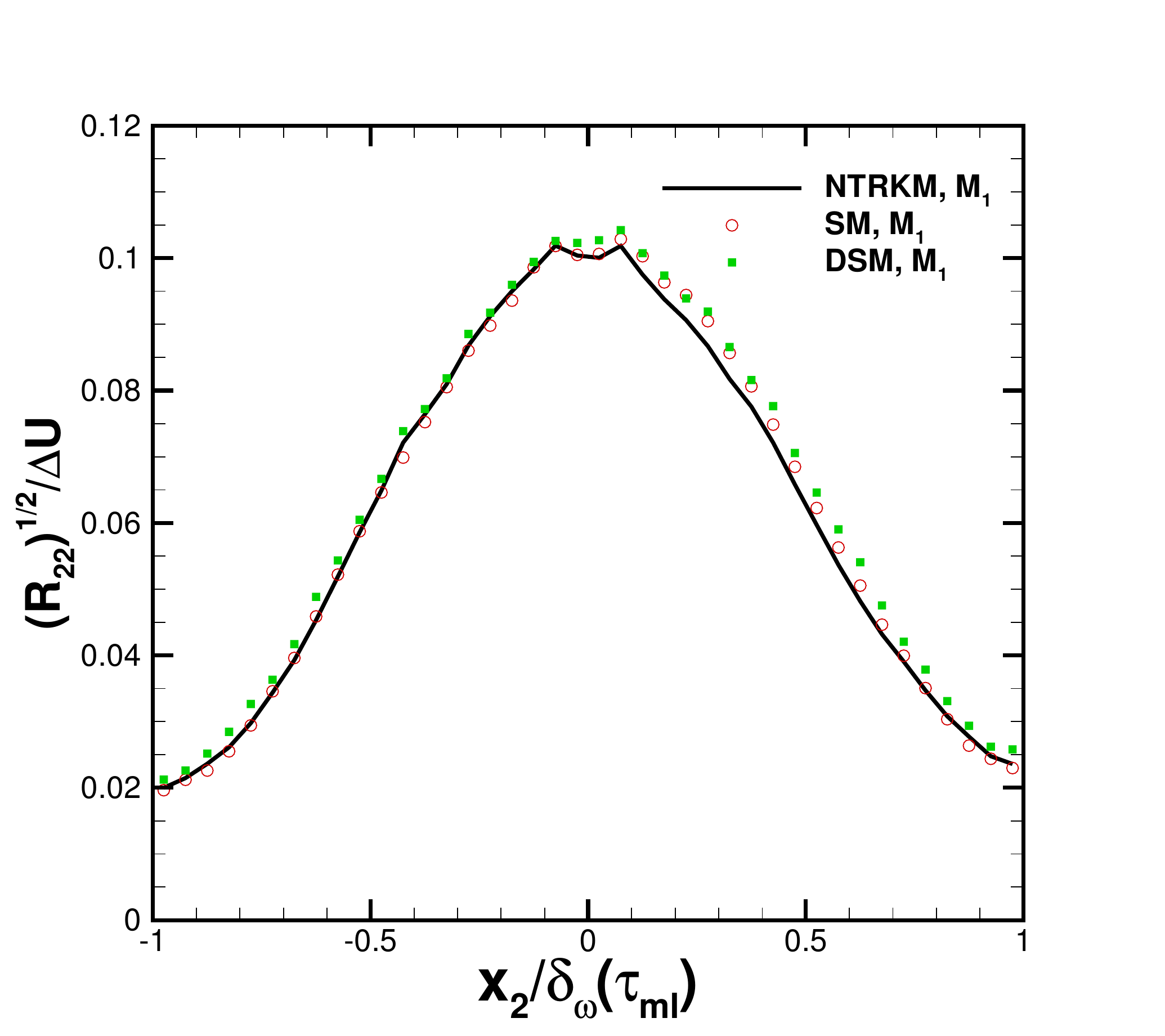}
		\put(-2,75){$(b)$}
	\end{overpic}
	\begin{overpic}[width=0.495\textwidth]{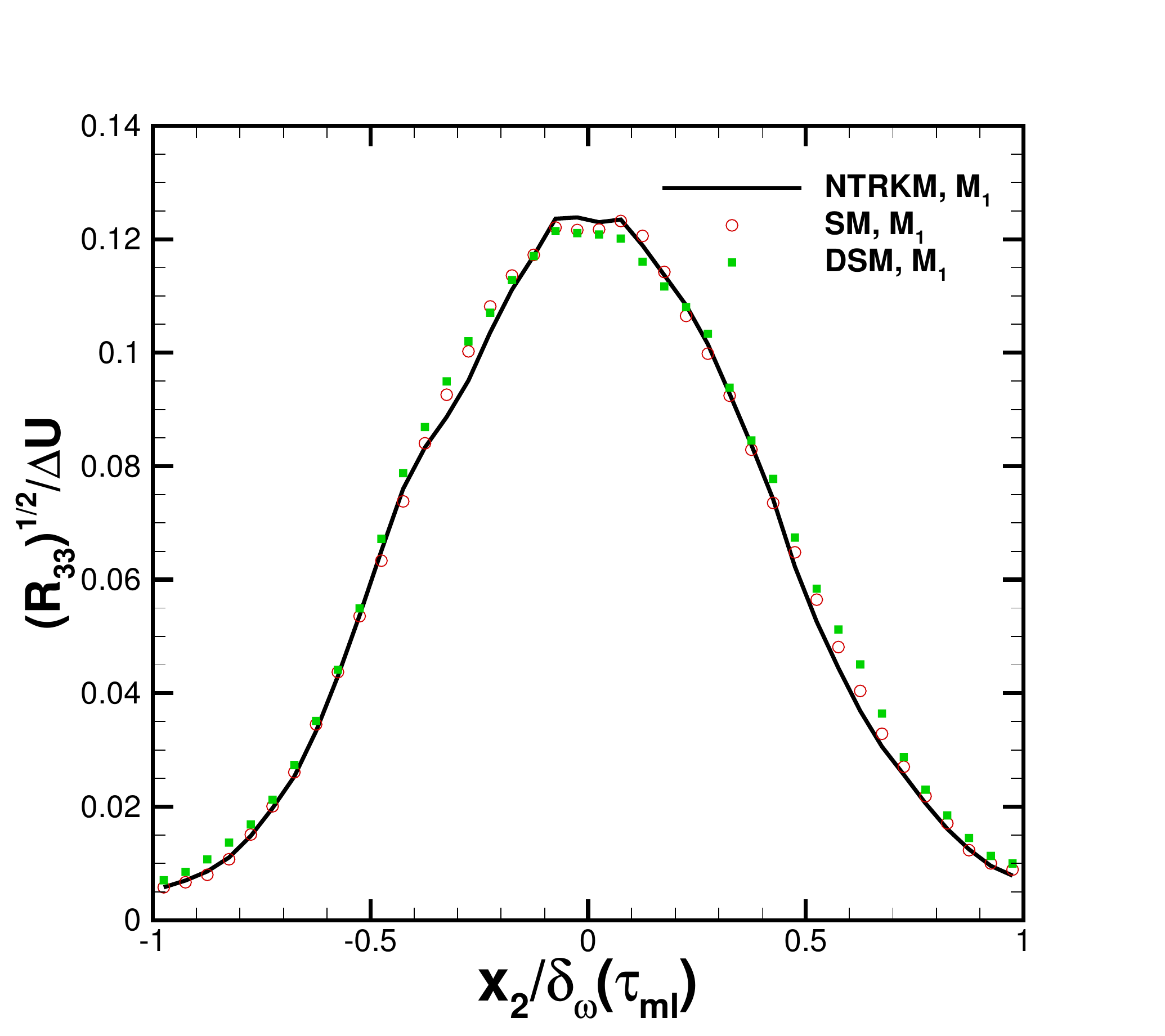}
		\put(-2,75){$(c)$}
	\end{overpic}
	\begin{overpic}[width=0.495\textwidth]{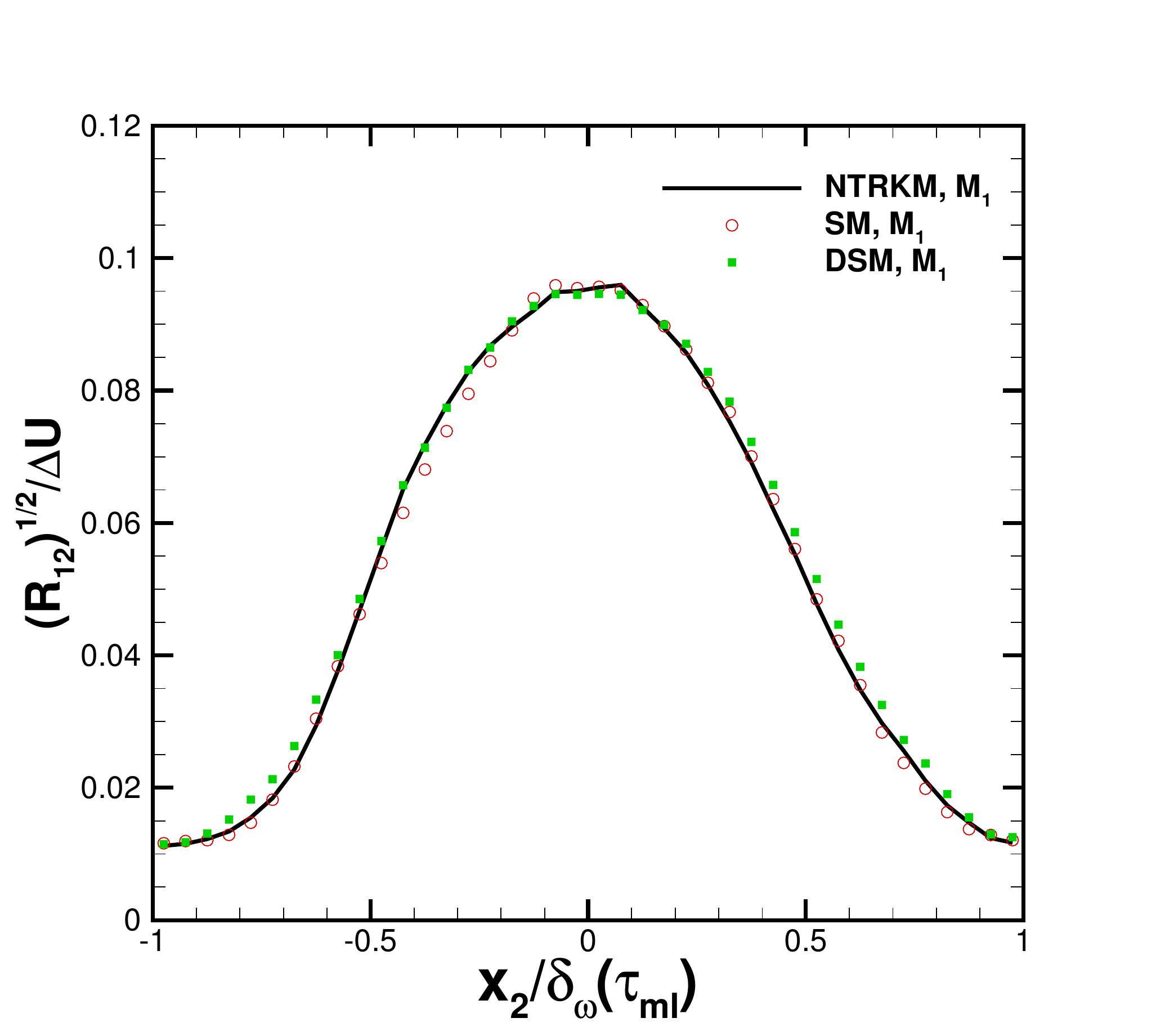}
		\put(-2,75){$(d)$}
	\end{overpic}
	\caption{\label{cml_les_r11122233}  Profiles of normalized turbulent stress $(a)$ $(R_{11})^{1/2}/\Delta U$, $(b)$ $(R_{22})^{1/2}/\Delta U$, $(c)$ $(R_{33})^{1/2}/\Delta U$ and $(d)$ $(R_{12})^{1/2}/\Delta U$ for case M$_1$.}
\end{figure}

Figure \ref{cml_les_r11122233} presents the profiles of normalized turbulent stress $(R_{11})^{1/2}/\Delta U$, $(R_{22})^{1/2}/\Delta U$, $(R_{33})^{1/2}/\Delta U$, and $(R_{12})^{1/2}/\Delta U$ (see equation \eqref{cml_reynolds_anisotropic2}).
The normalized turbulent stress from three models shows quite small deviations. 
Against the DNS solution as shown in figure \ref{cml_dns_reynoldstress}, the peak magnitude of turbulent stress as $(R_{11})^{1/2}/\Delta U$ and $(R_{22})^{1/2}/\Delta U$ are smaller in LES simulations, indicating that LES on unresolved grids underestimate the turbulent fluctuations.
In TCPML, the performance of key turbulent quantities up to second-order statistics confirm that current NTRKM is comparable with the widely-used eddy-viscosity models.
The results of NTRKM and DSM are much closer to DNS solution, and outweighing the SM guided with recommended modeling coefficient.

\section{Concluding remarks}\label{sec:NTRKM_conclusion}
We propose the non-equilibrium time-relaxation kinetic model for compressible turbulence modeling for the first time. 
The key idea is constructing the non-trivial turbulent quantities and its corresponding dynamic evolution quantitatively on unresolved grids.
Within the non-equilibrium time-relaxation framework, NTRKM is constructed in the form of modified BGK model on unresolved grids.
Based on the $1$st-order Chapman-Enskog expansion, NTRKM connects with the six-variable macroscopic governing equations. 
Phenomenologically, the unknown turbulent relaxation time and source term in NTRKM are determined by gradient-type assumption and dynamic modeling approach.
Therefore, the non-equilibrium kinetic model provides an profound mesoscopic understanding for transport equation of the compressible SGS turbulence kinetic energy. 
To solve the NTRKM accurately and robustly, finite volume non-equilibrium gas-kinetic scheme is developed in the spirt of well-established gas-kinetic scheme.
DCIT and TCPML are simulated as benchmarks to evaluate current non-equilibrium kinetic model and non-equilibrium gas-kinetic scheme. 
The performance of key statistical turbulent quantities up to second-order statistics confirms that current NTRKM is comparable with the widely-used eddy-viscosity models. 
As expected, the performance of NTRKM is much closer with DSM and outperforming SM.  
Present work not only points an alternative way for compressible turbulence modeling on unresolved grids, but also opens the great possibilities to simulate multi-scale flow physics within the non-equilibrium time-relaxation framework.

NTRKM will be further implemented for the more practical compressible turbulent flows to validate its strength, such as compressible wall-bounded turbulent flows and shock-boundary layer interaction \citep{chen2012reynolds, chen2017constrained}.
Compared with incompressible turbulence, compressible turbulent flows are more complex due to the non-linear coupling of velocity, density and pressure fields. 
Consequently, compressible LES models are much difficult to construct than the incompressible ones.
Be of scientific interest, the optimal physical distribution of $K_{utke}$ and modeling the source term directly as equation \eqref{source_time_relaxation} on unresolved grids, still requires the continuous effort in understanding the non-equilibrium properties of turbulence. 
To that extent, a complete time-relaxation compressible turbulence modeling without any ad-hoc technique from macroscopic turbulence models may be achieved.
In addition, if NTRKM can be solved by multi-scale kinetic solver, such as UGKS \citep{xu2010unified}, the non-equilibrium multi-scale fluxes may improve its performance correspondingly.
These challenging topics deserve to be explored in the subsequent studies.

~\\
\begin{flushleft}
\textbf{Acknowledgements.}
Thank Z.Y. Wang and Dr. W. Xu at HKUST for helpful discussions. 
The authors would like to thank TaiYi supercomputers in the SUSTech for providing high performance computational resources.
\end{flushleft}
~\\
\begin{flushleft}
\textbf{Funding.}
This research is supported by the National Natural Science Foundation of China (91852114, 11772281 and 11701038), the National Numerical Windtunnel project, the Department of Science and Technology of Guangdong Province (2020B1212030001), and the Fundamental Research Funds for the Central Universities.
\end{flushleft}
~\\
\begin{flushleft}
\textbf{Declaration of interests.}
The authors report no conflict of interest.
\end{flushleft}

\appendix
\section{Connection between non-equilibrium time-relaxation kinetic model and macro governing equations}\label{appA}
This appendix provides the details for the derivation of corresponding macroscopic governing equations on unresolved grids based on the NTRKM. 
Derivation of the Euler equations and the NS equations from the BGK model can be found in the Appendix B \citep{xu2015direct}. 
Similar as molecular relaxation time $\tau = \epsilon \hat{\tau}$ in refereed derivation, the turbulent relaxation time is rewritten as $\tau_t = \epsilon \hat{\tau_t}$ in which $\epsilon$ is a small dimensionless quantity. 
Suppose that $f^{eq}$ has a Taylor series expansion about point ($\boldsymbol{x}$, $t$). Since $\tau$ and $\tau_t$ depends on the local thermodynamic variables and unresolved $K_{utke}$, and these depend on the moments of $f^{eq}$, we may assume that $\tau$ and $\tau_t$ are consequently $\hat{\tau}$ and $\hat{\tau_t}$ can be expanded about the point ($\boldsymbol{x}$, $t$). 
Now consider the formal solution of the NTRKM for $f(\boldsymbol{x},t,\boldsymbol{u},\xi, k_u)$, supposing that $f^{eq}$ is known, it can be shown that $f(\boldsymbol{x},t,\boldsymbol{u},\xi, k_u)$ has an expansion in powers of $\epsilon$. 

We derive the terms in this expansion from the formal solution of $f(\boldsymbol{x},t,\boldsymbol{u},\xi, k_u)$, by putting the power expansion $f = f_0 + \epsilon f_1 + \epsilon^2 f_2 + \cdots$, and $(\tau + \tau_t) = \epsilon (\hat{\tau} + \hat{\tau_t})$ into the NTRKM directly.
Let $D_{\boldsymbol{u}} =  \partial / \partial t + u_i \partial / \partial x_i$, and write the NTRKM as $\epsilon (\hat{\tau} + \hat{\tau_t}) D_{\boldsymbol{u}} f + f - f^{eq} = 0$, with handling the source term $Q_s$ in equation \eqref{boltamann_bgk_neq} splitly. 
An expansion of this equation in powers of $\epsilon$ yields 
\begin{equation}
	f = g - \epsilon (\hat{\tau} + \hat{\tau_t}) D_{\boldsymbol{u}} f^{eq} + \epsilon^2 (\hat{\tau} + \hat{\tau_t}) D_{\boldsymbol{u}}((\hat{\tau} + \hat{\tau_t}) D_{\boldsymbol{u}} f^{eq}) + \cdots.
\end{equation}
According to compatibility condition, after dividing by $\epsilon (\hat{\tau} + \hat{\tau_t})$, gives
\begin{equation}\label{Appendix_1}
	\begin{aligned}
		\int \boldsymbol{\psi} D_{\boldsymbol{u}} f^{eq} \text{d} \Xi = \epsilon \int \boldsymbol{\psi} D_{\boldsymbol{u}} [(\hat{\tau} + \hat{\tau_t}) D_{\boldsymbol{u}} f^{eq}] \text{d} \Xi + \mathcal{O}{(\epsilon^2)},
	\end{aligned}
\end{equation}
where $\boldsymbol{\psi} = (1, u_1, u_2, u_3, \displaystyle\frac{1}{2}(u_1^2 + u_2^2 + u_3^2 + \xi^2) + k_u, k_u)^T$, and $\text{d} \Xi = \text{d}u_1 \text{d}u_2 \text{d}u_3  \text{d}\xi \text{d}k_u$.
The integral on the left-hand-side $\mathcal{L}_{\alpha}$ and the right-hand-side $\mathcal{R}_{\alpha}$ of equation \eqref{Appendix_1} can be defined as
\begin{equation}\label{Appendix_2}
	\begin{aligned}
		\mathcal{L}_{\alpha} = \epsilon \mathcal{R}_{\alpha} + \mathcal{O}{(\epsilon^2)},
	\end{aligned}
\end{equation}
which shows that $\mathcal{L}_{\alpha}$ is at least $\mathcal{O}{(\epsilon)}$, obviously. 
Therefore, in reducing the $\mathcal{R}_{\alpha}$ on the right side of equation \eqref{Appendix_2}, which is already $\mathcal{O}{(\epsilon)}$, we can drop $\mathcal{O}{(\epsilon)}$ quantities and their derives.
To simplify the notation, let
\begin{equation}
\begin{aligned}
	<\psi_{\alpha} (\cdot)> \equiv \int \psi_{\alpha} f^{eq} \text{d} \Xi, \quad \alpha = 1, 2, 3, 4, 5, 6,
\end{aligned}
\end{equation}
where $<\psi_{\alpha} (\cdot)>$ denotes the moments of $f^{eq}$ on $\psi_{\alpha}$, and $\psi_{\alpha}$ is the component of $\boldsymbol{\psi}$. 
Therefore, $\mathcal{L}_{\alpha}$ and $\mathcal{R}_{\alpha}$ are rewritten as
\begin{equation}
	\begin{aligned}
	\mathcal{L}_{\alpha} &= <\psi_{\alpha}>_{,t} + <\psi_{\alpha} u_l>_{,l}, \\
	\mathcal{R}_{\alpha} &= \{(\hat{\tau} + \hat{\tau_t}) [<\psi_{\alpha} u_k>_{,t} + <\psi_{\alpha} u_k u_l>_{,l}]\}_{,k} + \mathcal{O}{(\epsilon)},
\end{aligned}
\end{equation}
where $k$ and $l$ are subscript of molecular velocity, taking from $1$ to $3$ for three-dimension derivation. 
The macroscopic governing equations in Euler-type and NS-type can be obtained by truncating the equation \eqref{Appendix_2} up to the order of $\mathcal{O}{(1)}$ and $\mathcal{O}{(\epsilon)}$, respectively. 
The Euler-type macroscopic equations can be derived straightforwardly by the Chapman-Enskog expansion up to zeroth order, which are used to simplify the derivation of NS-type equations.
While, this appendix focuses on the NS-type macroscopic equations by the Chapman-Enskog expansion up to first order truncation of $\tau + \tau_t$.

\subsection{Continuity equation}
Continuity equation is derived straightforwardly as 
\begin{equation}\label{Appendix_3}
	\begin{aligned}
		\rho_{,t} + (\rho U_k)_{,k} = 0, 
	\end{aligned}
\end{equation}
which can be used to simplify the momentum equations, the total energy equations, and the unresolved TKE equation.

\subsection{Momentum equation}
To simplify the time derivative of pressure in the momentum equations, we introduce the following procedure firstly. 
For total energy equation, the left side $\mathcal{L}_5$ of equation \eqref{Appendix_2} can be grouped as
\begin{equation}
	\begin{aligned} 
	\mathcal{L}_5 &= \frac{1}{2}U_n^2[{\rho_{,t} + (\rho U_k)_{,k}}] + \rho U_n U_{n,t} + \rho U_k U_n U_{n,k} + U_k p_{,k} \\
	&+ \frac{N + 3}{2} [p_{,t} + U_k p_{,k}] + \frac{N + 5}{2} p U_{k,k} + {(\rho K_{utke})_{,t} + (\rho K_{utke} U_k)_{,k}},
\end{aligned}
\end{equation}
where $N$ is the total number of degrees of freedom in $\xi$.
The first term is $U_n^2 \mathcal{L}_1/2$ which is $\mathcal{O}(\epsilon^2)$, and next three are $U_n \mathcal{L}_n$, and are therefore $\mathcal{O}(\epsilon)$. 
Then $\mathcal{L}_5$ can be rewritten as 
\begin{equation} 
	\begin{aligned} \label{Appendix_4}
		\mathcal{L}_5 &= \frac{N + 3}{2} [p_{,t} + U_k p_{,k}] + \frac{N + 5}{2} p U_{k,k} + U_n \mathcal{L}_n \\
		&+ {(\rho K_{utke})_{,t} + (\rho K_{utke} U_k)_{,k}} + \mathcal{O}{(\epsilon^2)}. 
	\end{aligned}
\end{equation}
Based on the Chapman-Enskog expansion up to $0$th order, unresolved TKE equation $\mathcal{L}_6$ in Euler-type can be written as
\begin{equation}\label{Appendix_5}
	\begin{aligned}
		\mathcal{L}_6 = (\rho K_{utke})_{,t} + (\rho K_{utke} U_k)_{,k} - S_t  + \mathcal{O}{(\epsilon)},
	\end{aligned}
\end{equation}
which can be used to simplify the time derivative of $(\rho K_{utke})_{,t}$ in the following derivation. 
Combining equation \eqref{Appendix_4} and equation \eqref{Appendix_5}, we get
\begin{equation}\label{Appendix_6}
	\begin{aligned}
		p_{,t} + U_k p_{,k} = -\frac{N + 5}{N + 3} p U_{k,k} - \frac{2 S_t}{N + 3} + \mathcal{O}(\epsilon),
	\end{aligned}
\end{equation}
which can be used to simplify the time derivative of pressure $p_{,t}$ in the following derivation.

For the right hand sides of the momentum equations in equation \eqref{Appendix_2}, considering $\mathcal{R}_j = [(\hat{\tau} + \hat{\tau_t}) F_{jk}]_{,k}$, we get 
\begin{equation}
	\begin{aligned}
	F_{jk} &= <u_j u_k>_{,t} + <u_j u_k u_l>_{,l} \\
	&= U_j[{{(\rho U_k)_{,t} + [(\rho U_k U_l) + p \delta_{kl}]_{,l}}}] + \rho U_k U_{j,t} + (p \delta_{jk})_{,t} \\
	& + (\rho U_k U_l + p \delta_{kl}) U_{j,l} + (U_l p \delta_{jk} + U_k p \delta_{jl})_{,l},
	\end{aligned}
\end{equation}
where the fact that all odd moments in $w_k$ vanish has been used. Here $w_k = u_k - U_k$ is the peculiar velocity. 
The term in square brackets multiplying $U_j$ is $\mathcal{L}_k$, i.e. it is $\mathcal{O}(\epsilon)$, and can therefore be ignored. 
Then, after gathering terms with coefficients $U_k$ and $p$, we have
\begin{equation}
	\begin{aligned}
	F_{jk} = U_k[{{\rho U_{j,t} + \rho U_l U_{j,l} + p_{,j}}}] + p[U_{k,j} + U_{j,k} + U_{l,l} \delta_{jk}] + \delta_{jk}[{{p_{,t} + U_l p_{,l}}}] + \mathcal{O}(\epsilon). 
	\end{aligned}
\end{equation}
The coefficient of $U_k$ can be simplified by the following equation
\begin{equation}\label{Appendix_7}
	\begin{aligned}
		\mathcal{L}_i = \rho U_{i,t} + \rho U_k U_{i,k} + p_{,i} + \mathcal{O}(\epsilon^2), 
	\end{aligned}
\end{equation}
where equation \eqref{Appendix_7} is derived by multiplying the continuity equation by $U_i$ and subtracting the result from $\mathcal{L}_i$ ($i = 2,3,4$). 
To eliminate $p_{,t}$ from the last term we use the equation \eqref{Appendix_6} for $\mathcal{L}_5$. 
Finally, decompose the tensor $U_{k,j}$ into its dilation and shear parts in the usual way, which gives
\begin{equation}\label{Appendix_8}
	\begin{aligned}
		F_{jk} = p[U_{k,j} + U_{j,k} - \frac{2}{3} U_{l,l} \delta_{jk}] + \frac{2}{3}\frac{N}{(N + 3)}p U_{l,l} \delta_{jk} - \frac{2 S_t}{N + 3} \delta_{jk} + \mathcal{O}(\epsilon). 
	\end{aligned}
\end{equation}
The second term is due to bulk viscosity involves energy sharing between transnational and internal degrees of freedom of the molecular \citep{xu2015direct, cramer2012numerical}, and the last term resulted from the energy interaction between the resolved kinetic energy and the unresolved TKE.

\subsection{Total energy equation}
Analogy to derive the NS total energy equation, we write $\mathcal{R}_{5} = \{(\hat{\tau} + \hat{\tau_t}) N_k\}_{, k}$ with
\begin{equation}
	\begin{aligned}
	N_k = <u_k (\frac{u_n^2 + \xi^2}{2} + k_u)>_{,t} + <u_k u_l (\frac{u_n^2 + \xi^2}{2} + k_u)>_{,l},
	\end{aligned}
\end{equation}
where the $k_u$ is the sample-space variable corresponding to unresolved $K_{utke}$.
$N_k$ can be decomposed into $N_k = N_k^{(1)} + N_k^{(2)}$, where 
\begin{equation}
	\begin{aligned}
	N_k^{(1)} = [U_k (\frac{u_n^2 + \xi^2}{2} + k_u)]_{,t} + [U_k <u_l (\frac{u_n^2 + \xi^2}{2} + k_u)>]_{,l},
	\end{aligned}
\end{equation}
and
\begin{equation}
	\begin{aligned}
	N_k^{(2)} = <w_k (\frac{u_n^2 + \xi^2}{2} + k_u)>_{,t} + <w_k u_l (\frac{u_n^2 + \xi^2}{2} + k_u)>_{,l}.
	\end{aligned}
\end{equation}

For $N_k^{(1)}$, we have
\begin{equation}
	\begin{aligned}
	N_k^{(1)} &= U_k[{{<\frac{u_n^2 + \xi^2}{2} + k_u>_{, t} +  <u_l (\frac{u_n^2 + \xi^2}{2} + k_u)>_{, l} }}] \\
	&+[\frac{1}{2} \rho U_n^2 + \frac{N + 3}{2} p + \rho K_{utke}] U_{k,t} + [U_l (\frac{1}{2}\rho U_n^2 + \frac{N + 5}{2} p + \rho K_{utke})] U_{k,l}.
	\end{aligned}
\end{equation}
The coefficient of $U_k$ in the equation above is $\mathcal{L}_5$, and therefore can be dropped, and the remaining terms can be rewritten as
\begin{equation}
	\begin{aligned}
	N_k^{(1)} = [\frac{1}{2} \rho U_n^2 + \frac{N + 3}{2} p + \rho K_{utke}] [U_{k,t} + U_l  U_{k,l}] + p U_l U_{k,l} + \mathcal{O}(\epsilon).
	\end{aligned}
\end{equation}
According to equation \eqref{Appendix_7} to replace $U_{k,t}$, we get
\begin{equation}\label{Appendix_9}
	\begin{aligned}
		N_k^{(1)} &= - [\frac{1}{2} U_n^2 + \frac{N + 3}{2} \frac{p}{\rho} + K_{utke}] p_{,k} + p U_l U_{k,l} + \mathcal{O}(\epsilon). 
	\end{aligned}
\end{equation}

For $N_k^{(2)}$, remembering that moments odd in $w_k$ vanish, we have
\begin{equation}
	\begin{aligned}
	N_k^{(2)} &= <U_n w_n w_k>_{,t} + <U_l U_n w_n w_k>_{,l} \\
	&+ \frac{1}{2} <U_n^2 w_k w_l>_{,l} + <w_k w_l(\frac{w_n^2 + \xi^2}{2} + k_u)>_{,l} \\
	&= (p U_k)_{,t} + (p U_k U_l)_{,l} + \frac{1}{2}(U_n^2 p)_{,k} + \frac{N + 5}{2} (\frac{p^2}{\rho})_{,k} + (p K_{utke})_{,k}. 
	\end{aligned}
\end{equation}
$N_k^{(2)}$ can be rewritten as
\begin{equation}
	\begin{aligned}
	N_k^{(2)} &= p [U_{k,t} + U_l U_{k,l} + U_k U_{l,l} + U_l U_{l,k}] \\
	&+ U_k(p_{,t} + U_l p_{,l}) + \frac{1}{2} U_n^2 p_{,k} + \frac{N + 5}{2} (\frac{p^2}{\rho})_{,k} + (p K_{utke})_{,k}.
	\end{aligned}
\end{equation}
The $p_{,t}$ and $U_{k,t}$ can be replaced by equation \eqref{Appendix_6} and equation \eqref{Appendix_7}, respectively. 
Hence
\begin{equation}\label{Appendix_10}
	\begin{aligned}
		N_k^{(2)} &= p [- \frac{p_{,k}}{\rho} + U_k U_{l,l} + U_l U_{l,k}] + U_k[{ {- \frac{N + 5}{N + 3}p U_{l, l} - \frac{2 S_t}{N + 3}} }] \\
		&+ \frac{1}{2} U_n^2 p_{,k} + \frac{N + 5}{2} (\frac{p^2}{\rho})_{,k} + (p K_{utke})_{,k} + \mathcal{O}(\epsilon). 
	\end{aligned}
\end{equation}

Finally, $N_k$ can be obtained by summing $N_k^{(1)}$ and $N_k^{(2)}$ up
\begin{equation}\label{Appendix_11}
	\begin{aligned}
		N_k &=  p [U_l (U_{k,l} + U_{l,k}) - \frac{2}{N + 3} U_k U_{l,l} ] - U_k \frac{2S_t}{N + 3}  \\
		&+  {\frac{N + 5}{2} p (\frac{p}{\rho})_{,k} + (p K_{utke})_{,k}} + \mathcal{O}(\epsilon). 
	\end{aligned}
\end{equation}

\subsection{Unresolved turbulence kinetic energy equation}
For unresolved $K_{utke}$ equation, we write $\mathcal{R}_{6} = \{(\hat{\tau} + \hat{\tau_t}) Z\}_{, k}$, where
\begin{equation}
	\begin{aligned}
	Z &= <k u_k>_{, t} + <k u_k u_l>_{, l} \\
	&= K_{utke} [(\rho U_k)_{,t} + (\rho U_k U_l + p \delta_{kl})_{,l}] + \rho U_k K_{utke,t} + K_{utke,l} [\rho U_k U_l + p \delta_{kl}].
	\end{aligned}
\end{equation}
The term in square brackets is $\mathcal{L}_k$, i.e. $\mathcal{O}(\epsilon)$, and can be dropped. 
Equation \eqref{Appendix_5} subtracts the multiplication of the continuity  equation \eqref{Appendix_3} by $K_{utke}$ gives
\begin{equation}\label{Appendix_12}
	\begin{aligned}
		\mathcal{L}_6 = \rho K_{utke, t} + \rho U_l K_{utke,l} - S_t + \mathcal{O}(\epsilon^2). 
	\end{aligned}
\end{equation}
Gathering terms with coefficients $U_k$ and $p$, and replacing $K_{utke, t}$ through equation \eqref{Appendix_12}, we have
\begin{equation}\label{Appendix_13}
	\begin{aligned}
		Z = U_k S_t + p K_{utke, k} + \mathcal{O}(\epsilon).
	\end{aligned}
\end{equation}

All time derivatives have now been removed from $\mathcal{R}_{\alpha}$, and the remaining steps in deriving corresponding macroscopic governing equations for NTRKM may be summarized briefly as
\begin{itemize}
	\item Drop $\mathcal{O}(\epsilon^2)$ in equation \eqref{Appendix_2},
	\item Combine $\epsilon$ and  $\hat{\tau} + \hat{\tau_t}$ to recover $\tau + \tau_t = \epsilon (\hat{\tau} + \hat{\tau_t})$,
	\item Define the molecular dynamic viscosity as $\mu = \tau p$, and turbulent eddy viscosity is recovered by $\mu_t = \tau_t p$,
	\item Define the molecular thermal conductivity $\kappa = (N + 5) \tau p/2 $, and the turbulent thermal conductivity $\kappa_t = (N + 5) \tau_t p/2$.	
\end{itemize}
Finally, corresponding macroscopic governing equations to NTRKM can be rewritten as equations \eqref{tgks_macro_formula1}-\eqref{tgks_macro_formula4} in \S \ref{subsec:closure}.

\section{Dynamic approach to determine modeling coefficients}\label{appB}
Model coefficients $C_s$, $Pr_t$, and $C_{\Pi}$ can be dynamically computed through Germano identity \citep{germano1991dynamic, lilly1992proposed}, which assumes the similarity of SGS quantities between the grid filter width $\overline{\Delta}$ and the test filter width $\widehat{\overline\Delta}$. 
In finite volume framework, explicit filter is not used, while the grid length of control volume itself acts as the grid filter width, and the projection process when updating the macroscopic variables can be regarded as the filtering process.
For any term $\phi = \overline{\beta_1 \beta_2} - \overline{\beta}_1  \overline{\beta}_2$ on grid filter level, assuming that $\Phi = \widehat{\overline{\beta_1 \beta_2}} - \widehat{\overline{\beta}}_1 \widehat{\overline{\beta}}_2$ still holds on test filter level. 
Then, the resolved tensor (or vector/scalar) is defined as $L = \Phi - \widehat{\phi}$. 
Assume $\phi$ is modeled by the linear constitutive relationship $\phi = C m$, where $m$ is a function of the resolved quantities. 
At the test filter level, $\Phi = C M$, $M$ takes similar form to $m$ but is a function of the test-filtered quantities. 
Plugging the linear model for $\Phi$ and $\phi$, the Germano identity reads
\begin{equation}\label{germano_identity}
	\begin{aligned} 
		L = \widehat{\overline{\beta}_1  \overline{\beta}_2} - \widehat{\overline{\beta}}_1 \widehat{\overline{\beta}}_2 = C (M - \widehat{m}).
	\end{aligned}
\end{equation}
To avoid computational instability, $C$ can be optimized by the least-square method \citep{lilly1992proposed}.

\subsection{Dynamic coefficient for density-weighted Smagorinsky model}\label{appB_1}
For dynamic density-weighted SM \citep{moin1991dynamic}, the model coefficient $C_{dsm}$ is determined by
\begin{equation}\label{dynamic_cm_dsm}
\left\{
	\begin{aligned}	
		L_{ij}^* &= -2C_{dsm} M_{ij},\\
		M_{ij} &= \widehat{\overline\Delta}^2 \widehat{\overline\rho} |\widehat{\widetilde{S}}| \widehat{\widetilde{S}}^*_{ij} - {\overline\Delta}^2 \reallywidehat{\overline{\rho} |\widetilde{S}| \widetilde{S}^*_{ij}}, \\
		C_{dsm} &= - \frac{L_{ij}^*M_{ij}}{2{M_{ij}M_{ij}}},
	\end{aligned}
\right.
\end{equation}
where  $L_{ij} = \reallywidehat{{{\overline{\rho{U_i}} \, \overline{\rho{U_j}}}}/{\overline{\rho}}} - \widehat{\overline{\rho{U_i}}} \, \widehat{\overline{\rho{U_j}}}/{\widehat{\overline{\rho}}}$, $L_{ij}^* = L_{ij} - L_{kk} \delta_{ij}/3$, $\widetilde{S}_{ij} = (\widetilde{U}_{i,j} + \widetilde{U}_{j,i})/2$, $\widetilde{S}_{ij}^{\ast} = \widetilde{S}_{ij} - \widetilde{S}_{kk} \delta_{ij}/3$, $|\widetilde{S}| = (2 \widetilde{S}_{ij} \widetilde{S}_{ij})^{\frac{1}{2}}$, $|\widehat{\widetilde{S}}| = (2 \widehat{\widetilde{S}}_{ij} \widehat{\widetilde{S}}_{ij})^{\frac{1}{2}}$.

\subsection{Dynamic coefficient for non-equilibrium time-relaxation kinetic model}\label{appB_2}
In NTRKM, $\tau_{ij}^{sgs}$ is modeled by the gradient-type eddy viscosity model as equation \eqref{mut_tauij_closure2}, and the coefficient $C_s$ is determined by
\begin{equation}\label{dynamic_cs_ii}
\left\{
	\begin{aligned}	
		L_{ij}^* &= -2C_sM_{ij},\\
		M_{ij} &= \widehat{\overline\Delta} \widehat{\overline\rho} \widehat{\overline{K}}_{utke}^{\frac{1}{2}} \widehat{\widetilde{S}}^*_{ij} - {\overline\Delta} \reallywidehat{\overline{\rho}K_{utke}^{\frac{1}{2}} \widetilde{S}^*_{ij}}, \\
		C_s &= - \frac{L_{ij}^*M_{ij}}{2{M_{ij}M_{ij}}},
	\end{aligned}
\right.
\end{equation}
where  $L_{ij} = \reallywidehat{{{\overline{\rho{U_i}} \, \overline{\rho{U_j}}}}/{\overline{\rho}}} - \widehat{\overline{\rho{U_i}}} \, \widehat{\overline{\rho{U_j}}}/{\widehat{\overline{\rho}}}$, $L_{ij}^* = L_{ij} - L_{kk} \delta_{ij}/3$, and $\widehat{\overline{K}}_{utke} = \widehat{K}_{utke} + (\widehat{\widetilde{U}_k \widetilde{U}_k} - \widehat{\widetilde{U}}_k \widehat{\widetilde{U}}_k)/2$.
When modifying the intrinsic turbulent Prandtl number $Pr_t = 1$ in NTRKM, the realistic turbulent Prandtl number can be computed dynamically as 
\begin{equation}\label{dynamic_prt_ii}
\left\{
	\begin{aligned}	
		L_{j} &= -C_s M_j/Pr_t,\\
		M_{j} &= \widehat{\overline{\Delta}} \widehat{\overline{\rho}} \widehat{\overline{K}}_{utke}^{\frac{1}{2}} \widehat{\widetilde{T}}_{,j} - \overline{\Delta} \reallywidehat{\overline \rho K_{utke}^{\frac{1}{2}} \widetilde{T}_{,j}},\\
		Pr_{t} &= - \frac{C_s M_jM_j}{L_jM_j},
	\end{aligned}
\right.
\end{equation}
where $L_{j} = R(\reallywidehat{{\overline{{\rho}T} \, \overline{{\rho}U_j}}/{\overline{\rho}}} - \widehat{\overline{{\rho}T}} \, \widehat{\overline{{\rho}U_j}}/{\widehat{\overline{\rho}}})$.
To model the pressure-dilation transfer, following the series expansion \citep{chai2012dynamic}, $C_{\Pi}$ reads
\begin{equation}\label{dynamic_cpi_ii}
\left\{
	\begin{aligned}	
		L &= C_\Pi M, \\
		M &= \widehat{\overline{\Delta}}^2 \widehat{\overline{p}}_{,j} (\widehat{\overline{\rho U_k}}/ {\widehat{\overline{\rho}}})_{j,k} - \overline{\Delta}^2 \reallywidehat{ \overline{p}_{,j} (\widetilde{U}_k)_{j,k}}, \\
		C_\Pi &= \frac{L}{M},
	\end{aligned}
\right.
\end{equation}
where $L = \reallywidehat{\overline{p} (\overline{\rho U_k}/{\overline \rho})_{,k}} - \widehat{\overline{p}} (\widehat{\overline{\rho U_k}}/\widehat {\overline \rho})_{,k}$.

The coefficients $C_{\epsilon s}$ and $C_{\epsilon d}$ are obtained by the analogy between the grid-filter-level SGS stress and the resolved stress across the test filter level \citep{menon1996high}, where $L = C M$ instead of $L = C (M - \widehat{m})$. 
To model the solenoidal dissipation rate, $C_{\epsilon s}$ is obtained by $ C_{\epsilon s} = L/M$, where 
\begin{equation}\label{dynamic_ceps_s}
\left\{
	\begin{aligned}	
		L  &= \widehat{\overline{\mu} \widetilde{w}_i \widetilde{w}_i} - \widehat{\overline{\mu}} \widehat{\widetilde{w}}_i \widehat{\widetilde{w}}_i, \\
		M &= \widehat{\overline{\rho}}\widehat{\overline{K}}_{utke}^\frac{3}{2}/{\widehat{\overline{\Delta}}}.
	\end{aligned}
\right.
\end{equation}
To model the dilational dissipation rate, $C_{\epsilon d}$ is obtained by $C_{\epsilon d} = L/M$, where
\begin{equation}\label{dynamic_ceps_d}
\left\{
	\begin{aligned}	
		L &= 4[\reallywidehat{\overline{\mu} \widetilde{U}_{k,k}^2}-\widehat{\overline{\mu}} \widehat{\widetilde{U}}_{k,k}^2]/3, \\
		M &= \widehat{\overline{\rho}}^2 \widehat{\overline{K}}_{utke}^\frac{5}{2}/({\gamma \widehat{\overline{p}} \, \widehat{\overline{\Delta}}}).
	\end{aligned}
\right.
\end{equation}
\begin{rmk}
	For DCIT, the global dynamic coefficient is obtained through the ensemble average in the whole computational domain.
	In terms of TCPML, the local dynamic coefficient is obtained through the plane averaging in the streamwise and spanwise directions.
\end{rmk}
\begin{rmk}
	When modeling SGS pressure-dilation transfer, the dynamic denominator in $C_{\Pi}$ may approach to a pretty tiny value, which cause a large value for model coefficient $C_{\Pi}$. 
	To smoothen the spurious behavior of $C_{\Pi}$, the $C_{\Pi}$ is limited in a bound $[-0.1, 0.1]$ to guarantee the numerical robustness.
	When $C_{\Pi}$ goes beyond the bound, it would be modified as $C_{\Pi} = C_{\Pi}^{lim} \boldsymbol{\cdot} sign(C_{\Pi})$, where $C_{\Pi}^{lim} = 0.05$ and $sign(\cdot)$ is the sign function.
	Both for DCIT and TCPML, numerical tests indicate that the key statistical quantities are not sensitive to the $C_{\Pi}^{lim}$, as long as the numerical simulations are stable.
\end{rmk}
\begin{rmk}
	For macroscopic varialbe $\beta$, the filtered macroscopic variable and Favre-filtered variable are denoted as $\overline{\beta}$ and $\tilde{\beta}$  in equations \eqref{rhok_les} - \eqref{chai_sgs_closure3} and Appendix \ref{appB}, respectively.
	By default, macroscopic variable $\beta$ in the rest part of this paper, such as the derivation of macroscopic governing equations from NTRKM in Appendix \ref{appA} represents resolved value without additional filtering symbols.
	Since resolved conservative variables on unresolved grids can be regarded as the filtered conservative variables, the corresponding Favre-filtered primitive variables also can be obtained.
	In summary, the \textbf{resolved variables} in current paper can be treated as the customary \textbf{filtered variables} equivalently.
\end{rmk}

\section{Connection between macroscopic variables and mesoscopic coefficients}\label{appC}
The connection between the spatial derivatives of macroscopic flow variables and the expansion of turbulence equilibrium distribution function $f^{eq}$ reads
\begin{equation}\label{micro_coefficients} 
	\frac{\partial \boldsymbol{Q}}{\partial x_i}
	=\int \frac{\partial f^{eq}}{\partial x_i}\boldsymbol{\psi}\text{d}\Xi
	\equiv \int af^{eq}\boldsymbol{\psi}\text{d}\Xi,
\end{equation}
where $a$ denotes the spatial mesoscopic coefficients in equation \eqref{formalsolution_neq} as
\begin{equation}
	a=\boldsymbol{a}^T \boldsymbol{\psi}=a_1+ a_2 u_1 + a_3 u_2 + a_4 u_3 + a_5 [\frac{1}{2}(u_1^2 + u_2^2 + u_3^2 + \xi^2) + k_u] + a_6 k_u.
\end{equation} 
Equation \eqref{micro_coefficients} can be rewritten into following linear system 
\begin{equation}\label{micro_coefficients_matrix} 
	\frac{1}{\rho}\frac{\partial \boldsymbol{Q}}{\partial x_i} =\Big(\frac{1}{\rho}\int\boldsymbol{\psi} \otimes \boldsymbol{\psi}^T f^{eq} \text{d}\Xi\Big)\boldsymbol{a}\triangleq\boldsymbol{M} \boldsymbol{a},
\end{equation}
Each component of $(a_1,…,a_6)^T$ in equation \eqref{micro_coefficients_matrix} can be determined uniquely 
\begin{equation}
	\left\{
	\begin{aligned}
		a_6 &= \frac{4}{\rho K_{utke}^2} B_5 - a_5, \\
		a_5 &= \frac{8 \lambda^2}{\rho (N + 3)} (B_4 -  U_1 B_1 - U_2 B_2 - U_3 B_3 - B_5), \\
		a_4 &= \frac{2 \lambda}{\rho} B_3 - U_3 a_5,  \\
		a_3 &= \frac{2 \lambda}{\rho} B_2 - U_2 a_5, \\
		a_2 &= \frac{2 \lambda}{\rho} B_1 - U_1 a_5, \\
		a_1 &= \frac{1}{\rho} \frac{\partial \rho}{\partial x_i} - U_1 a_2 - U_2 a_3 - U_3 a_4 - E a_5 - K_{utke} a_6,
	\end{aligned}
	\right.
\end{equation}
where
\begin{equation}
	\left\{
	\begin{aligned}
		B_1 &= \frac{\partial (\rho U_1)}{\partial x_i} - U_1 \frac{\partial \rho}{\partial x_i},\\
		B_2 &= \frac{\partial (\rho U_2)}{\partial x_i} - U_2 \frac{\partial \rho}{\partial x_i},\\
		B_3 &= \frac{\partial (\rho U_3)}{\partial x_i} - U_3 \frac{\partial \rho}{\partial x_i},\\
		B_4 &= \frac{\partial (\rho E)}{\partial x_i} - E \frac{\partial \rho}{\partial x_i},\\
		B_5 &= \frac{\partial (\rho K_{utke})}{\partial x_i} - K_{utke} \frac{\partial \rho}{\partial x_i}.
	\end{aligned}
	\right.
\end{equation}
For the temporal mesoscopic coefficient in equation \eqref{formalsolution_neq}, the relation between temporal derivatives of macroscopic variables and turbulence equilibrium distribution can be written as
\begin{equation}\label{micro_coefficients_temp} 
	\frac{\partial \boldsymbol{Q}}{\partial t}=\int \frac{\partial f^{eq}}{\partial t}\boldsymbol{\psi}\text{d}\Xi
	\equiv \int Af^{eq}\boldsymbol{\psi}\text{d}\Xi,
\end{equation}
where 
\begin{equation}
	A = \boldsymbol{A}^T \boldsymbol{\psi}=A_1+ A_2 u_1 + A_3 u_2 + A_4 u_3 + A_5 [\frac{1}{2}(u_1^2 + u_2^2 + u_3^2 + \xi^2) + k_u] + A_6 k_u.
\end{equation} 
The temporal derivatives of macroscopic variables can be given according to the compatibility condition as
\begin{equation}
	\begin{aligned}
		\int (\frac{\partial{f^{eq}}}{\partial t} + u_i \frac{\partial f^{eq}}{\partial x_i}) \boldsymbol{\psi} \text{d}\Xi = \boldsymbol{0}.
	\end{aligned}
\end{equation}
In a similar way, the above components $(A_1,…,A_6)^T$ in equation \eqref{micro_coefficients_temp} can be determined uniquely.

\section{Eigenstructure for characteristic reconstruction}\label{appD}
The Jacobian matrix $\boldsymbol{J}$ for the flux $\boldsymbol{F}(\boldsymbol{Q})$ of the hyperbolic part in equations \eqref{tgks_macro_formula1}-\eqref{tgks_macro_formula4} is given by
\begin{equation}\label{jacobian}
	\begin{aligned}	
		&\boldsymbol{J} =
		\begin{pmatrix}
			0  &1 &0 &0 &0 &0\\
			\frac{\hat{\gamma}}{2}\boldsymbol{U^2} - U_1^2  &(3 - \gamma)U_1 &-\hat{\gamma}U_2 &-\hat{\gamma}U_3 &\hat{\gamma} &-\hat{\gamma} \\
			-U_1 U_2 &U_2 &U_1 &0 &0 &0\\
			-U_1 U_3 &U_3 &0 &U_1 &0 &0\\
			U_1(\frac{\gamma - 2}{2}\boldsymbol{U^2} - \frac{c^2}{\gamma - 1} - K_{utke}) &H - \hat{\gamma}U_1^2 + K_{utke} &-\hat{\gamma}U_1 U_2 &-\hat{\gamma}U_1 U_3 &\gamma U_1 &-\hat{\gamma} U_1\\
			-U_1 K_{utke}  &K_{utke} &0 &0 &0 &U_1\\
		\end{pmatrix},
	\end{aligned}
\end{equation}
where $H = \boldsymbol{U^2}/2 + c^2/(\gamma - 1)$, and $\hat{\gamma} = \gamma - 1$. 
The eigenvalues of the quasi one-dimensional system are
\begin{equation}
	\lambda_1 = U_1 - c, \quad \lambda_2 = U_1, \quad \lambda_3 = U_1, \quad\lambda_4 = U_1, \quad\lambda_5 = U_1, \quad \lambda_6 = U_1 + c,
\end{equation}
where $c$ is the local sound speed.
The matrix of corresponding right eigenmatrix is
\begin{equation}
	\begin{aligned}	
		\boldsymbol{R}  =
		\begin{pmatrix}
			1  &1 &0 &0 &0 &1\\
			U_1 - c  &U_1 &0 &0 &0 & U_1 + c \\
			U_2 &U_2 &1 &0 &0 &U_2\\
			U_3 &U_3 &0 &1 &0 &U_3\\
			H - U_1 c + K_{utke} &\frac{1}{2}\boldsymbol{U^2} &U_2 &U_3 &1 &H + U_1c + K_{utke}\\
			K_{utke}  &0 &0 &0 &1 &K_{utke}\\
		\end{pmatrix},
	\end{aligned}
\end{equation}
and the inverse matrix of $\boldsymbol{R}$ is given by
\begin{equation}
	\begin{aligned}	
		\boldsymbol{R}^{-1} = \frac{\hat{\gamma}}{2 c^2}
		\begin{pmatrix}
			\frac{1}{2}\boldsymbol{U^2} + \frac{U_1 c}{\hat{\gamma}}  &-(U_1 + \frac{c}{\hat{\gamma}}) &-U_2 &-U_3 &1 &-1\\
			-\boldsymbol{U^2} + \frac{2 c^2}{\hat{\gamma}}  &2U_1 &2U_2 &2U_3 &-2 &2 \\
			-\frac{2 U_2 c^2}{\hat{\gamma}} &0 &\frac{2 c^2}{\hat{\gamma}} &0 &0 &0\\
			-\frac{2 U_3 c^2}{\hat{\gamma}} &0 &0 &\frac{2 c^2}{\hat{\gamma}} &0 &0\\
			-\boldsymbol{U^2} K_{utke}   &2 U_1 K_{utke} &2 U_2 K_{utke}  &2 U_3 K_{utke}  &-2 K_{utke} &2 K_{utke} + \frac{2 c^2}{\hat{\gamma}}\\
			\frac{1}{2}\boldsymbol{U^2} - \frac{U_1 c}{\hat{\gamma}}  &-U_1 + \frac{c}{\hat{\gamma}} &-U_2 &-U_3 &1 &-1
		\end{pmatrix}.
	\end{aligned}
\end{equation}

\bibliographystyle{jfm}
\bibliography{caogybib}
\end{document}